%% file: paper.tex
\documentclass[11pt]{article}

\usepackage{verbatim}
\usepackage{amsmath}
\usepackage{amsfonts}
\usepackage{amssymb}
\usepackage{bbm}
\usepackage{latexsym}
\usepackage{lscape} 
\usepackage{epsfig}
\usepackage{color}
\usepackage{graphicx}

\usepackage{subfig}
\usepackage{mathbbol}
\usepackage{slashbox}
\usepackage{multirow}
\usepackage{mathrsfs}

\usepackage{amscd}
\usepackage{cite}

\input{texdefs}

\begin{document}
\newcommand{\stretcharry}{1.3}

\thispagestyle{empty}

\begin{flushright}
LMU-ASC 49/10 
\end{flushright}
\vskip 1cm
\begin{center}
{\Large {\bf Heterotic MSSM on a Resolved Orbifold}}
\\[0pt]

\bigskip
\bigskip 
{\large
{\bf Michael Blaszczyk$^{a,}$\footnote{
{{ {\ {\ {\ E-mail: michael@th.physik.uni-bonn.de}}}}}}},
{\bf Stefan Groot Nibbelink$^{b,}$\footnote{
{{ {\ {\ {\ E-mail: Groot.Nibbelink@physik.uni-muenchen.de}}}}}}},
{\bf Fabian Ruehle$^{a,}$\footnote{
{{ {\ {\ {\ E-mail: ruehle@th.physik.uni-bonn.de}}}}}}},\\
{\bf Michele Trapletti$^{a,}$\footnote{
{{ {\ {\ {\ E-mail: mtraplet@th.physik.uni-bonn.de}}}}}}} 
{\bf and} 
{\bf Patrick K.S.~Vaudrevange$^{b,}$\footnote{
{{ {\ {\ {\ E-mail: Patrick.Vaudrevange@physik.uni-muenchen.de}}}}}}
\bigskip }\\[1ex]
${}^a$ {\it Bethe Center for Theoretical Physics and
Physikalisches Institut der Universit\"at Bonn, 
\mbox{Nussallee 12}, 53115 Bonn, Germany
}\\[1ex] 
${}^b$ {\it 
Arnold-Sommerfeld-Center for Theoretical Physics,  
Department f\"ur Physik, Ludwig-Maximilians-Universit\"at M\"unchen, 
Theresienstra\ss e 37, 80333 M\"unchen, Germany 
\\[1ex]  }
}
\bigskip
\end{center}
%
%
%
%
%
%
%
%
%
%
%
%
%
%
%
%
%
%
%
%
%
%
%
%
%
%
%
%
%
%
%
%
%
%
\thispagestyle{empty} 

\vskip 1cm

\subsection*{\centering Abstract}

We construct an MSSM with three generations from the heterotic string compactified on a smooth 6D internal manifold using Abelian gauge fluxes only. The compactification space is obtained as a resolution of the $T^6/\ztwo \times \zf$ orbifold. The \zf\ involution of such a resolution breaks the $\text{SU}(5)$ GUT group down to the SM gauge group using a suitably chosen (freely acting) Wilson line. Surprisingly, the spectrum on a given resolution is larger than the one on the corresponding orbifold taking into account the branching and Higgsing due to the blow-up modes. The existence of extra resolution states is closely related to the fact that the resolution procedure is not unique. Rather, the various resolutions are connected to each other by flop transitions.

\newpage

\setcounter{page}{1}
\setcounter{footnote}{0} 

\tableofcontents

\section{Introduction}
\label{ch:Introduction}


One of the central objectives of string phenomenology is to obtain the Standard Model (SM) or its minimal supersymmetric extension (MSSM) from a consistent string construction. In the context of the heterotic string this requires compactification of six extra dimensions on a Calabi-Yau (CY) space \cite{Candelas:1985}, or on some singular limit thereof (e.g. on an orbifold \cite{Dixon:1985, Dixon:1986}). In the case of a CY manifold a stable gauge bundle \cite{Donaldson:1985,Uhlenbeck:1986} has to be constructed, breaking the \eeight gauge group\footnote{The equally interesting possibility offered by the $\text{SO}(32)$ heterotic string will be neglected in the present paper.} to some unified gauge group like $\text{SU}(5)$, and a suitable four dimensional chiral spectrum has to be obtained \cite{Andreas:1999ty,Donagi:2000zs}. The unified gauge group can subsequently be broken to the SM group by the embedding of a freely acting CY involution into the gauge degrees of freedom \cite{Candelas:1985en,Witten:1985xc,Donagi:2000fw}. 
The gauge bundle stability and the existence of a suitable involution on a given CY are highly non-trivial requirements \cite{Donagi:2000zf}, and therefore only a few MSSM-like models have been built up to now in this way, see e.g. \cite{Braun:2005,Bouchard:2005}.


A major limitation of a smooth CY construction comes from the fact that the quantization of the heterotic string on this compactification space is essentially out of reach, as this requires the ability of quantizing an interacting conformal field theory (CFT) with an explicitly known CY metric. Therefore, one is forced to consider a supergravity (SUGRA) truncation of the full string theory where some of the stringy effects are encoded in $\ga'$ corrections to the SUGRA Lagrangian. Moreover, even the SUGRA field equations lead to back-reactions that introduce a non-trivial background for the antisymmetric field driving the geometry away from being CY \cite{Strominger:1986uh}.


The complete string dynamics is fully computable only at very special points of enhanced symmetries where an exact CFT description is known. Orbifolds \cite{Dixon:1985, Dixon:1986} are one of the most commonly considered examples of this, because they still allow for a straightforward geometrical interpretation. (Other examples are Gepner models \cite{Dijkstra:2004cc,Dijkstra:2004ym,GatoRivera:2009yt,GatoRivera:2010gv} and free fermionic constructions \cite{Faraggi:1989ka,Kiritsis:2008ry}.) Being exact CFTs the consequences of stringy consistency conditions, most notably modular invariance, can be worked out and solved. The full string spectra and, in particular, the ones of the low energy effective theory in four dimensions can be determined systematically \cite{Ibanez:1986,Ibanez:1987sn,Ibanez:1987pj}. Recently, phenomenologically guided search strategies combined with computer assisted searches have resulted in a large class of MSSM-like models \cite{Buchmuller:2005,Buchmuller:2006,Lebedev:2006kn,Lebedev:2008}.


Even though orbifolds provide a fully consistent approach to heterotic string phenomenology, one should be aware that they only correspond to very special (non-generic) points in moduli space. To see the bigger picture around (and far from) an orbifold point, the singular geometry should be deformed to a smooth space, the so-called resolution or blow-up. In particular, when the orbifold theory provides a so-called Fayet-Iliopoulos (FI) D-term, this is essentially unavoidable, because at least some fields in the effective theory need to acquire non-vanishing vacuum expectation values (VEVs) to cancel the FI D-term leading to a (partial) blow-up. For the reasons mentioned above, this suggests that one should pass from the CFT description to the SUGRA approximation of string theory. In addition, in this way one might have more control on moduli stabilization issues in orbifold inspired models. Reversely, a SUGRA construction would greatly benefit from having an exact CFT description in some specific moduli limit: For example, one can clarify the full string consistency of the construction and in principle can have access to the complete (i.e.\ not only massless) spectrum and the interactions of the theory. Therefore, it is clear that having control on a model both from a CFT and a SUGRA perspective would greatly improve both pictures.


In most MSSM-like orbifold models the GUT group is broken locally by the embedding of the orbifold action into the gauge degrees of freedom. The result is that in full blow-up, where a SUGRA description applies, the GUT group is broken by gauge fluxes inducing an anomalous hypercharge \cite{SGN:2009}. This does not necessarily spoil the nice phenomenological properties of the orbifold models studied in \cite{Buchmuller:2005,Buchmuller:2006,Lebedev:2006kn,Lebedev:2008}, because an FI D-term does not enforce a full blow-up of all singularities. As long as those singularities on which only SM charged states live are not resolved, the hypercharge remains non-anomalous and hence unbroken. However, such models can clearly not be considered from a purely SUGRA perspective.

\subsection*{Model building on a resolved orbifold with an involution}

With this in mind we are ready to state the main objectives of the present paper. We would like to obtain MSSM-like models from the heterotic string in the SUGRA regime which allow for an exact orbifold CFT description at a specific point in moduli space.
In order to avoid having an anomalous and hence broken hypercharge the GUT breaking of $\text{SU}(5)$ to the SM gauge group should not be achieved by some gauge flux on the resolution, but rather by the embedding of some geometrical, freely acting involution into the gauge degrees of freedom, as mentioned above. Moreover, since the breaking is delocalized in the internal space, the scale of GUT breaking is set by the compactification moduli, and thus is decoupled from the string scale. This alleviates the known hierarchy problem between the Planck and the GUT symmetry breaking scale in orbifold models, see \cite{Witten:1996mz} and \cite{Hebecker:2003we,Hebecker:2004ce}. For this reason we choose to study the resolved $T^6/\ztwo$ geometry, which is equipped with such an $\zf$ involution \cite{Donagi:2008}. In this paper we show that this involution can be maintained in the blow-up. Concretely, we investigate resolutions of an orbifold CFT model with an MSSM-like spectrum, similar to the one introduced recently in \cite{Blaszczyk:2009in}, see also \cite{Kappl:2010inprep}.


The fact that we have an underlying exact CFT orbifold description has some important consequences. First of all we can investigate the extra truly ``stringy'' consistency requirements on the Wilson line associated with the involution \cite{Blaszczyk:2009in}. This means that, contrary to conventional CY model building, we are guaranteed that the SUGRA construction also lifts to full string theory. In addition, we can make use of the well-developed classification machinery for heterotic orbifolds to systematically search for viable models and determine their spectra and interactions.


As the construction of stable bundles is a major obstacle in heterotic SUGRA model building, we are primarily concerned with Abelian gauge fluxes in this paper. The requirement of bundle stability simplifies considerably for purely Abelian bundles, because they do not contain subsheaves. Hence, the stability condition boils down to the condition of vanishing D-terms for the $\text{U}(1)$'s along which the line bundles are embedded in the \eeight algebra \cite{Blumenhagen:2005pm,Blumenhagen:2005}. These D-term cancellations require specific relations between the volumes of the divisors on which the gauge fluxes are wrapped. Alternatively, additional charged fields may take non-vanishing VEVs and deform the Abelian bundle to a non-Abelian one.\footnote{In addition these VEVs could have great relevance in the generation of hierarchically suppressed Yukawa couplings \cite{Anderson:2010tc}: The Yukawa couplings forbidden by a $\text{U}(1)$ symmetry could still be present but suppressed when this symmetry is anomalous.} Flux quantization and the integrated Bianchi identities furnish stringent consistency requirements on the Abelian gauge fluxes resulting in relations between the Abelian gauge fluxes on the resolution and the gauge shifts and Wilson lines on the orbifold. Using these relations we obtain novel MSSM models on CY manifolds that are resolved orbifolds.


The identification of the Abelian gauge fluxes with the embedding of the orbifold action into the gauge degrees of freedom (i.e. with shifts and Wilson lines) is a particular example of a matching of properties between orbifolds and their resolutions. In table \ref{tb:matching} we give an overview of various matching pairs which we have been able to investigate to a certain level. However, an exact matching of CFT and SUGRA models is complicated. One reason is that the $T^6/\ztwo$ singularities admit many topologically inequivalent resolutions, so that there is a huge number of corresponding smooth CY's for a given orbifold, i.e. of the order $\sim10^{33}$, due to the choice of triangulation at each $\ztwo$ singularity. Also on the orbifold side there seems to be some level of arbitrariness, as models characterized by gauge shifts and Wilson lines that differ only by lattice vectors can nevertheless result in models with different chiral spectra \cite{Ploger:2007} (which can be reformulated by introducing generalized discrete torsion). Due to this we have to leave some of the details of the identification of matching pairs to future work.

\begin{table}[t] 
\centering
\begin{tabular}{|p{5.5cm}cp{5.5cm}|p{2.5cm}|}
\hline
Orbifold & & Resolution & Reference \\
\hline
\hline
Gauge embedding of space group (shifts and Wilson lines)         & $\Leftrightarrow$ & flux quantization condition 
&  subsection \ref{sc:LineBundleOrbiShift}, appendix \ref{sc:FluxQuant} \\
\hline
Shifted (gauge) momentum $P_{sh}$ of blow-up mode & $\Leftrightarrow$ & local $\text{U}(1)$ gauge background $V_{k,\rho\sigma}$ 
& subsection \ref{sc:OpenIssues}
 \\
\hline
Mass-equation for massless blow-up mode & $\Rightarrow$     & Bianchi identity  
& ref.\ \cite{SGN:2009} 
\\
$P_{sh}^2=\frac{3}{2}$                  &                   & $\sum\limits_{\rho,\sigma=1}^{4} V_{k,\rho\sigma}^{2} = 24$ & \\
\hline
(Abelian) D-flatness                  & $\approx$         & Donaldson-Uhlenbeck-Yau equations: 
 $\text{vol}(E_{r})\, V_{r} = \gx_{1L}$      
 & subsection \ref{sc:DUY}
 \\
\hline
Vacuum Expectation Value (VEV)  of blow-up mode              & $\Leftrightarrow$ & K\"ahler modulus $b_r$                 
& ref.\ \cite{GrootNibbelink:2007ew} 
\\
\hline
Gauge group after VEVs                   & $\Leftrightarrow$ & 
Abelian gauge flux commutant                   & \\
(including anomalous $\text{U}(1)$'s)   &                   & (including anomalous $\text{U}(1)$'s)  
& 
\\
\hline
Twisted matter spectrum after field redefinition                   & $\approx$         & Matter from \eeight gauginos                   
& subsection \ref{sc:NovelStates}, 
ref.\ \cite{GrootNibbelink:2007ew} 
\\
\hline
\hline
\multicolumn{4}{|l|}{speculative} \\
\hline
Ratio of VEVs at intersection of three fixed tori& $\stackrel{?}{\Leftrightarrow}$ & Triangulation of the $\ztwo$ singularity                             
& subsection \ref{sc:OpenIssues}
\\
\hline
F-flatness $F=0$                                   & $\stackrel{?}{\Leftrightarrow}$ & Bianchi identity + triangulation & \\
\hline
\end{tabular}
\caption{\label{tb:matching}
This table gives an overview of various properties of orbifolds and their resolutions that can be matched in detail. The bottom part of the table gives identifications which are of speculative nature at present and require further investigations.}
\end{table}

\subsection*{Paper overview}

The paper is organized as follows: 
In section \ref{ch:Het_Sugra_on Resolution} we discuss the geometry of the $T^6/\ztwo$ orbifold and of its resolutions, and investigate the consequences of modding out the $\zf$ involution. 
In particular we describe the resolutions as hypersurfaces of particular toric varieties. 
Section \ref{ch:Preliminaries} reviews the general properties of SUGRA compactifications on resolved orbifolds with line bundles and explains how the chiral spectra can be computed in these cases. 
The various results are illustrated with examples of resolutions with the same choice of triangulation at all 64 $\ztwo$ singularities. 
In section \ref{sec:Relation_Orbifold_CY} we describe the matching between orbifold and resolution models. In particular we show that the input data of smooth line bundle models and the one of heterotic orbifold CFTs are identical up to lattice vectors. Furthermore, we illustrate the surprising but generic result that the massless chiral spectra on resolutions can be larger than the ones obtained from the orbifold CFT. 
Section \ref{ch:MSSM_in_Blowup} contains the main results of this work where we give an explicit example of a MSSM-like resolution model. The conclusions of this work can be found in section \ref{sec:Conclusions}.

The appendices provide further background information and more technical results. 
In particular, in appendix \ref{sc:Weierstrass} we review how one can use the Weierstrass function to describe the torus as an elliptic curve. 
Appendix \ref{sc:NonFact} provides a different perspective on the resolution of the $T^6/\ztwo\times\zf$ orbifold, namely that it can be obtained as the resolution of a non-factorizable $\ztwo$ orbifold. 
Using this it is shown that the spectrum on a resolution of $T^6/\ztwo \times \zf$ is effectively the one on the corresponding resolution of $T^6/\ztwo$ divided by two. 
Appendix \ref{sc:FluxQuant} shows that the 48 vectors that characterize the embedding of line bundles in the \eeight gauge group are determined in terms of only eight independent ones (two shifts and six Wilson lines), up to lattice vectors. 
Finally, in appendix \ref{ch:Derivation_MI_Free_WL} we describe the partition function of $\ztwo$ orbifold models that admit a $\zf$ involution and derive a novel modular invariance condition on the $\zf$ associated Wilson line. In addition, we argue that the $T^6/\ztwo\times \zf$ orbifold CFT has additional twisted winding modes. 

%
%
%
%
%
%
%
%
%
%
%
%
%
%
%
%
%
%
%
%
%
%
%
%
%
%
%
%
%
%
%
%
%
%
\section{The resolved orbifold}
\label{ch:Het_Sugra_on Resolution}

\begin{figure}[t]
\centering
\includegraphics[width=0.85\textwidth]{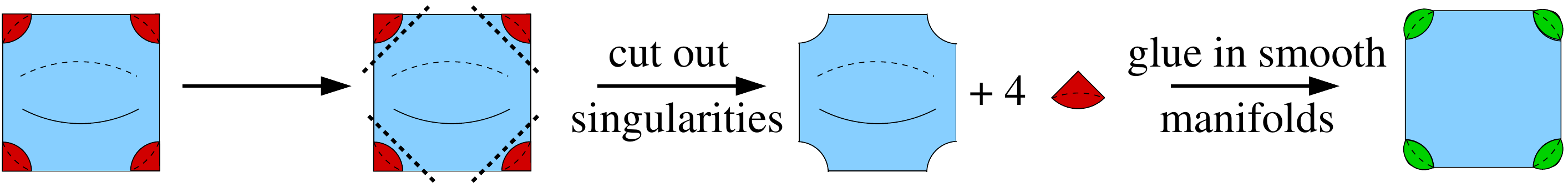}
\caption{Outline of the blow-up procedure. The orbifold singularities 
(red) are cut out and replaced by smooth manifolds (green).
\label{fig:Blowup_Construction}}
\end{figure}

In this section we lay the foundation of the present paper by giving a general introduction to the toroidal orbifold $T^6/\ztwo$ geometry \cite{Dixon:1985,Dixon:1986}. Furthermore, we describe the resolution procedure that smoothens out the singularities of the orbifold space: we cut out the singularities and glue smooth hypersurfaces (divisors) into the orbifold instead, yielding a smooth manifold. The idea is schematically illustrated in figure \ref{fig:Blowup_Construction}. Many details of the resolution procedure have already been worked out in \cite{Held:2009, Lust:2006, Reffert:2006, SGN:2007, SGN:2007xq, SGN:2008, SGN:2009}, so we only briefly recall the specific elements needed for the discussion of the resolution of $T^6/\ztwo$. To give further motivation for these methods we recall that this orbifold allows for a concrete description using the representation of two-tori as elliptic curves, as e.g.\ used in \cite{Vafa:1994}.

The essential tool to investigate the local resolutions is toric geometry, see \cite{Fulton}. A basic mathematical background is provided in \cite{Nakahara, Griffiths}. First we explain how to resolve local fixed points of non-compact $\Cplx^3/\ztwo$ orbifolds.  After that we explain how we can make use of the description of two-tori as elliptic curves to combine the local information we gained at each \ztwo fixed point to obtain a resolved compact orbifold using linear equivalence relations. With this at hand, we calculate the intersection numbers of the divisors, Hodge numbers and Chern classes of the resolution. By introducing the K\"ahler form also the volumes of the resolution and of its hypersurfaces can be determined.

We complete this section by introducing a freely-acting $\zf$ involution on the $T^6/\ztwo$ orbifold, which under certain conditions also descends to the resolution.

\subsection[The $T^6/\ztwo$ orbifold]{The $\boldsymbol{T^6/\ztwo}$ orbifold}
\label{sec:Orbifolds}

The $T^6/\ztwo$ orbifold can be introduced in the following steps: First, we generate a six-dimensional torus $T^{6} :=\mathbbm{C}^{3} / \Gamma_{\rm fac}$ by choosing a (factorizable) lattice $\Gamma_{\rm fac}$ that is spanned by six orthonormal basis vectors $e_{p}$, $p = 1,\ldots 6$, i.e.
\begin{equation}
\label{eq:Torus_Lattice}
\Gamma_{\rm fac} = \Big\{\sum\limits_{p=1}^6 n^{p} e_{p},~n^{p} \in \mathbbm{Z}\Big\}\;. 
\end{equation}
In particular we assume that the torus factorizes as $T^{6}=T^{2} \otimes T^{2} \otimes T^{2}$. A fundamental domain can be chosen as 
\equ{
T^6 = \Big\{ \sum\limits_{p=1}^6
x_p \, e_p, ~0 \leq x_p < 1 \Big\}\;.
}
The complex coordinate $z_i = x_{2i-1}+i x_{2i} \in\mathbbm{C}$ of the $i$-th two-torus has periodicities coming from the lattice identifications. Next, we define the discrete symmetry group $\ztwo \equiv \{\Id,\gth_1,\gth_2,\gth_3 \}$, including the identity element and the ``twist'' elements $\gth_1$, $\gth_2$ and $\gth_3=\gth_1\gth_2$, with the action
\begin{equation}
\label{eq:Z2xZ2Action}
\theta_i:\quad  (z_{1},z_{2},z_{3}) \mapsto 
(e^{2 \pi i (\varphi_{i}){}^{1}}\, z_{1},
e^{2 \pi i (\varphi_{i}){}^{2}}\, z_{2},
e^{2 \pi i (\varphi_{i}){}^{3}}\, z_{3})\;,
\end{equation}
with
\begin{equation}
\label{eq:Z2_TwistVectors}
\varphi_{1} = \left(0,\frac{1}{2},-\frac{1}{2}\right),~~
\varphi_{2} = \left(-\frac{1}{2},0,\frac{1}{2}\right),\quad\text{and}\quad
\varphi_{3} = \varphi_{1} + \varphi_{2} = \left(-\frac{1}{2},\frac{1}{2},0\right). 
\end{equation}
The $\ztwo$ action fulfills the CY condition  ensuring $\mathcal{N}=1$ supersymmetry in four dimensions. The orbifold $T^6/\ztwo$ is obtained by dividing out the $\ztwo$ point group from the torus $T^6$.

Orbifolds are singular spaces with singularities at the fixed points of the orbifold action. To analyze these singularities it is convenient to define space group elements $g = (\gth, l)$ as combinations of rotations $\gth \in \{\Id, \gth_1,\gth_2,\gth_3\}$ and lattice translations $l \in \gG_{\rm fac}$. In this approach, the orbifold is defined as $\mathbb{C}^{3}$ with equivalence relations $z \sim gz$ for all $g$. In more detail, a space group element $g$ acts on $z \in \mathbb{C}^{3}$ as
\begin{equation}
gz = (\theta, l) z \defi \theta \,z+l\;.
\end{equation}
This orbifold action has (singular) fixed sets $z_f = g z_f$ for non-trivial rotations $\gth \neq \Id$. As the action of $\gth_{i}$ leaves the $i$-th coordinate invariant while acting non-freely on the other two coordinates, the orbifold singularities of $\gth_i$ are fixed tori. The combinations of indices $(\gb,\gg)$, $(\ga,\gg)$ and $(\ga,\gb)$ (with $\ga, \gb, \gg = 1,\ldots,4$) denote the locations of the fixed tori in the $(z_2,z_3)$, $(z_1,z_3)$ and $(z_1,z_2)$ complex planes associated to the orbifold elements $\gth_1$, $\gth_2$ and $\gth_3$, respectively. In table~\ref{tb:Z2_sectors} we indicate the fixed tori structure of the \ztwo orbifold on the covering space $T^6$ by specifying the lattice shifts
\equ{
\gth_1:~~l_{1,\gb\gg} = n^p_{1,\gb\gg} e_p\;, 
\qquad 
\gth_2:~~l_{2,\ga\gg} = n^p_{2,\ga\gg} e_p\;, 
\qquad 
\gth_3:~~l_{3,\ga\gb} = n^p_{3,\ga\gb} e_p\;, 
\label{Def_n}
}
that bring the fixed tori back to themselves after the associated $\gth_i$ rotation, e.g. $\gth_1 z_f + l_{1,\gb\gg} = z_f$. Note that two fixed tori of different twists intersect at points in the orbifold. There are 64 of such intersection points within the orbifold, called \ztwo fixed points and labeled by $(\ga,\gb,\gg)$. Even though these \ztwo fixed points do not correspond to any single space group element, they play a crucial role in the resolution process.

The volumes of the orbifold and its hypersurfaces can be determined straightforwardly from those of the underlying torus. We denote the four-cycle obtained by setting one complex coordinate $z_i$ of the orbifold equal to a fixed value $c_i$ by $R_i$, i.e.\ on the covering space $T^6$ it is given by ${R_i = \{z_i = c_i\} \cup \{z_i = - c_i\}}$ due to the $\ztwo$ action. Consequently, $R_i R_j$ with $j\neq i$ defines a two dimensional real hypersurface within the orbifold, i.e.\ the union of four hypersurfaces in the covering space $T^6$. For simplicity we fix the complex structure of $T^6$ such that it factorizes in three square two-tori with radii $r_i$. The volumes of the orbifold and its hypersurfaces are equal to those of the underlying torus divided by the order of the orbifold group, taking into account that on the covering space $R_i$ consists of two disjoint connected parts: 
\equ{
\text{vol}(R_iR_j) = r_k^2\;, 
\qquad
\text{vol}(R_i) = \frac 12 \, r_j^2r_k^2\;, 
\qquad 
\text{vol}(T^6/\ztwo) = \frac 14\, r_1^2r_2^2r_3^2\;, 
\label{Volumes_Orbifold} 
}
where $i\neq j \neq k$. For example $\text{vol}(R_1 R_2) = r_3^2$ because in the covering space both $R_1$ and $R_2$ consists of two parts that get mapped to each other under the orbifold action, but the whole third torus remains.

\begin{table}[p!]
\begin{center} 
\hskip1.5cm
\raisebox{1cm}{\Large $\gth_1$-sector:} 
\qquad  \includegraphics[width=0.5\textwidth]{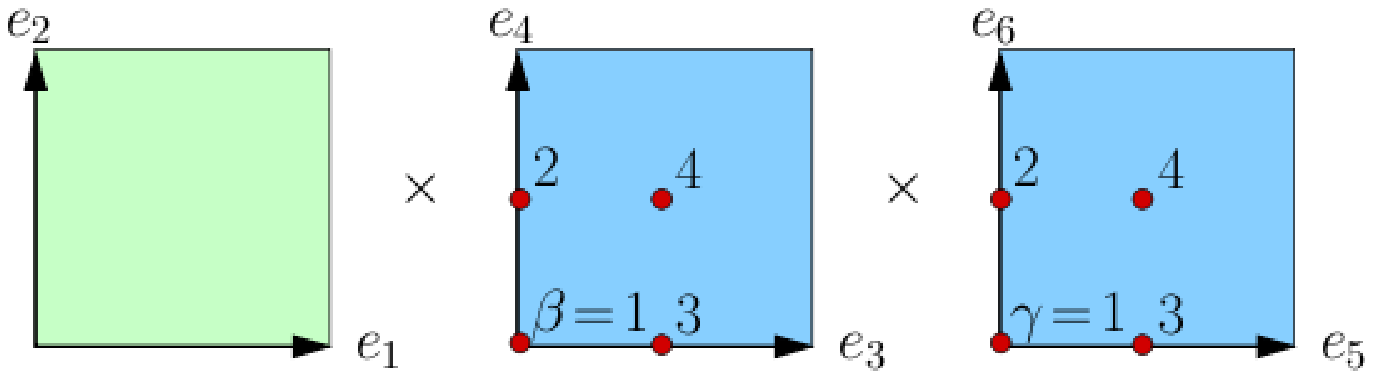}\vspace{0.5cm} 
\\[-1ex] 
\begin{tabular}{|c||c|c|c|c|}
\hline
\multicolumn{5}{|c|}{torus shifts $l_{1,\beta\gamma}$ in the $\theta_{1}$-sector} \\
\hline
\backslashbox{$\beta$}{$\gamma$} & $1$ & $2$ & $3$ & $4$ \\
\hline\hline 
$1$ & $0$ 			& $e_{6}$ 				& $e_{5}$ 				& $e_{5}+e_{6}$ \\
$2$ & $e_{4}$ 		& $e_{4}+e_{6}$ 			& $e_{4}+e_{5}$ 			& $e_{4}+e_{5}+e_{6}$ \\
$3$ & $e_{3}$ 		& $e_{3}+e_{6}$ 			& $e_{3}+e_{5}$ 			& $e_{3}+e_{5}+e_{6}$ \\
$4$ & $e_{3}+e_{4}$ 	& $e_{3}+e_{4}+e_{6}$ 	& $e_{3}+e_{4}+e_{5}$ 	& $e_{3}+e_{4}+e_{5}+e_{6}$ \\
\hline
\end{tabular}
\\[2ex] 
\hskip1.5cm 
\raisebox{1cm}{\Large $\gth_2$-sector:} 
\qquad  \includegraphics[width=0.5\textwidth]{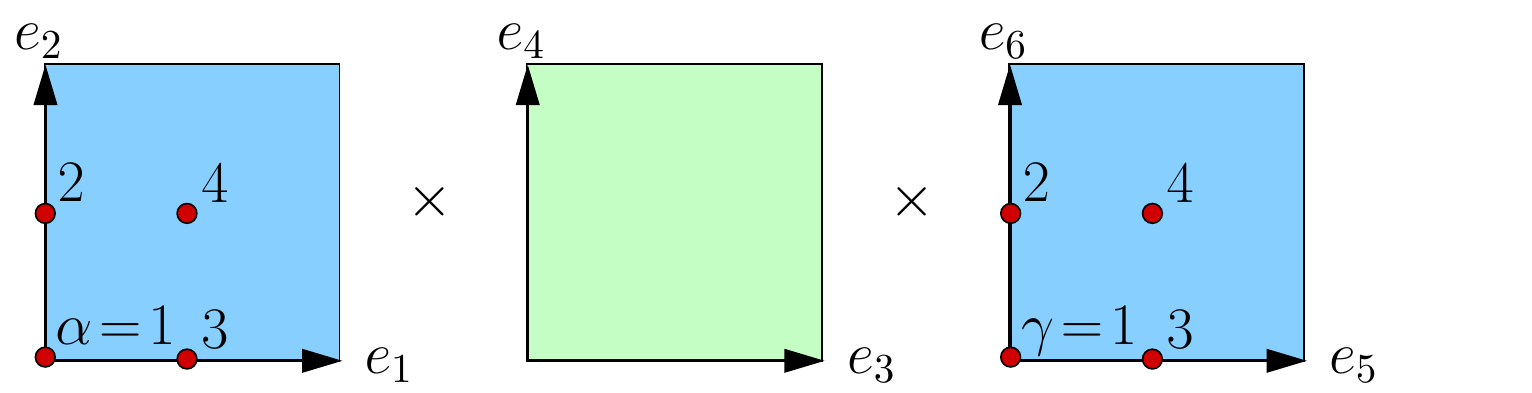}\vspace{0.5cm}
\\[-1ex] 
\begin{tabular}{|c||c|c|c|c|}
\hline
\multicolumn{5}{|c|}{torus shifts $l_{2,\alpha\gamma}$ in the $\theta_{2}$-sector} \\
\hline
\backslashbox{$\alpha$}{$\gamma$} & $1$ & $2$ & $3$ & $4$ \\
\hline\hline 
$1$ & $0$ 			& $e_{6}$ 				& $e_{5}$ 				& $e_{5}+e_{6}$ \\
$2$ & $e_{2}$ 		& $e_{2}+e_{6}$ 			& $e_{2}+e_{5}$ 			& $e_{2}+e_{5}+e_{6}$ \\
$3$ & $e_{1}$ 		& $e_{1}+e_{6}$ 			& $e_{1}+e_{5}$ 			& $e_{1}+e_{5}+e_{6}$ \\
$4$ & $e_{1}+e_{2}$ 	& $e_{1}+e_{2}+e_{6}$ 	& $e_{1}+e_{2}+e_{5}$ 	& $e_{1}+e_{2}+e_{5}+e_{6}$ \\
\hline
\end{tabular}
\\[2ex] 
\hskip1.5cm 
\raisebox{1cm}{\Large $\gth_3$-sector:} 
\qquad  \includegraphics[width=0.5\textwidth]{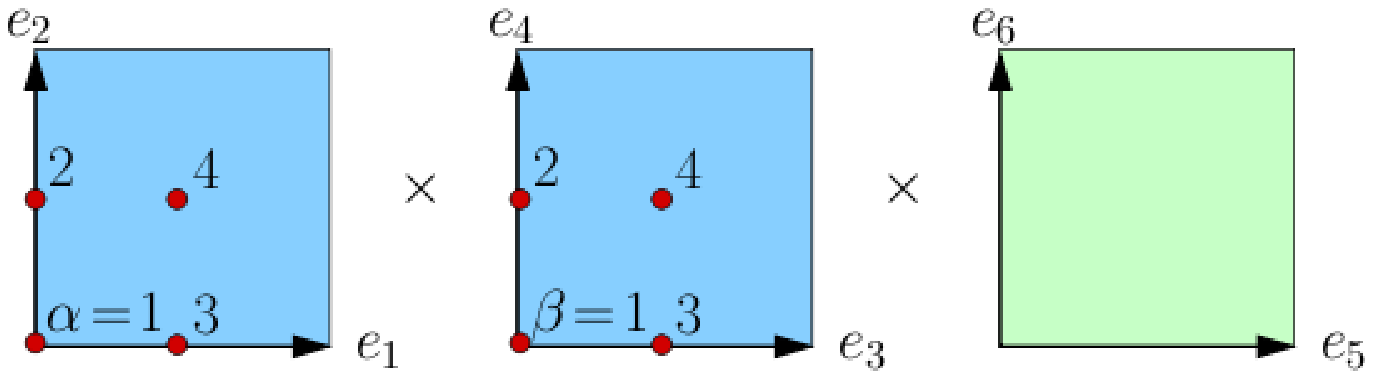}\vspace{0.5cm}
\begin{tabular}{|c||c|c|c|c|}
\hline
\multicolumn{5}{|c|}{torus shifts $l_{3,\alpha\beta}$ in the $\theta_{1}\theta_{2}$-sector} \\
\hline
\backslashbox{$\alpha$}{$\beta$} & $1$ & $2$ & $3$ & $4$ \\
\hline\hline 
$1$ & $0$ 			& $e_{4}$ 				& $e_{3}$ 				& $e_{3}+e_{4}$ \\
$2$ & $e_{2}$ 		& $e_{2}+e_{4}$ 			& $e_{2}+e_{3}$ 			& $e_{2}+e_{3}+e_{4}$ \\
$3$ & $e_{1}$ 		& $e_{1}+e_{4}$ 			& $e_{1}+e_{3}$ 			& $e_{1}+e_{3}+e_{4}$ \\
$4$ & $e_{1}+e_{2}$ 	& $e_{1}+e_{2}+e_{4}$ 	& $e_{1}+e_{2}+e_{3}$ 	& $e_{1}+e_{2}+e_{3}+e_{4}$ \\
\hline
\end{tabular}
\end{center} 
\caption{This table gives the fixed points of the three sectors, $\gth_1$, $\gth_2$ and $\gth_3$, and the corresponding torus shifts $l_{1,\beta\gamma}$, $l_{2,\alpha\gamma}$ and $l_{3,\alpha\beta}$, respectively, see equation \eqref{Def_n}. 
\label{tb:Z2_sectors} 
}
\end{table}

\subsection[The $T^6/\ztwo\times\zf$ orbifold]{The $\boldsymbol{T^6/\ztwo\times\zf}$ orbifold}

The $T^6/\ztwo$ orbifold admits a freely-acting involution $\zf$, generated by the translation 
\begin{equation}
\tau = \frac{1}{2}(e_2 + e_4 + e_6)\;,
\end{equation}
over half a lattice vector in all three imaginary directions simultaneously. 
Since $2\tau = e_2 + e_4 + e_6 \in \Gamma_{\rm fac}$, $\tau$ has order 2. As this is a symmetry of the $T^6/\ztwo$ orbifold, we may divide it out. In addition, there are no new fixed points corresponding to the new element $\tau$ or a combination of $\tau$ and the other space group elements, i.e. $\tau$ acts freely. Instead, the 48 fixed tori of the $T^6/\ztwo$ orbifold are identified pairwise resulting in 24 fixed tori in the $T^6/\ztwo\times\zf$ case. Comparing to table~\ref{tb:Z2_sectors}, we can choose them to correspond to the $8+8+8=24$ torus shifts $l_{1,\beta\gamma}$ (with $\beta=1,3$ and $\gamma=1,\ldots,4$), $l_{2,\alpha\gamma}$ (with $\alpha=1,3$ and $\gamma=1,\ldots,4$) and $l_{3,\alpha\beta}$ (with $\alpha=1,3$ and $\beta=1,\ldots,4$).

\subsection[Local resolution of the non-compact $\Cplx^3/\ztwo$ orbifold]{Local resolution of the non-compact $\boldsymbol{\Cplx^3/\ztwo}$ orbifold}
\label{sc:CZ22Res}

We now focus on the resolution of a single $\Cplx^3/\ztwo$ singularity. In the resolution process one identifies two kinds of two-dimensional complex hypersurfaces: ordinary divisors $D_i$, that are also present on the orbifold, and exceptional divisors $E_j$, that hide inside the singularity. In the following we will discuss how these divisors arise and how their intersection properties can be read off from the toric diagram. 

Starting from the non-compact \ztwo orbifold with coordinates $z = (z_1, z_2, z_3) \in \mathbbm{C}^{3}$ of the covering space, one can define new, so-called local, coordinates $(u_1, u_2, u_3)$ that are invariant under the orbifold action (\ref{eq:Z2xZ2Action}). We choose
\begin{equation}
(u_1, u_2, u_3) = (z_1^2, z_2^2, z_3^2)\;.
\end{equation}
For each coordinate $z_i$ one associates a three-dimensional vector $\mathrm{v}_i$ that specifies these local coordinates in the following way: The $j$-th component $(\mathrm{v}_i)_j$ of the vector $\mathrm{v}_i$ is given by the exponent of $z_i$ for the local coordinate $u_j=z_1^{(\mathrm{v}_1)_j} z_2^{(\mathrm{v}_2)_j} z_3^{(\mathrm{v}_3)_j}$, i.e.
\begin{equation}
\label{eq:Z2_Ordinary_Divisor_Positions}
\mathrm{v}_{1}=\left(\begin{array}{c} 2\\0\\0\end{array}\right),~
\mathrm{v}_{2}=\left(\begin{array}{c} 0\\2\\0\end{array}\right),~
\mathrm{v}_{3}=\left(\begin{array}{c} 0\\0\\2\end{array}\right). 
\end{equation}
Next, we define ordinary divisors by $D_i := \{z_i = 0\}$ and associate the vector $\mathrm{v}_i$ to the divisor $D_i$. 

\begin{figure}[t]
\centering 
\tabu{cc}{ 
\subfloat[\label{fig:Z2_Triangulation1}\footnotesize{Diagrams for triangulation ``$E_1$".}]
{
\includegraphics[width=0.16\textwidth]{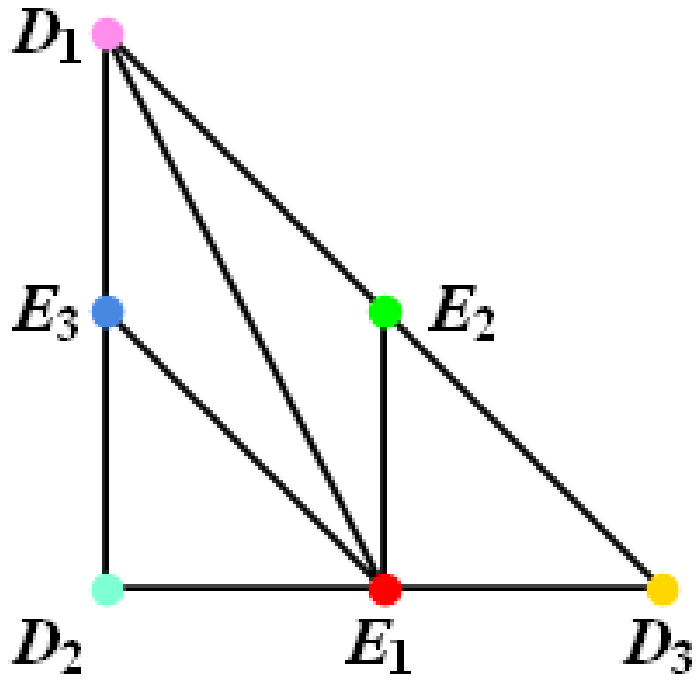}\hskip.6cm
\raisebox{2ex}{
\includegraphics[width=0.2\textwidth]{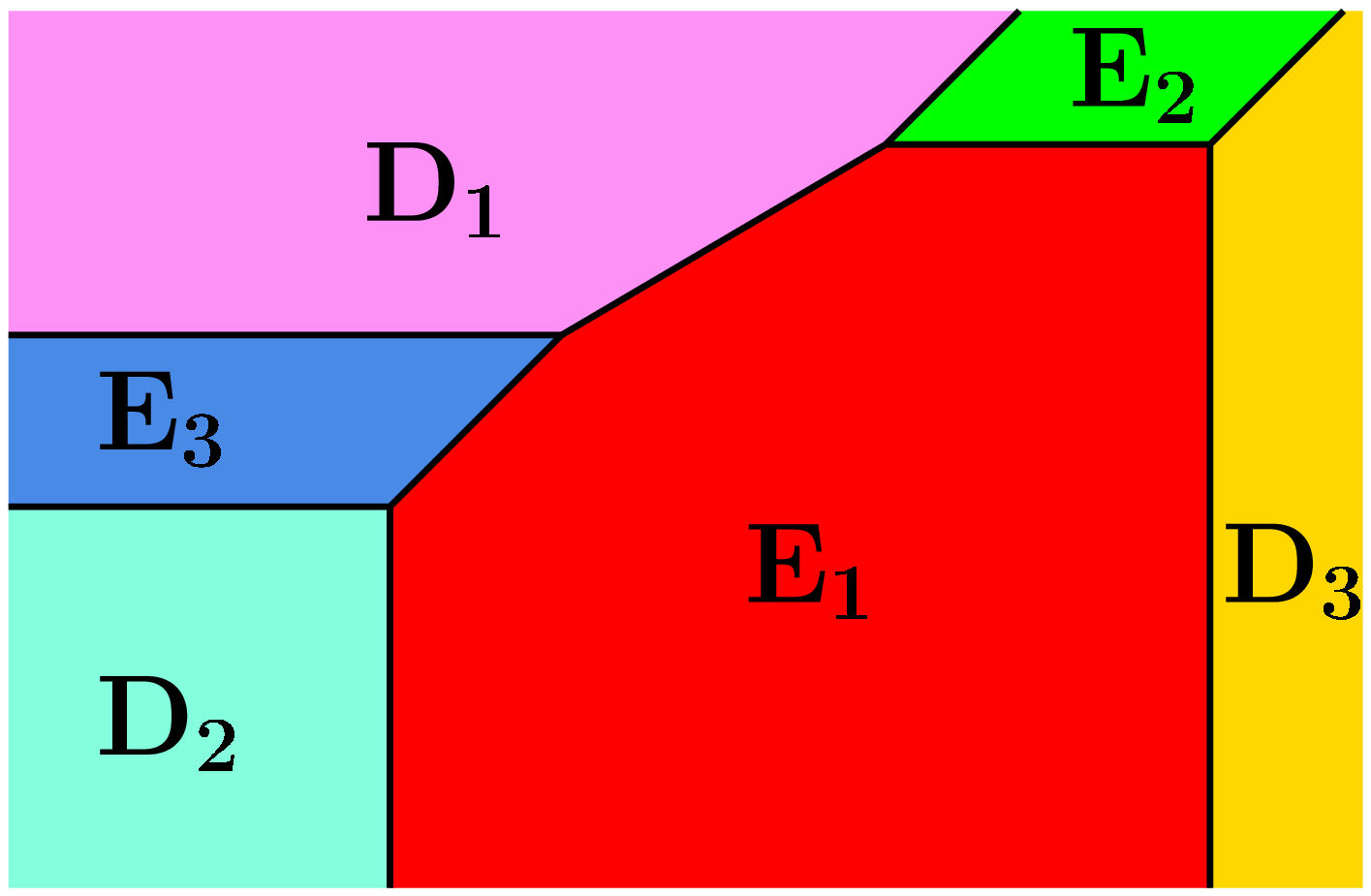}}
}
\qquad & \qquad 
\subfloat[\label{fig:Z2_Triangulation2}\footnotesize{Diagrams for triangulation ``$E_2$".}]
{
\includegraphics[width=0.16\textwidth]{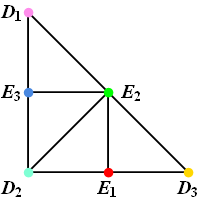}\hskip.6cm
\raisebox{2ex}{\includegraphics[width=0.2\textwidth]{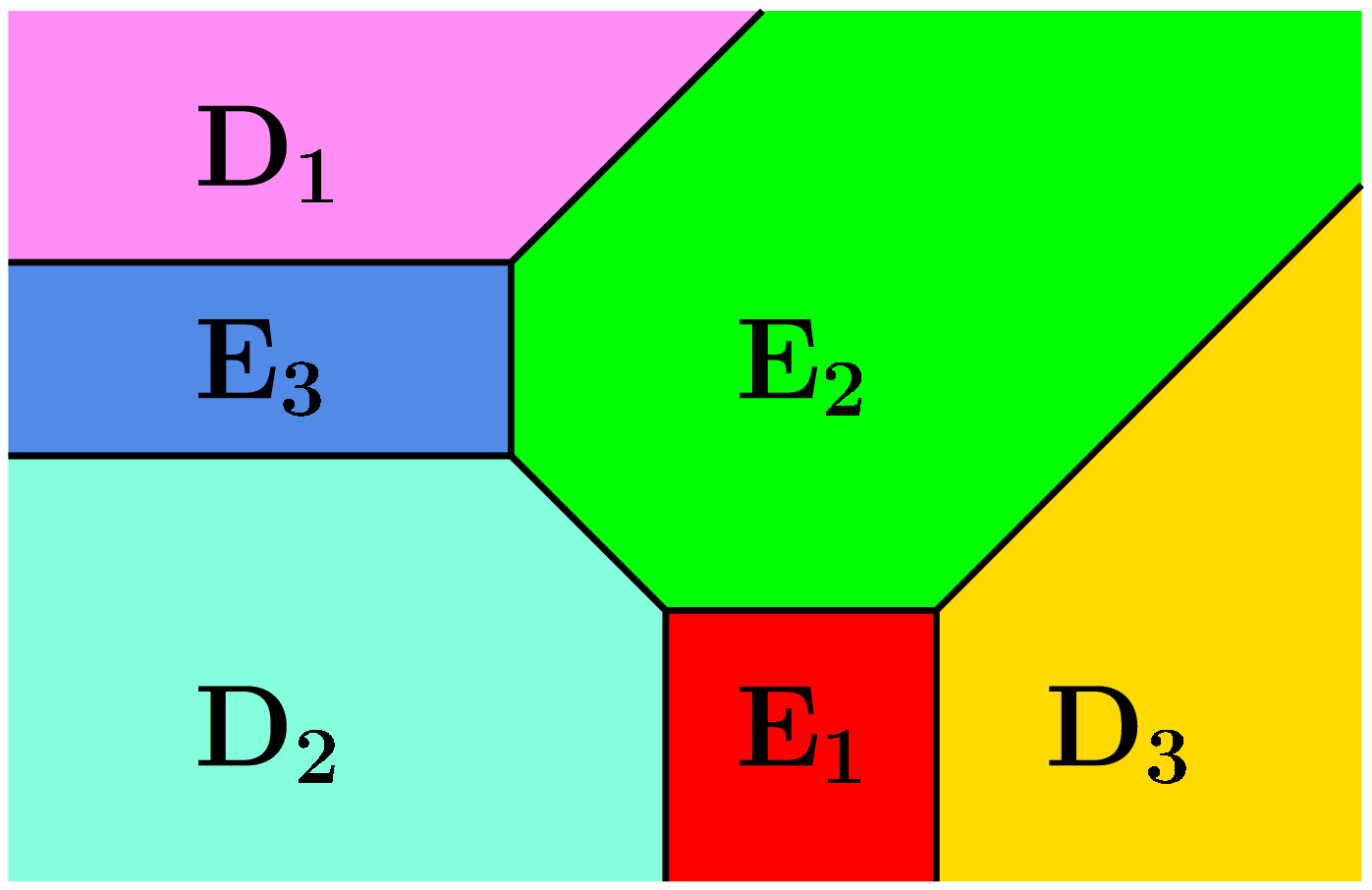}}
}
\\[2ex] 
\subfloat[\label{fig:Z2_Triangulation3}\footnotesize{Diagrams for triangulation ``$E_3$".}]
{
\includegraphics[width=0.16\textwidth]{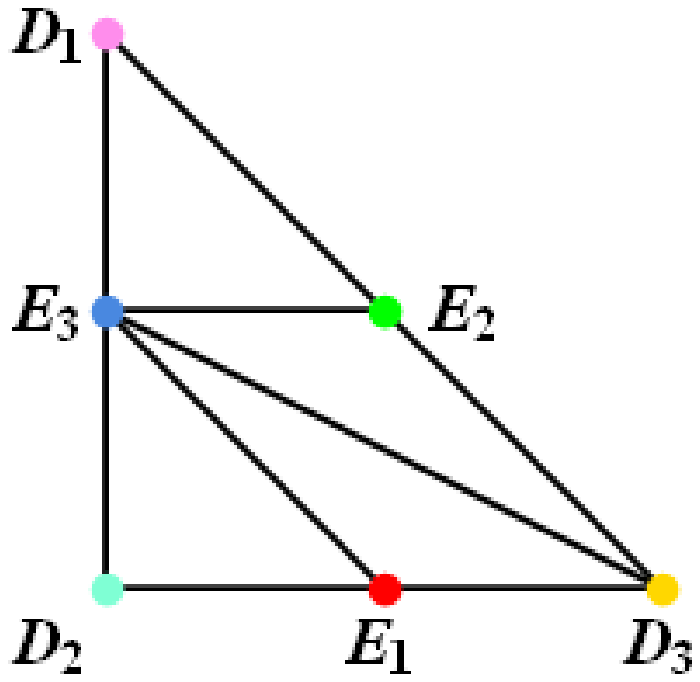}\hskip.6cm
\raisebox{2ex}{\includegraphics[width=0.2\textwidth]{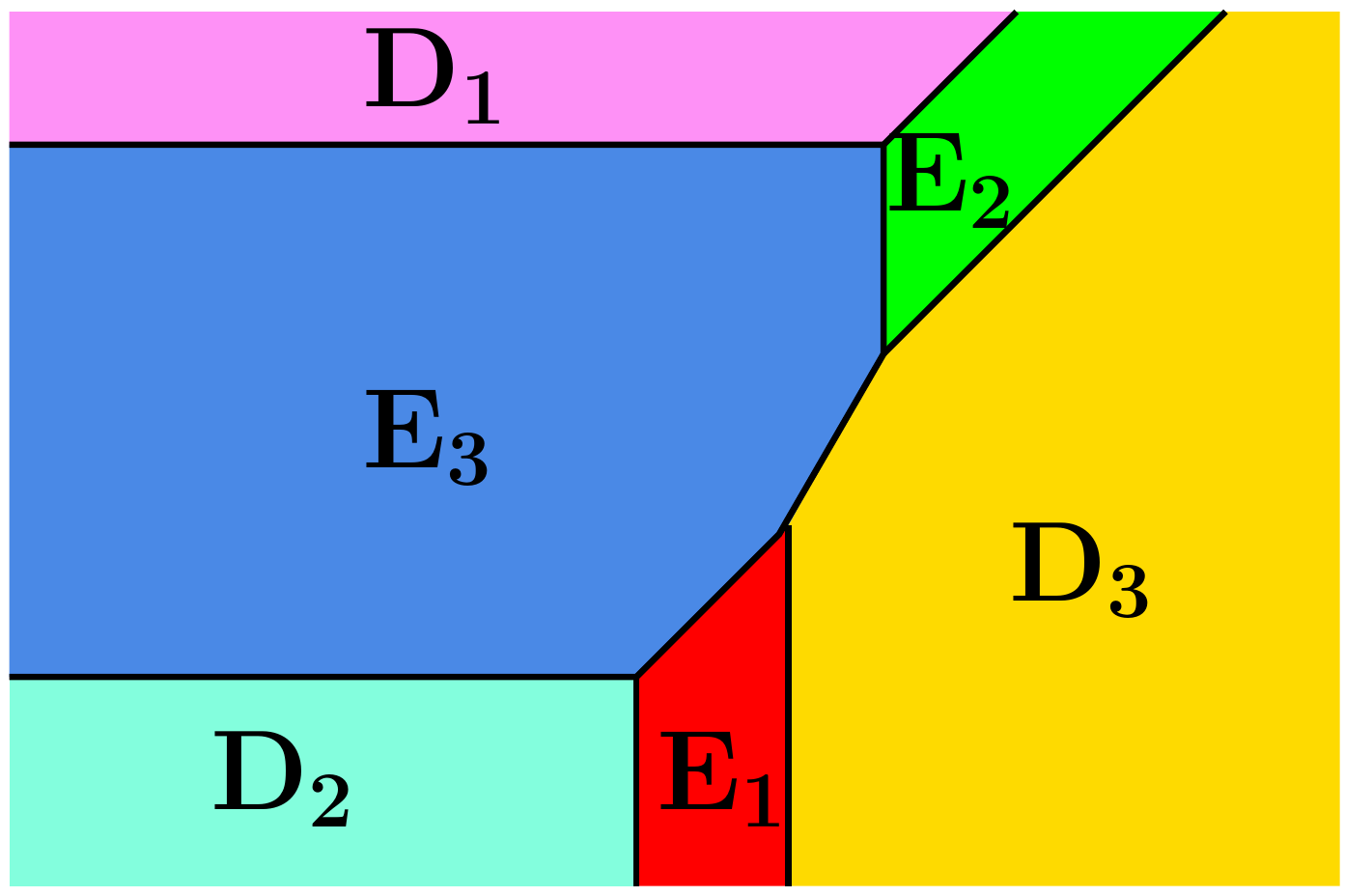}}
}
\qquad & \qquad 
\subfloat[\label{fig:Z2_Triangulation4}\footnotesize{Diagrams for triangulation ``$S$".}]
{
\includegraphics[width=0.16\textwidth]{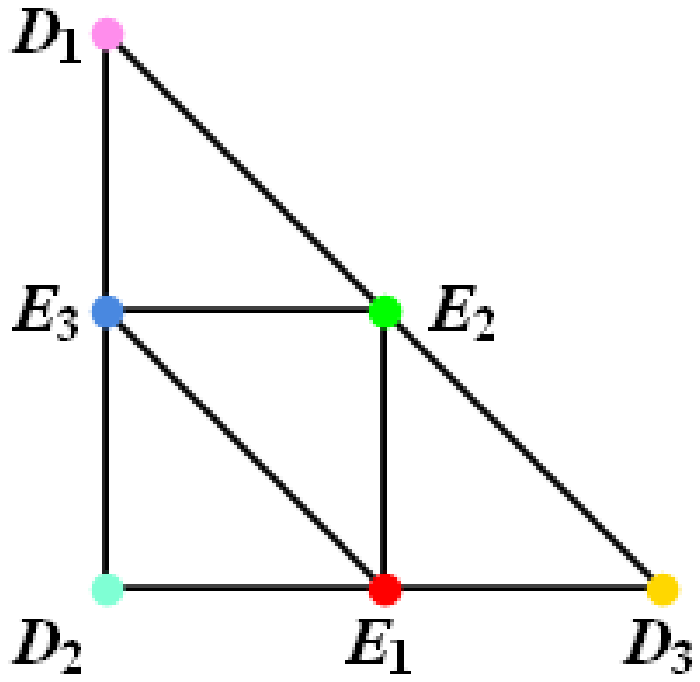}\hskip.6cm
\raisebox{2ex}{\includegraphics[width=0.2\textwidth]{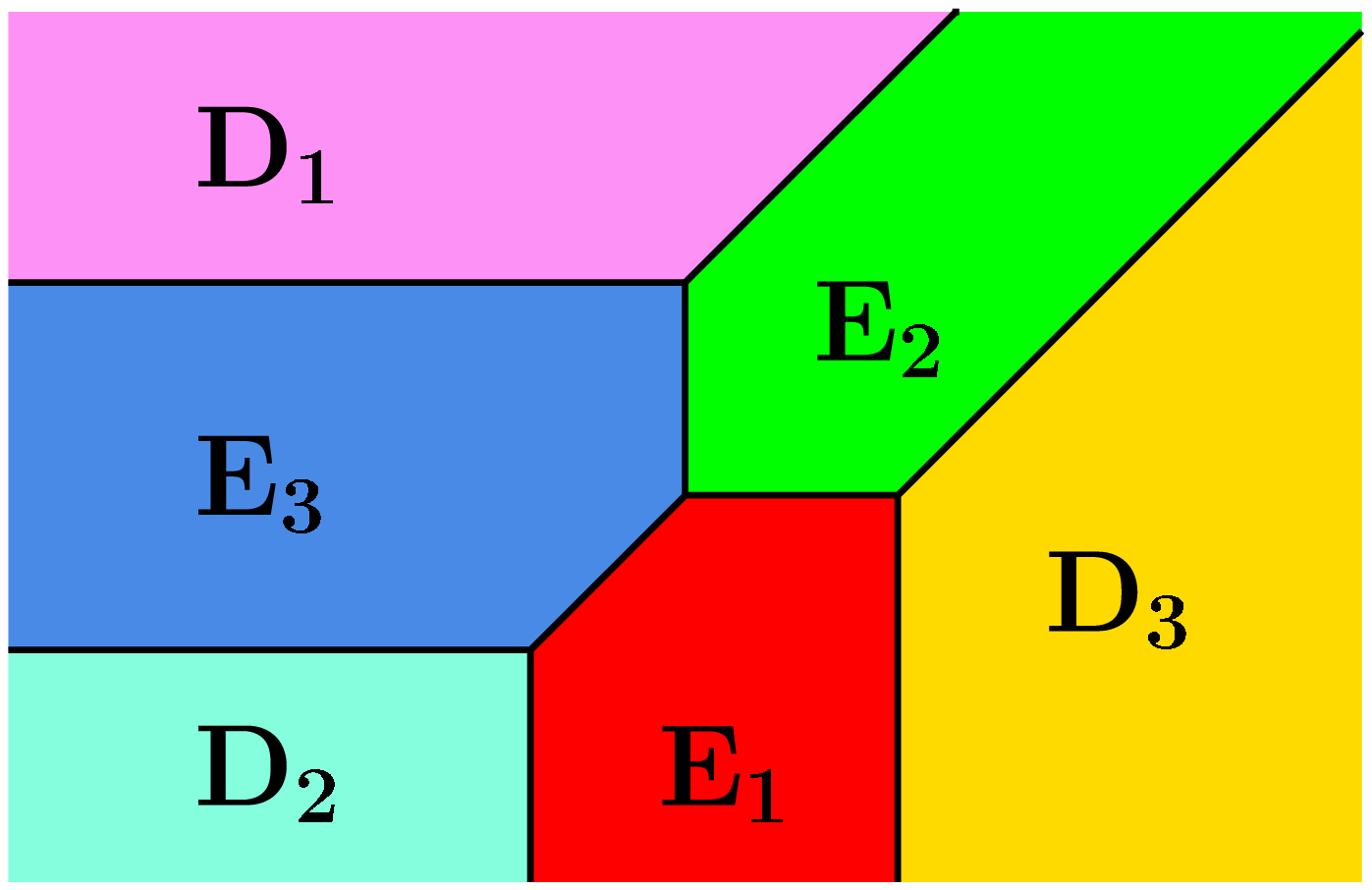}}
}
}
\caption{The toric diagrams and their dual web graphs for the four triangulation possibilities of a $\mathbbm{C}^{3}/\ztwo$ fixed point are depicted.}
\label{fig:Z2_Noncompact_Toric_Diagrams}
\end{figure}

To construct the resolution of $\Cplx^3/\ztwo$ we need to include additional complex coordinates $x_k$, $k=1,2,3$, together with three $\Cplx^* := \Cplx-\{0\}$ actions
\begin{equation}
\label{eq:Cstaraction}
(z_1, z_2, z_3, x_1, x_2, x_3) \sim (\lambda_1\lambda_3 z_1, \lambda_1\lambda_2 z_2, \lambda_2\lambda_3 z_3, \lambda_2^{-2} x_1, \lambda_3^{-2} x_2, \lambda_1^{-2} x_3)\;,
\end{equation}
for $\lambda_i \in \mathbbm{C}^*$, $i=1,2,3$. Starting from the homogeneous coordinates $(z_1, z_2, z_3, x_1, x_2, x_3) \in \mathbbm{C}^6$ and imposing three $\mathbbm{C}^*$ actions one obtains a complex three-dimensional toric variety -- the resolved $\Cplx^3/\ztwo$ orbifold. At $x_k \neq 0$, one can use the $\mathbbm{C}^*$ actions $\lambda_1 = \pm \sqrt{x_3}$, $\lambda_2 = \pm \sqrt{x_1}$ and $\lambda_3 = \pm \sqrt{x_2}$ to set the additional coordinates $x_k$ to $1$, i.e. $(z_1, z_2, z_3, 1, 1, 1)$. Due to the possible choice of $\pm 1$ in these $\mathbbm{C}^*$ actions, a residual \ztwo action remains. Hence, for $x_k \neq 0$ the resolution looks like the original $\Cplx^3/\ztwo$ orbifold. We define exceptional divisors $E_k := \{x_k = 0\}$ which are hiding inside the orbifold singularity. As these divisors are smooth spaces, the $\Cplx^3/\ztwo$ singularity has been resolved. Each divisor $E_k$ is associated to a vector $\mathrm{w}_k$ that is calculated according to 
\begin{equation}
\label{eq:CY_Exceptional_Divisors}
\mathrm{w}_k = (\tgvf_k)^j \, \mathrm{v}_j\;, 
\end{equation} 
where $(\tilde\varphi_i)^j = (\varphi_i)^j \mod 1$, such that $0 \leq (\tilde\varphi_i)^j < 1$, with $\varphi_i$ given in \eqref{eq:Z2_TwistVectors}. This ensures that the vectors ($\mathrm{v}_i; \mathrm{w}_i)$, $i=1,2,3$, are orthogonal to the scalings defined in \eqref{eq:Cstaraction}. This gives
\begin{equation}
\label{eq:Z2_Exceptional_Divisor_Positions}
\mathrm{w}_{1}=\left(\begin{array}{c} 0\\ 1\\1\end{array}\right)\;,~~
\mathrm{w}_{2}=\left(\begin{array}{c} 1\\ 0\\1\end{array}\right)\;,~~
\mathrm{w}_{3}=\left(\begin{array}{c} 1\\ 1\\0\end{array}\right) 
\end{equation}
for the choice \eqref{eq:Z2_Ordinary_Divisor_Positions}, e.g. for $(\mathrm{v}_1;\mathrm{w}_1)$, and for $\lambda_1$-scaling we have $(2,0,0;0,1,1)\cdot (1,1,0;0,0,-2) = 0$. Having specified the ordinary and exceptional divisors, $D_i$ and $E_k$, the toric diagram is characterized by the associated vectors $\mathrm{v}_i$ and $\mathrm{w}_k$.  The solutions (\ref{eq:Z2_Ordinary_Divisor_Positions}) and (\ref{eq:Z2_Exceptional_Divisor_Positions}) are chosen such that all vectors $\mathrm{v}_i$ and $\mathrm{w}_k$ lie within a single plane. This ensures that the CY condition is fulfilled. As a consequence, one can draw the projection of the toric diagram in two dimensions, see figure \ref{fig:Z2_Noncompact_Toric_Diagrams} for a schematic illustration. Similar to the singular orbifold case, the vectors $\mathrm{v}_i$ and $\mathrm{w}_k$ specify how one can obtain local coordinates on the resolutions out of the homogeneous coordinates
\equ{
\label{inv_mon}
u_{j} = \prod_{i,k}  z_{i}^{(\mathrm{v}_{i})_{j}} x_{k}^{(\mathrm{w}_{k})_{j}}\;,
}
which, with the choice \eqref{eq:Z2_Ordinary_Divisor_Positions}, reduces to 
\equ{ 
\label{inv_mon_concrete} 
u_1 =  z_1^2 x_2 x_3\;,
\qquad 
 u_2 =  z_2^2 x_1 x_3\;,\quad\text{and} \quad u_3 = z_3^2 x_1 x_2\;.
}
The three independent $\Cplx^*$-scalings under which the homogeneous coordinates $z_i$ and $x_k$ transform are chosen such that the local coordinates $u_j$ are invariant, cf. equation~(\ref{eq:Cstaraction}). As the local coordinates $u_i$ are $\Cplx^*$-invariant, i.e.\ trivial line bundle sections, one infers the following linear equivalences between the ordinary and exceptional divisors \cite{Lust:2006}
\begin{equation}
0 \sim 2D_{1} + E_{2} + E_{3}\;,\hskip5mm
0 \sim 2D_{2} + E_{1} + E_{3}\;,\hskip5mm
0 \sim 2D_{3} + E_{1} + E_{2}\;.\hskip5mm
\label{eq:Z2_Noncompact_Equivalences}
\end{equation}
These equivalence relations encode part of the topological properties of the toric variety.

The other crucial part in the specification of the topology of a resolution is provided by the set of intersection numbers. The intersection numbers of the divisors are fixed by triangulating the toric diagram, which is done by introducing lines between all divisors such that no lines cross each other and that no additional line that does not cross another line can be added. Note that in the language of toric geometry, the triangulation specifies the so-called exclusion set, necessary for the $\Cplx^*$ actions to be free of fixed sets and thus the toric variety to be non-singular. For the \ztwo\ singularity this procedure is ambiguous: There are four independent triangulations of the toric diagram, i.e. there are four distinct ways of drawing these lines. The corresponding four projected toric diagrams and the dual graphs are given in figure \ref{fig:Z2_Noncompact_Toric_Diagrams}.

\subsubsection{Auxiliary polyhedra}
\label{sect:auxpoly}
\begin{figure}[t]
\centering
\subfloat[\label{fig:Z2_Auxiliary_Polyhedron1}\footnotesize{Triangulation ``$E_1$".}]
{
  \includegraphics[width=0.20\textwidth]{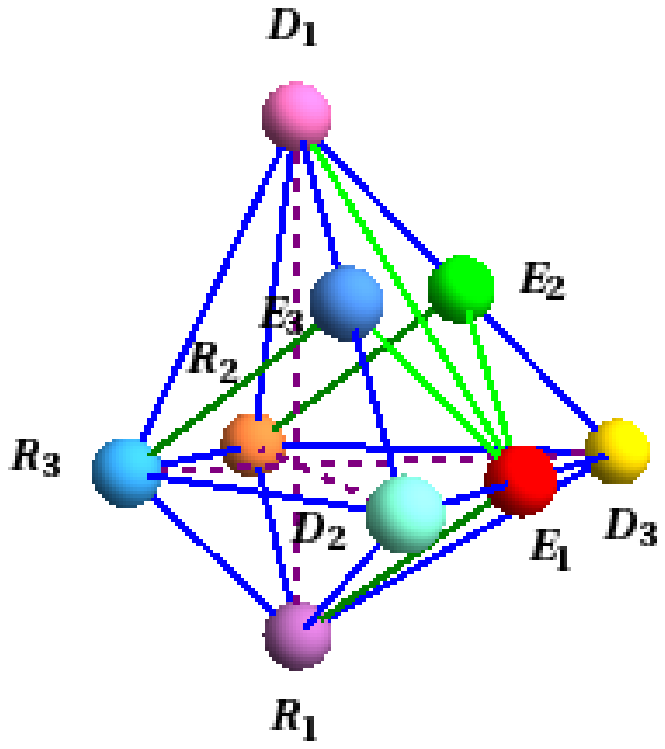}
}
\qquad 
\subfloat[\label{fig:Z2_Auxiliary_Polyhedron2}\footnotesize{Triangulation ``$E_2$".}]
{
  \includegraphics[width=0.20\textwidth]{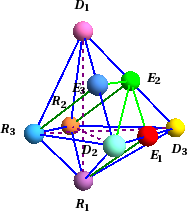}
}
\qquad 
\subfloat[\label{fig:Z2_Auxiliary_Polyhedron3}\footnotesize{Triangulation ``$E_3$".}]
{
  \includegraphics[width=0.20\textwidth]{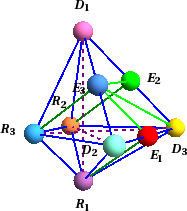}
}
\qquad 
\subfloat[\label{fig:Z2_Auxiliary_Polyhedron4}\footnotesize{Triangulation ``$S$".}]
{
  \includegraphics[width=0.20\textwidth]{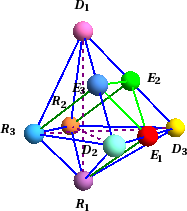}
}
\caption{The auxiliary polyhedra for the four possible triangulations of a local $\Cplx^3/\ztwo$ singularity.}
\label{fig:Z2_Auxiliary_Polyhedra}
\end{figure}

A resolved singularity is locally described as a non-compact toric variety. This implies that intersections can be trusted only when they are between compact subspaces. It is possible to build a compact toric variety that includes the resolution of a single $\ztwo$ singularity. This compactification is by itself not unique because it depends on the boundary conditions chosen at infinity and is therefore not determined by the local singularity alone. In what follows we present a choice of boundary conditions so as to match those of the globally resolved $T^6/\ztwo$ space defined below. This ensures that in the new space all (self-)intersection numbers are not only available, but that they also match those obtained for the global resolved $T^6/\ztwo$ space. We want to stress that the variety obtained in this way is compact but it is not a Calabi-Yau space, since it is not the resolution of the global $T^6/\ztwo$ space, hence, we use this ``auxiliary'', non-Calabi-Yau toric variety as a convenient tool to compute the intersection numbers of the resolved $T^6/\ztwo$ space.

The toric diagram of the variety is usually called ``auxiliary polyhedron'' and is obtained based on the key idea of promoting the local and homogeneous coordinates $u_i$ and $z_i$ to be coordinates of a weighted projective space. The scalings defining such a space encode, as mentioned above, all the informations concerning the boundary conditions at infinity, i.e.\ {\it both} the way in which the orbifold group acts on $\Cplx^3$ {\it and} the way how $T^6$ is built out of $\Cplx^3$. In the specific case at hand ($\ztwo$ orbifold of a factorizable $T^6$) we have
\begin{equation}
\label{newpro}
(u_i,z_i)\sim(\mu_iu_i,\mu_i^{1/2}z_i)\;,
\end{equation}
where $\gm_i \in \Cplx^*$ for $i=1,2,3$. Given this, the new variety includes a new set of divisors $R_i:=\{u_i=0\}$. These divisors are strictly speaking not the same as those defined on the compact orbifold in section \ref{sec:Orbifolds} or the ones on the global resolution defined below \ref{sec:Z2_Blowup_Properties}. However, as mentioned above their intersection numbers with the local divisors $D_i$ and $E_i$ are identical. (For this reason we allow ourselves this slight abuse of notation.)

To find the linear equivalence relations that govern the divisors $D_i, E_i$ and $R_i$, we note that the combinations $z_1^2 x_2 x_3/u_1$, $z_2^2 x_3 x_1/u_2$, and $z_3^2 x_1 x_2/u_3$ are invariant under the scaling \eqref{eq:Cstaraction} and \eqref{newpro}. Consequently, the local linear relations in the non-compact case \eqref{eq:Z2_Noncompact_Equivalences} are promoted to linear relations
\begin{equation}
R_1\sim 2 D_1+E_2+E_3\;,\qquad R_2\sim 2 D_2+E_3+E_1\;,\qquad R_3\sim 2 D_3+E_1+E_2
\end{equation}
of the compactification of the resolved local $\Cplx^3/\ztwo$ singularity. 
The toric diagram of the new variety is given by extending the diagram of the resolved $\Cplx^3/\ztwo$ singularity to include the new divisors. The position of the latter is encoded in three $s_i$ vectors defined such that they are perpendicular to the charges of the $\Cplx^*$ scalings, namely $s_i+v_i/2=0$, so that
\begin{equation}
\mathrm{s}_{1}=\left(\begin{array}{c} -1\\ 0\\0\end{array}\right)\;,~
\mathrm{s}_{2}=\left(\begin{array}{c} 0\\ -1\\0\end{array}\right)\;,~
\mathrm{s}_{3}=\left(\begin{array}{c} 0\\ 0\\-1\end{array}\right)\;. 
\end{equation}
In Fig.~\ref{fig:Z2_Auxiliary_Polyhedra} we visualize the diagrams dropping the origin and the lines connecting with it. We give four diagrams reflecting the triangulation dependence of the resolution.

\subsubsection{Visualization of flop transitions}

\begin{figure}[t]
\centering
\subfloat[\label{fig:Z2_Unprojected_Dual_Graph1}\footnotesize{Triangulation ``$E_1$".}]
{
  \includegraphics[width=0.20\textwidth]{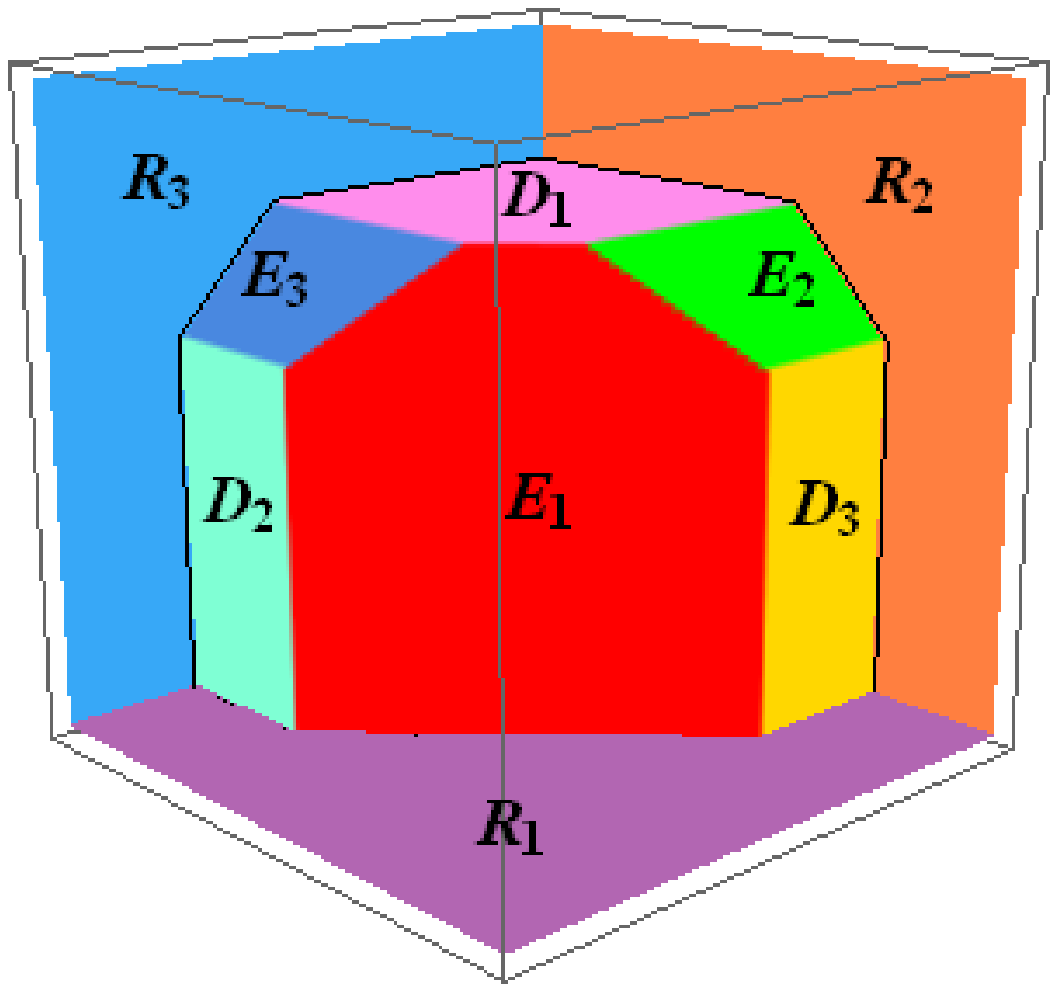}
}
\qquad 
\subfloat[\label{fig:Z2_Unprojected_Dual_Graph2}\footnotesize{Triangulation ``$E_2$".}]
{
  \includegraphics[width=0.20\textwidth]{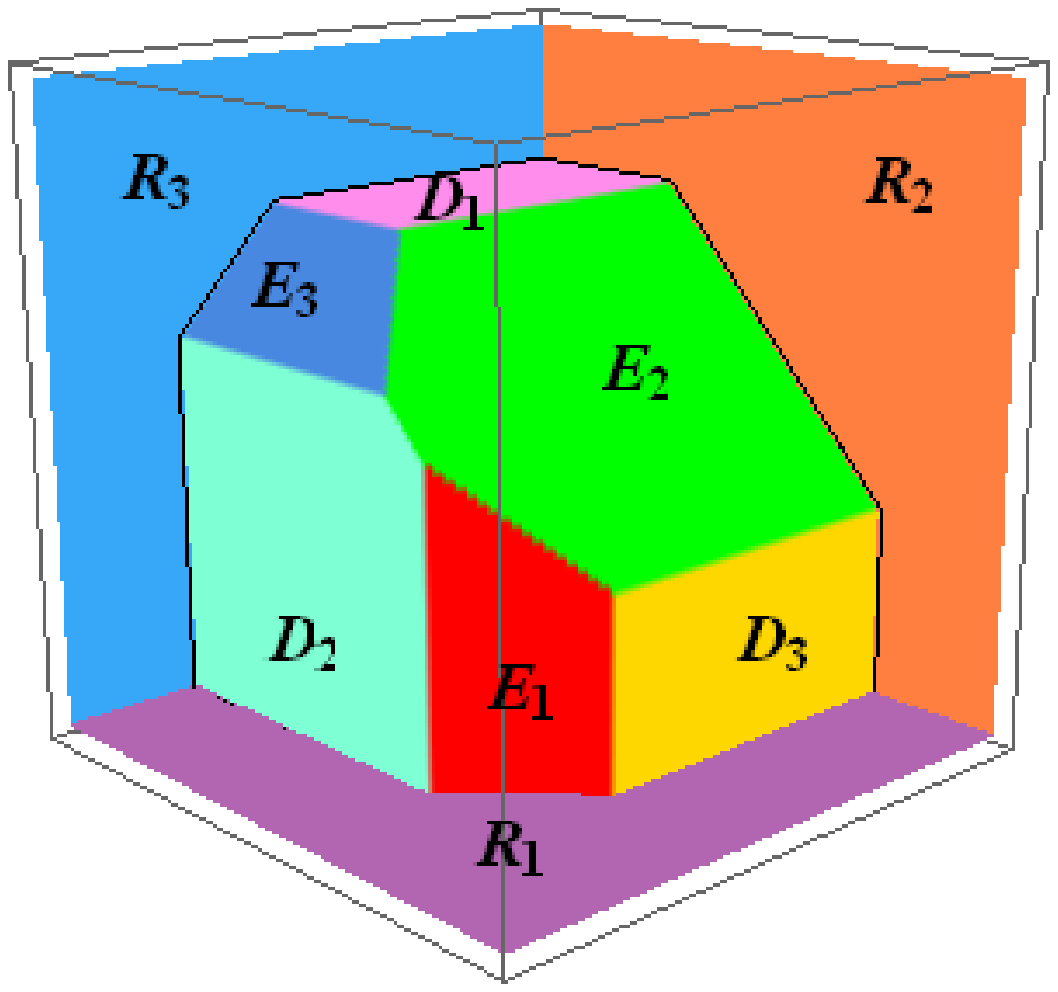}
}
\qquad 
\subfloat[\label{fig:Z2_Unprojected_Dual_Graph3}\footnotesize{Triangulation ``$E_3$".}]
{
  \includegraphics[width=0.20\textwidth]{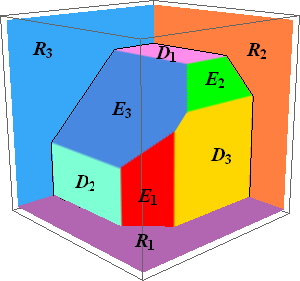}
}
\qquad 
\subfloat[\label{fig:Z2_Unprojected_Dual_Graph4}\footnotesize{Triangulation ``$S$".}]
{
  \includegraphics[width=0.20\textwidth]{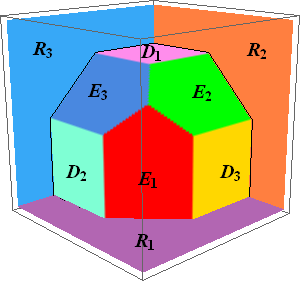}
}
\caption{The unprojected dual toric graphs for the four possible local  triangulations.}
\label{fig:Z2_Unprojected_Dual_Graphs}
\end{figure}

The four different triangulations are related to each other via so-called flop transitions. The triangulation ``$E_i$'' can be obtained from the triangulation ``$S$'' by the flop that exchanges the curve $E_jE_k$ with the curve $D_iE_i$ ($i \neq j \neq k$), see e.g.\ figure \ref{fig:Z2_Noncompact_Toric_Diagrams}. From these pictures one can infer that for the transition from triangulation ``$E_i$'' to ``$E_j$'' ($j\neq i$) two flop transitions are required: First triangulation ``$E_i$'' is converted to ``$S$'', which is after that transformed into triangulation ``$E_j$''.

In order to visualize such flop transitions and to show that there is a continuous underlying process linking the different triangulations, we introduce the unprojected dual toric or web graphs. They are a pictorial simplification of the diagrams that are the duals of the auxiliary polyhedra, making the visualization of the latter easier. Their construction can be sketched as follows: We associate the three real coordinate planes with the three divisors $R_{i}$. The ordinary divisors $D_{i}$ are represented by planes placed parallel to the coordinate planes at distances $a_{i} > 0$. This results in a cuboid. The planes that correspond to the exceptional divisors are defined by the equations 
\begin{equation}
E_{k}: ~ (\varphi_{k}){}^{1} x + (\varphi_{k}){}^{2} y + (\varphi_{k}){}^{3} z = (\varphi_{k}){}^{1} a_{1} + (\varphi_{k}){}^{2} a_{2} + (\varphi_{k}){}^{3} a_{3} - b_{k}\;.
\end{equation}
All planes are cut off at the lines where they intersect. The pictures in figure \ref{fig:Z2_Unprojected_Dual_Graphs} give the graphs for the \ztwo resolutions.

One can change from one triangulation to another by continuously adjusting the distance parameters $a_{i}$ and $b_{k}$. In other words, this provides means to describe flop transitions, which appear to be discrete in the toric diagrams, via smooth variations of continuous parameters. For example, if one moves the plane associated to $E_1$ in figure \ref{fig:Z2_Unprojected_Dual_Graph1} away from the $z$-axis by making $b_1$ small, then ultimately the picture changes over into that of figure \ref{fig:Z2_Unprojected_Dual_Graph4}. This indicates that even though the different triangulations are topologically distinct, as they differ by their intersection numbers, they are in fact continuously related to each other from a possible stringy CFT description point of view \cite{Aspinwall:1993,Witten:1993}. Moreover, it is also possible to link the continuous parameters introduced here to the K\"ahler moduli of the resolved space, linking flops to specific (continuous) choices of the metric properties of the resolved space (see section~\ref{metricprops} for details.)

\subsection[Compact $T^6/\ztwo$ resolutions]{Compact $\boldsymbol{T^6/\ztwo}$ resolutions}
\label{sec:Z2_Blowup_Properties}

In this subsection we describe some properties of compact resolutions of the orbifold $T^6/\ztwo$. First, we show how to obtain $\text{Res}(T^6/\ztwo)$ as an intersection of hypersurfaces in a toric variety. Then we discuss the properties of the divisors of the resolved space, and finally we discuss the intersection numbers among the divisors.

\subsubsection[$\text{Res}(T^6/\ztwo)$ as an intersection of hypersurfaces in a toric variety]{$\text{Res}(\boldsymbol{T^6/\ztwo})$ as an intersection of hypersurfaces in a toric variety}

Following \cite{Vafa:1994} we may describe the orbifold $T^6/\ztwo$ as a hypersurface in the $\cO(2,2,2)$ bundle over $(\CP^1)^3$. This description goes as follows:

First we describe each torus as an elliptic curve. Denote the homogeneous coordinates of the three $\CP^1$ by $U_i= (u_i, v_i)$ and let $y_i$ be the three fiber coordinates in the $\cO(2)$ bundles over each of these $\CP^1$'s. The $\Cplx^*$-scalings of each of the $\CP^1$'s then read 
\equ{
U_i ~\ra~ \gm_i\, U_i\;, 
\qquad 
y_i ~\ra~ \gm_i^2\, y_i\;, 
\label{CPscalings}
}
with $\gm_i \in \Cplx^*$. As explained in the appendix \ref{sc:Weierstrass} each elliptic curve defines a two-torus, described by the equation
\equ{
y_i^2 = P_i(U_i)\;, 
\label{EllipticCurves}
}
where each $P_i(U_i)$ is a homogeneous polynomial of degree four in $u_i, v_i$. A homogeneous polynomial of degree four is characterized by five complex numbers. However, their overall complex scales can be absorbed in $y$ and by $SL(2,\Cplx)$ transformations three additional complex parameters can be removed. Hence each $P_i$ characterizes a single complex number $\gr$, the complex structure of the $i$-th two-torus. For simplicity we have fixed these three complex structures to be equal to $\gr$ in this paper. In appendix \ref{sc:Weierstrass} we show that we can give an explicit form of these polynomials, 
\equ{
P_i(U_i) =  4 \prod\limits_{\ga=1}^4 N_\ga \cdot U_i\;,
\label{BundleHyperSurface} 
}
where the $N_\ga$ are vectors given by $N_1 = (0, 1)$, $N_{2}= (1, - \gve_2)$, $N_{3}= (1, - \gve_1)$ and $N_{4}= (1, - \gve_3)$. The points $N_\ga$ on the elliptic curve correspond to the four $\Intr_2$-fixed points on the torus $T^2$ under the action $z_i \ra - z_i$ on the torus coordinates. Parameterizing the elliptic curve using the Weierstrass function and its derivative given in \eqref{WeierstrassPOLES} of appendix \ref{sc:Weierstrass} we infer that the homogeneous coordinates $U_i = (u_i, v_i)$ and $y_i$ transform under this $\Intr_2$ 
action as 
\equ{
U_i \ra U_i\;, \qquad y_i \ra - y_i\;. 
} 

We are not interested in an alternative description of $(T^2)^3$ as the product of three elliptic curves; rather, we would like to find an algebraic description of the orbifold $T^6/\ztwo$. To this end we observe that under the orbifold elements $\gth_i$ one has $y_j \ra - y_j$, for $j \neq i$. Consequently, the section $y = y_1y_2y_3$ of the $\cO(2,2,2)$ bundle over $(\CP^1)^3$ is $\ztwo$ invariant. Hence if we want to describe the orbifold $T^6/\ztwo$ rather than $T^6$, we need to multiply the three equations \eqref{EllipticCurves} to obtain an equation in terms of the $\ztwo$-invariant coordinate $y$ only 
\equ{
y^2 = \prod\limits_{i} P_i(U_i)\;, 
\label{SingularityEquation}
}
In this description two combinations of the three $\Intr_2$'s that act on each of the elliptic curves (i.e.\ tori) separately have been modded out. The remaining diagonal $\Intr_2$ ($z_i \ra - z_i$) is realized as $y \ra -y$.

The $\Cplx^3/\ztwo$ singularities are located at the positions where all three polynomials $P_i$ vanish simultaneously. Near such a singularity we have the following description. For concreteness take the \ztwo singularity at $(0,0,0)$. Consider the coordinate patch where all $v_i \neq 0$, hence by the $\Cplx^*$-scalings we can set them all equal to unity. Expanding equation \eqref{SingularityEquation} in this patch around $(0,0,0)$ gives 
\equ{
y^2 \simeq u_1 u_2 u_3\;, 
}
where $\simeq$ denotes equal up to a complex factor. This equation can be solved by \eqref{inv_mon_concrete} and 
\equ{
y \simeq z_1z_2z_3\, x_1x_2x_3
}
by some local coordinates $z_i, x_i$ for the neighborhood of the resolved singularity. As in the local description above, the additional coordinates $x_1,x_2,x_3$ come again at the expense of three additional $\Cplx^*$-scalings.

In order to obtain a similar description for all 64 \ztwo fixed points simultaneously we write 
\begin{subequations}
\equa{
N_\ga\cdot U_1 = z_{1,\ga}^2 \, 
\prod\limits_\gg x_{2,\ga\gg} 
\prod\limits_\gb x_{3,\ga\gb}\;, 
\quad & \quad 
N_\gb\cdot U_2 = z_{2,\gb}^2 \, 
\prod\limits_\gg x_{1,\gb\gg} 
\prod\limits_\ga x_{3,\ga\gb}\;, 
\\
N_\gg\cdot U_3 = z_{3,\gg}^2 \, 
\prod\limits_\gb x_{1,\gb\gg} 
\prod\limits_\ga x_{2,\ga\gg}\;, 
\quad & \quad 
y \simeq \prod\limits_{i,\gr} z_{i,\gr} \, \prod\limits_{i,\gr\gs} x_{i,\gr\gs}\;, 
}
\label{GlobalCoordConstraints}
\end{subequations} 
where the indices indicate to which $\Intr_2$ singularity each coordinate corresponds, analogous to equation~\eqref{Def_n}.

Therefore, we have in total $3\cdot 2+1 + 3\cdot 4 + 3\cdot 4^2 = 67$ local coordinates $u_i, v_i, y, z_{i,\gr}$ and $x_{i,\gr\gs}$ with $3 \cdot 4+1 = 13$ constraints encoded in the equations \eqref{GlobalCoordConstraints}. This means that the space has complex dimension $67-13= 54$ rather than 3. To reduce to 3 dimensions, one consequently must be able to identify $51$ $\Cplx^*$-scalings. Three of them correspond to the $\Cplx^*$-scalings defining the $\CP^1$'s in \eqref{CPscalings}. Since the left hand sides of the first three equations in \eqref{GlobalCoordConstraints} transform under these scalings, also the right hand sides have to transform, resulting in a transformation of some of the local singularity coordinates under these three scalings. The unique choice for this is 
\equ{
\label{CPzScaling} 
z_{i,\gr} \ra \gm_i^{1/2} \, z_{i,\gr}\;, 
}
up to mixing with the other $\Cplx^*$-scalings that act on the $x_r$'s as well.

A basis for all the 51 $\Cplx^*$-scalings is defined by the charges given in table \ref{tb:C*scalings} via 
\equ{ 
t:\ Z \ra \gl(t) \, Z\;, 
\qquad 
\gl(t) = e^{t\cdot Q}\;, 
\label{CoordC*scalings}
}
where we collectively denote the coordinates by $Z = (u_i, v_i, z_{i,\gr}, x_{i,\gr\gs})$ and scaling parameters by $t = (t_i, t_{i,\gr\gs}) \in (\Cplx^*)^{51}$. These scalings encode the linear equivalence relations we are discussing in the next subsection. In the description here we have resolved the 48 $\Cplx^2/\Intr_2$ singularities, while the complex structure has not been deformed. As we are not considering torsion effects in this paper, this is sufficient for our purposes.

\begin{table} 
\[
\renewcommand{\arraystretch}{\stretcharry} 
\arry{|c||c | c | c | c | c  | c|}{ 
\hline 
& u_i & v_i & z_{1,\ga} & z_{2,\gb} & z_{3,\gg} & x_{i,\gr \gs} 
\\\hline\hline  
Q_{i'} &   \gd_{i'i}  &  \gd_{i'i} & \frac 12\,\gd_{i'1}   &  \frac 12\,\gd_{i'2}   &  \frac 12\,\gd_{i'3}  & 0  
\\\hline 
Q_{i', \gr' \gs'} & 0 & 0 
& \frac 12\, ( \gd_{i'2} + \gd_{i'3} ) \,  \gd_{\gr'\ga} 
& \frac 12\, ( \gd_{i'1} \,  \gd_{\gr'\gb} + \gd_{i'3} \,  \gd_{\gs'\gb} ) 
& \frac 12\, ( \gd_{i'1}    + \gd_{i'2} ) \,  \gd_{\gs'\gg}  
& - \gd_{i'i} \, \gd_{\gr'\gr} \, \gd_{\gs'\gs} 
\\\hline 
}
\renewcommand{\arraystretch}{1} 
\]
\caption{
This table summarizes the charges of the 66 homogeneous coordinates  of the resolution which define how the 51 $\Cplx^*$-scalings act. 
\label{tb:C*scalings}}
\end{table} 

\subsubsection{Divisors, linear equivalence relations}
\label{sc:DivsGlobal}

The divisors of the resolution of the orbifold $T^6/\ztwo$ correspond to the four-cycles obtained by setting one complex coordinate equal to zero. In detail, the local resolutions of the 48 fixed tori and 64 \ztwo singularities give rise to 12 ordinary divisors $D_{1,\alpha} :=\{z_{1,\ga} \!=\!0\}$, $D_{2,\beta} :=\{z_{2,\gb} \!=\!0\}$, and $D_{3,\gamma} :=\{z_{3,\gg} \!=\!0\}$. In addition we obtain 48 exceptional divisors $E_{1,\beta\gamma} :=\{x_{1,\gb\gg} \!=\!0\}$, $E_{2,\alpha\gamma} :=\{x_{2,\ga\gg} \!=\!0\}$, and $E_{3,\alpha\beta} :=\{x_{3,\ga\gb} \!=\!0\}$.  The global structure of the orbifold resolution is characterized by six inherited divisors $R_{i} := \{u_i\!=\!0\}$ and $R'_{i} := \{v_i\!=\!0\}$. Finally, $R_0 := \{ y\!=\!0 \}$ is redundant: The coordinate $y$ that was used to parameterize the singularity \eqref{SingularityEquation} has been replaced by the local singularity coordinates $z_\gk$ and $x_r$. Hence, we may disregard $R_0$ in the rest of the discussion. To shorten the notation, we  refer to the three types of divisors as $E_r, D_\gk$ and $R_\go$ and to all of them collectively as $S_u$.

Not all of these divisors define independent four-cycles because of various linear equivalence relations among them. These linear equivalences can be read off from \eqref{GlobalCoordConstraints} and extend the linear equivalences \eqref{eq:Z2_Noncompact_Equivalences} to 
\begin{subequations}
\begin{eqnarray}
2D_{1,\alpha}  \sim ~ R_{1}  - \sum\limits_{\gamma} E_{2,\alpha\gamma} -\sum\limits_{\beta} E_{3,\alpha\beta}\;, 
\quad & \quad 
2D_{2,\beta} \sim ~ R_{2} - \sum\limits_{\gamma} E_{1,\beta\gamma} - \sum\limits_{\alpha} E_{3,\alpha\beta}\;, 
\\
2D_{3,\gamma}  \sim ~ R_{3} - \sum\limits_{\beta} E_{1,\beta\gamma} - \sum\limits_{\alpha} E_{2,\alpha\gamma}\;, 
\quad  & \quad 
R'_i  \sim R_i\;.
\end{eqnarray}
\label{eq:Z2_Compact_Equivalences}
\end{subequations}

These relations tell us that the inherited divisors $R_i$ and $R'_i$ are all linearly dependent. In addition every ordinary divisor $D_{i,\gr}$ can be expressed through a linear combination of inherited and exceptional ones. Consequently the $R_{i}$ and $E_{r}$ provide via the Poincar\'e duality a basis of the real cohomology group, i.e.\ the $(1,1)$-forms, on the resolved manifold. The additional divisors $R'_i$ and $D_{\gk}$ are crucial to obtain an integral cohomology basis. The identification of the divisors and their linear equivalences constitutes the first step of the gluing procedure.

\subsubsection{Intersection numbers and triangulation dependence}
\label{sc:Int_Num}

The second step in the gluing procedure is the computation of the (self-)intersection numbers of the divisors of the resolution. In this process one needs to combine the global intersections of the inherited divisors with the local intersection properties of divisors of the local resolutions.

A method to perform this task is by employing  the auxiliary polyhedra \cite{Lust:2006} that combine local and global intersection data. The reason why this method works, i.e.\ that the auxiliary polyhedra constructed in subsection~\ref{sect:auxpoly} are compatible with the global structure of the resolution of the ${T^6/\ztwo}$ orbifold, is that the compactification of the resolved $\Cplx^3/\ztwo$ singularity defined by the auxiliary polyhedra have the appropriate boundary conditions at infinity. The technical way to see this is by noting that the scalings of $u_i$ and $z_{i,\gr}$ defined in equations \eqref{CPscalings} and \eqref {CPzScaling} of subsection \ref{sec:Z2_Blowup_Properties} are identical to the $\Cplx^*$-scalings for $u_i$ and $z_i$ defined in \eqref{newpro} of the auxiliary polyhedra. One convenient consequence of this is that all intersections of $D_{i,\gr}$, $E_{j,\gs\gt}$ and $R_k$ at the resolution of a given $\ztwo$-fixed point introduced in subsection \ref{sc:DivsGlobal} can simply be read off from the auxiliary polyhedra by making the identification between them and the divisors $D_i$, $E_j$ and $R_k$ defined in subsection \ref{sect:auxpoly} for the auxiliary polyhedra.

Because of the triangulation ambiguity of a local $\Cplx^3/\ztwo$ resolution there are four auxiliary polyhedra depicted in figure \ref{fig:Z2_Auxiliary_Polyhedra}. Hence to completely define a resolution of the orbifold $T^6/\ztwo$ one has to specify the local triangulation of all 64 $\ztwo$ singularities. Given such a choice of triangulations the intersection numbers are computed straightforwardly: All triple non-self-intersections can be read off from the auxiliary polyhedra as one of their fundamental cones; all other non-self-intersections vanish. Self-intersections can subsequently be computed via the linear equivalence relations \eqref{eq:Z2_Compact_Equivalences}. We have resorted to a computer code as for a given choice of triangulation there are of the order of hundred thousand (self-)intersection numbers to be computed.

\begin{table}[t]
\centering
\renewcommand{\arraystretch}{\stretcharry}
\begin{tabular}{|l||c|c|c|c|}
\hline
\backslashbox{$\text{Int}(S_{1}S_{2}S_{3})$}{Triangulation}		&``$E_1$" 	& ``$E_2$" 	& ``$E_3$" 	& ``$S$" 	\\
\hline\hline 
$E_{1,\beta\gamma}E_{2,\alpha\gamma}E_{3,\alpha\beta}$	& $0$ 	& $0$	& $0$	& $1$	\\ 
\hline
$E_{1,\beta\gamma}E_{2,\alpha\gamma}^{2}$, $E_{1,\beta\gamma}E_{3,\alpha\beta}^{2}$					& $-2$ 	& $0$	& $0$	& $-1$	\\ 
$E_{2,\alpha\gamma}E_{1,\beta\gamma}^{2}$, $E_{2,\alpha\gamma}E_{3,\alpha\beta}^{2}$					& $0$ 	& $-2$	& $0$	& $-1$	\\ 
$E_{3,\alpha\beta}E_{1,\beta\gamma}^{2}$, $E_{3,\alpha\beta}E_{2,\alpha\gamma}^{2}$					& $0$ 	& $0$	& $-2$	& $-1$	\\ 
\hline
$E_{1,\beta\gamma}^{3}$									& $0$	& $8$	& $8$	& $4$	\\ 
$E_{2,\alpha\gamma}^{3}$									& $8$	& $0$	& $8$	& $4$	\\ 
$E_{3,\alpha\beta}^{3}$									& $8$	& $8$	& $0$	& $4$	\\ 
\hline
\hline
$R_{1}R_{2}R_{3}$		 								& \multicolumn{4}{c|}{$2$}			\\	
$R_{1}E_{1,\beta\gamma}^{2}$, $R_{2}E_{2,\alpha\gamma}^{2}$, $R_{3}E_{3,\alpha\beta}^{2}$								& \multicolumn{4}{c|}{$-2$}			\\
\hline
\end{tabular}
\renewcommand{\arraystretch}{1}
\caption{The upper part gives the intersection numbers when using the same triangulation at all $64$ fixed points. The lower part gives the triangulation-independent intersection numbers.}
\label{table:Z2_Intersection_Numbers}
\end{table}

The triangulation dependence is a major complication since all intersection numbers and hence many topological quantities of physical interest, such as the Bianchi identities and the chiral spectrum, depend on it. In table~\ref{table:Z2_Intersection_Numbers} we give a complete overview of all not always vanishing intersection numbers involving the divisors $R_{i}$ and $E_{r}$ provided that the same triangulation is used at all \ztwo fixed points. Since the $R_{i}$ and $E_{r}$ form a basis in the real cohomology, all the other intersection numbers can be deduced from the given ones using the linear equivalence relations \eqref{eq:Z2_Compact_Equivalences}. The intersections that involve an inherited divisor $R_i$ are universal, while the other intersections depend strongly on the chosen triangulation.

To understand just how big the triangulation dependence is, we estimate the number of different resolutions of the \ztwo orbifold. 
The upper bound is given by 4 triangulation possibilities at 64 fixed
points, which yields $4^{64} \approx 3.40 \cdot 10^{38}$ inequivalent triangulations. However, due to the symmetry underlying the orbifold, this number is reduced  \cite{SGN:2009}. 
An accurate approximation of the number of independent triangulations is given by 
\begin{equation}
\frac{4^{64}}{3! 4!^3} \approx 4.10 \cdot 10^{33}\;. 
  \label{eq:NTriang_Z2}
\end{equation}
%
Here the factor $3!$ comes from permutations of the three two-tori and the factor $4!^3$ results from permutations of the fixed points within each two-torus.

\subsubsection{Hodge numbers and Chern classes}

From the compact linear equivalences and the intersection numbers, we can compute the topological invariants for the blown-up $T^6/\ztwo$ (resolved) manifold, starting with the Hodge numbers. As the $R_{i}$ and $E_{r}$ form a basis for $H^{1,1}$, we find $h^{1,1}=3+48=51$. We define the $(1,0)$-forms ${\rm d}z_{1}$, ${\rm d}z_{2}$ and ${\rm d}z_{3}$.  As they are not invariant by themselves, $h^{1,0}=0$. In this notation the $R_i$ can be represented as $\d \bz_i \wedge \d z_i$.  In addition we can build the invariant $(3,0)$ holomorphic volume form ${\rm d}z_{1}\wedge{\rm d}z_{2}\wedge{\rm d}z_{3}$ from these. Invariant $(2,1)$-forms are ${\rm d}{\overline{z}}_{\overline{1}}\wedge{\rm d}z_{2}\wedge{\rm d}z_{3}$, ${\rm d}z_{1}\wedge{\rm d}{\overline{z}}_{\overline{2}}\wedge{\rm d}z_{3}$, and ${\rm d}z_{1}\wedge{\rm d}z_{2}\wedge{\rm d}{\overline{z}}_{\overline{3}}$, so $h^{2,1}=3$. There are no further contributions from the twisted sector because we only have fixed lines with fixed points on them \cite{Lust:2006}. Hence the  Hodge numbers are given by
\begin{equation}
\label{eq:Z2_Hodge_Diamond}
\begin{array}{cccccccccccccccc}
&&&h^{3,3}&&& && &&&1&&&\\
&&h^{3,2}&&h^{2,3}&& && &&0&&0&&\\
&h^{3,1}&&h^{2,2}&&h^{1,3}& && &0&&3+48&&0&\\
h^{3,0}&&h^{2,1}&&h^{1,2}&&h^{0,3} &=& 1&&3&&3&&1\;,\\
&h^{2,0}&&h^{1,1}&&h^{0,2}& && &0&&3+48&&0&\\
&&h^{1,0}&&h^{0,1}&& && &&0&&0&&\\
&&&h^{0,0}&&& && &&&1&&&
\end{array}
\end{equation}
which has the CY structure and agrees with \cite{Lust:2006}. From this we compute the Euler number 
\begin{equation}
\label{eq:Z2_Euler_Number}
\chi(X) = 2(1+51-(1+3))=96\;.
\end{equation}

The total Chern class $c(X)$ of the resolution is determined by an expansion of the complex 66 dimensional classifying space of the resolution in terms of divisors:
\begin{equation}
c(X) = \prod\limits_u (1 + (-)^u S_u) = 
\prod\limits_{\gk=1}^{12}(1+D_{\gk})
\prod\limits_{r=1}^{48} (1+E_{r})
\prod\limits_{r=1}^{3} (1-R_{i}) (1-R_i')\;. 
\label{eq:Z2_Total_Chern_Class}
\end{equation}
The inherited divisors come with a minus sign as their coordinates are on the opposite side of the defining equations \eqref{GlobalCoordConstraints}. Expanding \eqref{eq:Z2_Total_Chern_Class} yields
\begin{subequations}
\begin{eqnarray}
c_{1}(X) & = & \sum\limits_{\gk=1}^{12}D_{\gk}+\sum\limits_{r=1}^{48} E_{r}-\sum\limits_{i=1}^{3}R_{i}-\sum\limits_{i=1}^{3}R_{i}'\;, 
\label{eq:Z2_Chern_Class1}
\\
c_{2}(X) & = & \frac{1}{2!} \sum\limits_{u}(c_{1}(X)-S_{u}) S_{u}\;,
 \label{eq:Z2_Chern_Class2}
 \\
c_{3}(X) & = & \frac{1}{3!} \sum\limits_{u,v}(c_{1}(X)-S_{u}-S_{v}) S_{u} S_{v}\;,
 \label{eq:Z2_Chern_Class3}
\end{eqnarray}
\label{eq:Z2_Chern_Classes}
\end{subequations}
Inserting the linear equivalences \eqref{eq:Z2_Compact_Equivalences} in \eqref{eq:Z2_Chern_Class1} shows that our resolution space is CY, i.e.\ $c_{1}(X)=0$.

The Euler number \eqref{eq:Z2_Euler_Number} can also be computed from the third Chern class \eqref{eq:Z2_Chern_Class3} by inserting the intersection numbers. The contributions containing an inherited divisor are triangulation independent and hence universal. One can easily count that they sum up to $-160$. The remaining part consists of all possible intersection numbers of three distinct ordinary and exceptional divisors. These intersection numbers are equal to one if the three divisors span a fundamental triangle on the front face of the relevant auxiliary polyhedra and zero otherwise. Together with the fact that in each triangulation there are exactly four fundamental triangles in each of the $64$ auxiliary polyhedra, we obtain $c_{3}(X)=64\cdot4-160=96$ in agreement with the Euler number computation in \eqref{eq:Z2_Euler_Number}.

\subsubsection{Volumes and the K\"ahler form}
\label{metricprops}

\begin{table}[h!t]
\[
\renewcommand{\arraystretch}{\stretcharry} 
\arry{|c|c||c|c|}{
\hline 
\multicolumn{4}{|c|}{\text{Volumes within triangulation ``$E_1$"}}
\\ \hline \hline  
\multicolumn{4}{|c|}{\text{Curves}}
\\ \hline\hline 
%
%
E_{1,\beta\gamma} E_{2,\alpha\gamma} & 2b_{2,\alpha\gamma}  
& 
D_{1,\alpha}E_{1,\beta\gamma} 	& 
b_{1,\beta\gamma}-b_{2,\alpha\gamma}-b_{3,\alpha\beta}
\\
E_{1,\beta\gamma} E_{3,\alpha\beta} & 2b_{3,\alpha\beta}
&
D_{1,\alpha}E_{2,\alpha\gamma} & 
a_{2}+4b_{2,\alpha\gamma}-\sum\limits_{\beta}b_{1,\beta\gamma}
\\
E_{2,\alpha\gamma} E_{3,\alpha\beta} & 0 	
&
D_{1,\alpha}E_{3,\alpha\beta} & 
a_{3}+4b_{3,\alpha\beta}-\sum\limits_{\gamma}b_{1,\beta\gamma}
\\\hline 
D_{2,\beta}E_{1,\beta\gamma} & 
a_{1}-\sum\limits_{\alpha} b_{3,\alpha\beta}
&
D_{3,\gamma}E_{1,\beta\gamma} &  
a_{1}-\sum\limits_{\alpha}b_{2,\alpha\gamma}
\\
D_{2,\beta}E_{2,\alpha\gamma} & 0 
&
D_{3,\gamma}E_{2,\alpha\gamma} & 
a_{2}-\sum\limits_{\beta}b_{1,\beta\gamma}
\\	
D_{2,\beta}E_{3,\alpha\beta} & 
a_{3}-\sum\limits_{\gamma}b_{1,\beta\gamma}
&
D_{3,\gamma}E_{3,\alpha\beta} & 0 
\\\hline 
R_{i}E_{i,\rho\sigma} & 2b_{i,\rho\sigma}
&
R_{i}R_{j} & 2a_{k}~~(i \neq j \neq k)	
\\
R_{i}D_{j,\rho} & 
a_{k}-\sum\limits_{\sigma}b_{i,\rho\sigma}
~~(i \neq j \neq k)
& &
\\ \hline\hline 
\multicolumn{4}{|c|}{\text{Divisors}}
\\ \hline\hline 
%
%
E_{1,\beta\gamma} & a_1 b_{1,\beta\gamma}  
 -\sum\limits_\alpha \left\lbrace 
b_{2,\alpha\gamma}^2 +  b_{3,\alpha\beta}^2  \right\rbrace 
& 
E_{3,\beta\gamma} & a_3 b_{3,\beta\gamma}  
 + \sum\limits_\alpha \left\lbrace
 b_{3,\beta\gamma}^2 - b_{3,\beta\gamma} b_{1,\alpha\gamma}  \right\rbrace 
\\ 
 E_{2,\beta\gamma} & a_2 b_{2,\beta\gamma}  
 + \sum\limits_\alpha \left\lbrace
 b_{2,\beta\gamma}^2 - b_{2,\beta\gamma} b_{1,\alpha\gamma}  \right\rbrace 
& 
R_i & 2a_j a_k - \sum\limits_{\gr,\gs} b_{i,\gr\gs}^2 
~~ (i\neq j \neq k)
\\ \hline 
D_{1,\alpha} & \multicolumn{3}{|c|}{ 
  a_2 a_3 - \sum\limits_\gamma a_2 b_{2,\alpha\gamma} - \sum\limits_\beta a_3 b_{3,\alpha\beta} + \sum\limits_{\beta,\gamma} \left\lbrace b_{1,\beta\gamma} \left(  b_{2,\alpha\gamma} + b_{3,\alpha\beta} \right)  - \frac{1}{2} \left(b_{1,\beta\gamma}^2 + b_{2,\alpha\gamma}^2 + b_{3,\alpha\beta}^2 \right) \right\rbrace }
 \\
D_{2,\beta} & \multicolumn{3}{|c|}{
  a_1 a_3 - \sum\limits_\gamma a_1 b_{1,\beta\gamma} - \sum\limits_\alpha a_3 b_{3,\alpha\beta} + \sum\limits_{\alpha,\gamma}  b_{1,\beta\gamma} b_{3,\alpha\beta} } 
  \\
D_{3,\gamma} & \multicolumn{3}{|c|}{
  a_2 a_3 - \sum\limits_\alpha a_2 b_{2,\alpha\gamma} - \sum\limits_\beta a_1 b_{1,\beta\gamma} + \sum\limits_{\alpha,\beta}  b_{1,\beta\gamma} b_{2,\alpha\gamma} } 
\\\hline\hline 
\multicolumn{4}{|c|}{\text{Full manifold}}
\\ \hline\hline
%
%
 X &  \multicolumn{3}{|c|}{
 2 a_1 a_2 a_3 - \sum\limits_{\beta,\gamma} a_1 b_{1,\beta\gamma}^2 - \sum\limits_{\alpha,\gamma} a_2 b_{2,\alpha\gamma}^2 - \sum\limits_{\alpha,\beta} a_3 b_{3,\alpha\beta}^2}
 \\ & \multicolumn{3}{|c|}{
 + \sum\limits_{\alpha,\beta,\gamma} \left\lbrace
b_{1,\beta\gamma} \left( b_{2,\alpha\gamma}^2  + b_{3,\alpha\beta}^2 \right) - \frac{1}{3} \left( b_{2,\alpha\gamma}^3 + b_{3,\alpha\beta}^3 \right)  \right\rbrace} 
\\\hline 
}
\]
\renewcommand{\arraystretch}{1} 
\caption{Volume of the curves, the divisors and the whole manifold   when using the triangulation ``$E_1$'' at all \ztwo orbifold fixed points.}
\label{table:Z2_Volume_E1}
\end{table}

After these purely topological quantities, we turn to geometrical properties. The volumes on the resolution space are determined by the K\"ahler form
\begin{equation}
\label{eq:Kaehlerform}
J=a_{i} R_{i} - b_{r} E_{r}\;, 
\end{equation}
expanded in terms of the untwisted and twisted K\"ahler moduli $a_{i}$ and $b_{r}$, respectively. The signs have been chosen such that a geometrical description applies when all $a_i$'s and $b_r$'s are positive. The volumes of curves $C$, divisors $S$, and the whole manifold $X$ read 
\begin{equation}
\label{eq:Volume_Definitions}
\text{vol}(C) \defi \int\limits_{C} J\;,
\hskip10mm 
\text{vol}(S) \defi \frac{1}{2!} \int\limits_{S} J \wedge J\;,
\hskip10mm 
\text{vol}(X) \defi \frac{1}{3!} \int\limits_{X} J \wedge J \wedge J\;. 
\end{equation}
Using these definitions, the volumes of curves, divisors and the whole resolution space can be computed given the complete intersection ring of the divisors. In tables \ref{table:Z2_Volume_E1} and \ref{table:Z2_Volume_S} we give the volumes of the curves, the divisors and the resolution manifold as a whole manifold using triangulations ``$E_1$'' and ``$S$'' everywhere, respectively. (The volumes for triangulations ``$E_2$'' or ``$E_3$'' can be obtained by cyclic permutations of the indices 1,2,3 from the results for triangulation ``$E_1$'' of table \ref{table:Z2_Volume_E1}.)

Even though the expressions for the volumes depend strongly on the chosen triangulation, some general patterns can be observed. There is a consistency requirement that the volumes of the existing curves and divisors in a given triangulation need to be all positive. This puts stringent and complicated requirements on the K\"ahler moduli $a_i$ and $b_{r}$. Taking the K\"ahler moduli $a_i$ parametrically larger than the $b_{r}$, ensures that the volumes of many curves, divisors and the whole manifold is positive. In particular in the geometrical blow-down limit in which all $b_r$ tend to zero, one finds 
\equ{
\text{vol}(R_iR_j) = 2 \, a_k\;, 
\qquad
\text{vol}(R_i) = 2 \, a_ja_k\;, 
\qquad 
\text{vol}(T^6/\ztwo) = 2\, a_1a_2a_3\;, 
\label{Volumes_Blow_Down} 
}
with $i \neq j \neq k$, in any triangulation because the intersections of the inherited divisors $R_i$ are universal.  These expressions are compatible with the volumes computed directly on the orbifold, see \eqref{Volumes_Orbifold}, with the identification $a_i = r_i^2/2$. The volumes generically decrease in the blow-up process. In particular the total space has the structure of a ``swiss-cheese'': As one increases the values of the K\"ahler parameters $b_{i,\gr\gs}$ the total volume in fact becomes smaller. 

The requirement of positivity for all volumes puts restrictions on the moduli space. First, we observe that in each triangulation the K\"ahler moduli $a_i$ and $b_r$ must be positive. In triangulation ``$E_1$'', positivity of the curve $E_{1}D_{1}$ implies that $b_{1} > b_{2} + b_{3}$ and we get analogous relations for triangulations ``$E_2$'' and ``$E_3$''. On the other hand, in the symmetric triangulation we find the restriction $b_i < b_j + b_k$ for all $i\neq j\neq k \neq i$. It is easy to see that we can take the separate K\"ahler moduli spaces of the four triangulations and glue them together at the surfaces given by $b_i = b_j + b_k$, thus obtaining one enhanced moduli space in which the triangulation is determined by the relations between the moduli. In this sense, the different triangulations can be seen as different phases of one theory with smooth flop transitions between them as was already observed in \cite{Aspinwall:1993}. Geometrically, going e.g.\ from triangulation ``$E_1$'' to triangulation ``$S$'' is done by shrinking the curve $E_{1}D_{1}$ to zero and then growing the curve $E_2E_3$ instead.

\begin{table}[h!t]
\[
\renewcommand{\arraystretch}{\stretcharry} 
\arry{|c|c||c|c|}{
\hline 
\multicolumn{4}{|c|}{\text{Volumes within triangulation ``$S$"}}
\\ \hline \hline  
\multicolumn{4}{|c|}{\text{Curves}}
\\ \hline\hline 
%
%
E_{1,\beta\gamma} E_{2,\alpha\gamma} & 
b_{1,\beta\gamma}+b_{2,\alpha\gamma}-b_{3,\alpha\beta}
& 
D_{1,\alpha}E_{1,\beta\gamma} & 0
\\
E_{1,\beta\gamma} E_{3,\alpha\beta} &  
b_{1,\beta\gamma}-b_{2,\alpha\gamma}+b_{3,\alpha\beta}
& 
D_{1,\alpha}E_{2,\alpha\gamma} & 
a_{2}-\sum\limits_{\beta}b_{3,\alpha\beta}
\\
E_{2,\alpha\gamma} E_{3,\alpha\beta} & 
-b_{1,\beta\gamma}+b_{2,\alpha\gamma}+b_{3,\alpha\beta}
& 
D_{1,\alpha}E_{3,\alpha\beta} & 
a_{3}-\sum\limits_{\gamma}b_{2,\alpha\gamma}
\\\hline
D_{2,\beta}E_{1,\beta\gamma} & 
a_{3}-\sum\limits_{\alpha} b_{3,\alpha\beta}
&
D_{3,\gamma}E_{1,\beta\gamma} & 
a_{1}-\sum\limits_{\alpha}b_{2,\alpha\gamma}
\\
D_{2,\beta}E_{2,\alpha\gamma} & 0
&
D_{3,\gamma}E_{2,\alpha\gamma} & 
a_{2}-\sum\limits_{\beta}b_{1,\beta\gamma}
\\
D_{2,\beta}E_{3,\alpha\beta} & 
a_{3}-\sum\limits_{\gamma}b_{1,\beta\gamma}
&
D_{3,\gamma}E_{3,\alpha\beta} & 0 	
\\\hline 
R_{i}E_{i,\rho\sigma} & 2b_{i,\rho\sigma}
& 
R_{i}R_{j}	& 2a_{k}~~(i \neq j \neq k)
\\
R_{i}D_{j,\rho} & 
a_{k}-\sum\limits_{\sigma}b_{i,\rho\sigma}
~~(i \neq j \neq k)	
&
&
%
%
\\\hline\hline 
\multicolumn{4}{|c|}{\text{Divisors}}
\\ \hline\hline 
 E_{1,\beta\gamma} & 
 \multicolumn{3}{|c|}{
  2a_1 b_{1,\beta\gamma}  
 + \sum\limits_\alpha \left\lbrace
b_{2,\alpha\gamma}b_{3,\alpha\beta} + \frac{1}{2} \left(b_{1,\beta\gamma}^2 - b_{2,\alpha\gamma}^2 - b_{3,\alpha\beta}^2 \right) - b_{1,\beta\gamma} \left(  b_{2,\alpha\gamma} + b_{3,\alpha\beta} \right)  \right\rbrace
}
\\
 E_{2,\ga\gg} & 
 \multicolumn{3}{|c|}{
 2a_2 b_{2,\alpha\gamma}  
 + \sum\limits_\beta \left\lbrace
b_{1,\beta\gamma}b_{3,\alpha\beta} + \frac{1}{2} \left(b_{2,\alpha\gamma}^2 - b_{1,\beta\gamma}^2 - b_{3,\alpha\beta}^2 \right) - b_{2,\alpha\gamma} \left(  b_{1,\beta\gamma} + b_{3,\alpha\beta} \right)  \right\rbrace
}
\\
 E_{3,\ga\gb} & 
 \multicolumn{3}{|c|}{
 2a_3 b_{3,\alpha\beta}  
 + \sum\limits_\gamma \left\lbrace
b_{1,\beta\gamma}b_{2,\alpha\gamma} + \frac{1}{2} \left(b_{3,\alpha\beta}^2 - b_{1,\beta\gamma}^2 - b_{2,\alpha\gamma}^2 \right) - b_{3,\alpha\beta} \left(  b_{1,\beta\gamma} + b_{2,\alpha\gamma} \right)  \right\rbrace 
} 
\\ 
D_{1,\alpha} & \multicolumn{3}{|c|}{
 a_2 a_3 - \sum\limits_\gamma a_2 b_{2,\alpha\gamma} - \sum\limits_\beta a_3 b_{3,\alpha\beta} + \sum\limits_{\beta,\gamma}  b_{2,\alpha\gamma} b_{3,\alpha\beta} } 
 \\
D_{2,\beta} & \multicolumn{3}{|c|}{
  a_1 a_3 - \sum\limits_\gamma a_1 b_{1,\beta\gamma} - \sum\limits_\alpha a_3 b_{3,\alpha\beta} + \sum\limits_{\alpha,\gamma}  b_{1,\beta\gamma} b_{3,\alpha\beta} } 
 \\
 D_{3,\gamma} & \multicolumn{3}{|c|}{
 a_2 a_3 - \sum\limits_\alpha a_2 b_{2,\alpha\gamma} - \sum\limits_\beta a_1 b_{1,\beta\gamma} + \sum\limits_{\alpha,\beta}  b_{1,\beta\gamma} b_{2,\alpha\gamma} } 
\\ 
R_i & 
\multicolumn{3}{|c|}{
2a_j a_k - \sum\limits_{\gr,\gs} b_{i,\gr\gs}^2 
~~ (i\neq j \neq k)}
\\\hline\hline
\multicolumn{4}{|c|}{\text{Full manifold}}
\\ \hline\hline 
%
%
X & 
\multicolumn{3}{|c|}{
2 a_1 a_2 a_3 - \sum\limits_{\beta,\gamma} a_1 b_{1,\beta\gamma}^2 - \sum\limits_{\alpha,\gamma} a_2 b_{2,\alpha\gamma}^2 - \sum\limits_{\alpha,\beta} a_3 b_{3,\alpha\beta}^2}
\\ & \multicolumn{3}{|c|}{
+ \sum\limits_{\alpha,\beta,\gamma} \left\lbrace 
\frac{1}{6}\left( b_{1,\beta\gamma} + b_{2,\alpha\gamma} + b_{3,\alpha\beta} \right)^3 - 2 b_{1,\beta\gamma} b_{2,\alpha\gamma} b_{3,\alpha\beta} - \frac{1}{3} \left( b_{1,\beta\gamma}^3 + b_{2,\alpha\gamma}^3 + b_{3,\alpha\beta}^3 \right) \right\rbrace } 
\\\hline 
}
\]
\renewcommand{\arraystretch}{1} 
\caption{Volume of the curves, the divisors and the whole manifold   when using the triangulation ``$S$" at all \ztwo orbifold fixed points.}
\label{table:Z2_Volume_S}
\end{table}

\subsection[The free $\zf$ involution on $\text{Res}(T^6/\ztwo)$]{The free $\boldsymbol{\zf}$ involution on $\boldsymbol{\text{Res}(T^6/\ztwo)}$}

In this subsection we briefly comment on the consequences of modding out the freely acting $\zf$ involution on the resolution of $T^6/\ztwo$; a more detailed and complete account of this can be found in appendix \ref{sc:NonFact}. In particular we show that the resulting 
geometry can be obtained equivalently as a $\ztwo$ orbifold of the non-factorizable lattice $\text{SU}(4) \times \text{SU}(2)^3$, cf. \cite{last,Faraggi:2006}. The $\zf$ action on the orbifold lifts to the resolution as well, provided that the \zf\ is compatible with the triangulation.

The description of the $T^6/\ztwo$ orbifold resolution \eqref{SingularityEquation} involves the vectors $N_\ga$ that parameterize the four $\Intr_2$ fixed points of the tori on the elliptic curves. The $\gt$ action on these vectors is given in equations \eqref{FixedPointMapping} of appendix \ref{sc:Weierstrass}. In order that the equations \eqref{GlobalCoordConstraints}, which define the local homogeneous resolution coordinates ($z_{i,\gs}$ and $x_{j,\gr\gs}$), are invariant, they have to transform as  
\begin{subequations} 
\equ{
x_{i,11} \leftrightarrow x_{i,22}\;,
\qquad\quad  
x_{i,12} \leftrightarrow x_{i,21}\;,
\qquad\quad  
x_{i,13} \leftrightarrow x_{i,24}\;,
\qquad\quad  
x_{i,14} \leftrightarrow x_{i,23}\;,
\\[2ex] 
x_{i,31} \leftrightarrow x_{i,42}\;,
\qquad\quad  
x_{i,32} \leftrightarrow x_{i,41}\;,
\qquad\quad  
x_{i,33} \leftrightarrow x_{i,44}\;,
\qquad\quad 
x_{i,34} \leftrightarrow x_{i,43}\;,
\\[2ex]
z_{i,1} \mapsto z_{i,2}\;, 
\quad 
z_{i,2} \mapsto   \sqrt{(\gve_1-\gve_2)(\gve_3-\gve_2)}\, z_{i,1}\;, 
\quad 
z_{i,3} \mapsto \sqrt{\gve_2-\gve_1}\, z_{i,4}\;,
\quad 
z_{i,4} \mapsto \sqrt{\gve_2-\gve_3}\, z_{i,3}\;, 
\\[2ex] 
y \mapsto - (\gve_1-\gve_2)^3(\gve_3-\gve_2)^3\, y\;. 
}
\label{GlobalCoordtau}
\end{subequations}

Hence, as expected, for the exceptional divisors the mapping under \zf\ is the same as the mapping of the fixed points. 
On the resolution the $\gt^2$ acts as the identity only after $\Cplx^*$ scalings in each of the three tori by $\gm_i = (\gve_1-\gve_2)(\gve_3-\gve_2)$, see below \eqref{CPtau} in appendix \ref{sc:Weierstrass}.

From equations \eqref{GlobalCoordtau} we infer that the resolution of $T^6/\ztwo$ orbifold only admits a corresponding freely acting \zf\ involution provided that the triangulations of  32 \ztwo fixed points and their 32 images under the \zf\ are pairwise equal. This reduces the number of triangulations to 
\equ{
\frac{4^{32}}{3! (8^3 / 2)} \approx 1.20 \cdot 10^{16}\;.
}
The factor $8^3/2$ 
comes about as follows: 
Remember the factors $4!$ in \eqref{eq:NTriang_Z2} came from the permutations in the indices $\alpha, \beta$, and $\gamma$. But there are permutations which transform a $\zf$ symmetric triangulation into a non symmetric one. The action of $\tau$ on the indices in permutation cycle language is $(12)(34)$. Now only those permutations are allowed that commute with $\tau$. They form a subgroup $D_4 \subset S_4$ which is generated by $(1324)$ and $(12)$. Since $|D_4|=8$ the number of inequivalent $\zf$ symmetric permutations of all three indices equal $8^3 / 2$.

%
%
%
%
%
%
%
%
%
%
%
%
%
%
%
%
%
%
%
%
%
%
%
%
%
%
%
%
%
%
%
%
%
%
\section{Heterotic supergravity on the resolved $\boldsymbol{T^6/\ztwo}$ orbifold}
\label{ch:Preliminaries}

We consider the compactification of ten dimensional heterotic supergravity on the resolved $T^6/\ztwo$ orbifold discussed in the previous section, in the presence of an Abelian gauge flux. We review topological consistency requirements such a compactification needs to fulfill and how the chiral matter spectrum can be computed. The section is concluded by an investigation of the consequences of the (loop corrected) Donaldson-Uhlenbeck-Yau equations. The general results are illustrated by explicit computations of two \ztwo resolutions using triangulation ``$E_1$" and ``$S$''.

\subsection{Topological consistency conditions}
\label{sc:Topological_consistency_conditions}

Compactifications of the heterotic string on CY spaces have to fulfill various consistency requirements. First of all the Green-Schwarz anomaly cancellation leads to constraints:  The field strength $H$ of the two-form field $B$ is globally defined by 
\begin{equation}
\label{eq:Bi_raw}
H=dB - \frac{\ga'}4\Big( \omega_{3,Y\!M} - \omega_{3,L} \Big)\;, 
\end{equation}
where $\omega_{3,Y\!M}$ and $\omega_{3,L}$ are the Yang-Mills and Lorentz Chern-Simons three-forms, respectively. Hence by acting on it with the exterior derivative one obtains a Bianchi identity. Integrating it over any closed 4-cycle $S$ gives the condition\footnote{For convention on traces see \cite{SGN:2009}.}
\begin{equation}
\label{eq:Bi_int}
0=\int_{\mathcal{S}} dH=\frac{\ga'}4 \, \int_{\mathcal{S}}  \TR(\mathcal{R}^2)-\TR(\mathcal{F}^2)\;.
\end{equation}

To lowest order in $\alpha^{\prime}$ the compactification geometry has to be a Calabi-Yau space, i.e.\ a complex K\"ahler manifold that is Ricci-flat. Higher order string corrections alters the Ricci-flatness with extra curvature corrections. Also the requirement of K\"ahlerianity is generically lost, because the non-integrated Bianchi identity can be written as 
\begin{equation}
i \bder \der J = \frac {\ga'}4 \Big( \TR(\cR^2) - \TR(\cF^2)\Big)\;. 
\end{equation} 
Consequently the K\"ahler form is not closed. The integrated Bianchi identities \eqref{eq:Bi_int} provide necessary conditions such that this condition can be solved globally. However, unless one employs the standard embedding (i.e. setting the gauge connection equal to the spin connection), this equation implies that the geometry is non-K\"ahler \cite{Strominger:1986uh}. As a result the metric and $B$-field background receive $\ga'$ corrections, see e.g.\ \cite{Hull:1997kk}. In this work we are mainly concerned with topological requirements and therefore we ignore such corrections.

In this paper we exclusively focus on Abelian gauge backgrounds. 
Such a gauge flux $\cF$ has to be properly quantized on any curve $C$ existing in the geometry  
\begin{equation}
\frac 1{2\gp } \int_C \cF = L_I H_I\;, 
\label{eq:quant_raw}
\end{equation} 
where $L \cong 0$, i.e.\ a vector in the \eeight root lattice $\gL$, and the $H_I$, $I=1,\ldots 16$ denote the Cartan generators of \eeight. These conditions, one for each independent curve, can be thought of as compatibility conditions for the gluing of the Abelian gauge bundles, initially defined on the various local resolutions only, over the whole resolution.

Given a consistent gauge bundle wrapped on a smooth orbifold resolution, we can compute the chiral spectrum by using index theorems. Following \cite{SGN:2007,SGN:2008} we start from the gaugino anomaly polynomial in ten dimensions and integrate out the six-dimensional internal space. This procedure corresponds to a representation dependent extension of the computation of Dirac indices on the resolution $X$. The resulting multiplicity operator is given by:
\begin{equation}
\label{eq:Multiplicity_Operator_Raw}
N=\int\limits_{X} \left\{ \frac{1}{6} \left(\frac{\mathcal{F}}{2 \pi}\right)^{3}-\frac{1}{24} \tr\left(\frac{\mathcal{R}}{2 \pi}\right)^2  \frac{\mathcal{F}}{2\pi}\right\}\;.
\end{equation}
As the integral is resolution dependent, so is the multiplicity operator. Due to the fact that $X$ is compact, $N$ takes only integral values. For each gaugino state of the ten-dimensional $\eeight$, the multiplicity operator gives the number of copies of states in the four dimensional effective theory.

\subsection{Line bundles on a resolution}
\label{sc:LineBundleRes}

In this paper we only consider Abelian gauge embeddings that disappear in the limit in which all exceptional divisors $E_r$ are shrunk to zero size, i.e.\ in this limit the gauge flux is only present inside the singularities. Gauge backgrounds are characterized by their transition functions. On toric varieties we do not need to give them explicitly as long as we describe how the $\Cplx^*$-scalings act on the gauge degrees of freedom.

To this end we introduce a holomorphic connection one-form $\gG_r$ for each of the $\Cplx^*$-scalings, that transforms as 
\equ{
\gG_r(e^{t_r Q_r} Z) = \gG_r(Z) - \der t_r\;,
} 
where the charges $Q_r$ are defined in table \ref{tb:C*scalings}. 
We can associate a field strength $E_r$ to this connection and its conjugate given by 
\equ{ 
E_r = \frac 1{2\gp} (\bder \gG_r - \der \bgG_r)\;. 
} 
As the notation suggests we identify this field strength with the first Chern class $c_1(E_r)$ of the divisor $E_r$. This can be motivated further by giving an explicit representative of the connection $\gG_r$ (in a specific gauge)
\equ{
\gG_r(Z) = \frac 1{2x_r}\, \d x_r\;. 
}
Using that $1/x_r$ is holomorphic on the complex plane up to a singularity at $x_r=0$, i.e.\ $\bder_{\bx_r} \, \frac1{x_r} = 2 \gp\, \gd^2(x_r,\bx_r)$, we find that 
$E_r = \gd^2(x_r,\bx_r) \, \d \bx_r \d x_r.$ 
Hence, this field strength forces the complex coordinate $x_r$ to $0$ as one expects for the divisor  $E_r := \{x_r = 0\}$ to do so.

Let $A= \cA + \bcA$ be a gauge connection splitted in a holomorphic and an anti-holomorphic part. In the orbifold theory the holomorphic part satisfies the orbifold boundary condition 
\equ{ 
\label{OrbiBound}
\cA(\gth_i z) = e^{- 2\gp i H_r} \cA(z) e^{2\gp i H_r}\;, 
\qquad 
H_r = V_r^I\,H_I\;.
}
Therefore, the $V_r^I$ can be thought of as the local orbifold gauge shifts defined up to lattice vectors of \eeight. On the resolution this local orbifold transformation is promoted to a $\Cplx^*$-scaling \eqref{CoordC*scalings} with parameter $t_r$ of the homogeneous coordinates $Z$ given by 
\equ{ 
\label{ResBound}
\cA(Z) \ra \cA(e^{t_r Q_r}Z) =  g_r(t_r) \inv \big( \cA(Z) + \der  \big) g_r(t_r)\;, 
\qquad 
g_r(t_r) = e^{t_r H_r}\;. 
}
The matrices $V_{r}^{I}$ here encode the embedding of the line bundle associated to the exceptional divisors $E_{r}$. Note that on the resolution they are no longer just defined up to lattice vectors. This is pure gauge on the 66 dimensional classifying space, but definitely not pure gauge on the resolution itself. The connection $\gG_r$ constructed above can be used as the background $\cA_0$ in $\cA$ and hence we expand it as 
\equ{
\cA = \cA_0 + \cB\;, 
\qquad 
\cA_0 = \gG_r \, H_r\;, 
}
where the perturbation $\cB(Z)$ transforms homogeneously under conjugation:
\(
\cB(Z) \ra \cB(y,e^{t_r\, Q_r} Z) = e^{- t_r\, H_r}\, \cB(y,Z)\, e^{t_r\, H_r}\;.
\)
The gauge background that combines the gauge fluxes supported at the exceptional divisors of the resolved singularities is then given by 
\begin{equation}
\label{eq:F_Definition}
\frac{\mathcal{F}}{2\pi }=
\frac 1{2\gp} \big( \bder \cA_0 - \der \bcA_0 \big) 
= E_{r} \,V_{r}^{I}\, H_{I}\;. 
\end{equation}

We conclude our general discussion by returning to the Bianchi identities . The curvature-dependent part of them is related to the second Chern class 
\( 
c_{2}(X)=-\tr{\mathcal{R}^{2}}/{(8\pi^{2})}. 
\) 
Therefore, we can use \eqref{eq:Z2_Chern_Class2} to express it as a product of all divisors, using that the first Chern class vanishes for a Calabi-Yau. This allows us to rewrite the Bianchi identities in terms of divisors as
\begin{equation}
\label{eq:BIs}
S \Big[\Big(\sum\limits_{r} E_{r} V_{r}\Big)^{2} - \sum\limits_{v} S_{v}^{2}\Big] = 0\;,
\end{equation}
for any 4-cycle $S$. Because a given CY contains many closed 4-cycles this leads to a large set of consistency requirements. This means that for a resolution of $T^6/\ztwo$ orbifold there are 51 independent conditions. When one enforces the resolution to be compatible with the \zf\ action some of these conditions become dependent and 3+24 = 27 conditions remain.

\begin{table}[t]
\[
\renewcommand{\arraystretch}{\stretcharry} 
\arry{|c|c||c|c|}{
\hline 
\multicolumn{4}{|c|}{\text{Flux quantization in triangulation ``$E_1$''}}
\\ \hline\hline 
%
%
E_{1,\beta\gamma} E_{2,\alpha\gamma} & 2V_{2,\alpha\gamma} \cong 0  
& 
D_{1,\alpha}E_{1,\beta\gamma} 	& 
V_{1,\beta\gamma}-V_{2,\alpha\gamma}-V_{3,\alpha\beta} \cong 0 
\\
E_{1,\beta\gamma} E_{3,\alpha\beta} & 2V_{3,\alpha\beta} \cong 0 
&
D_{1,\alpha}E_{2,\alpha\gamma} & 
4V_{2,\alpha\gamma}-\sum\limits_{\beta}V_{1,\beta\gamma} \cong 0 
\\
E_{2,\alpha\gamma} E_{3,\alpha\beta} & 0 	
&
D_{1,\alpha}E_{3,\alpha\beta} & 
4V_{3,\alpha\beta}-\sum\limits_{\gamma}V_{1,\beta\gamma} \cong 0 
\\\hline 
D_{2,\beta}E_{1,\beta\gamma} & 
-\sum\limits_{\alpha} V_{3,\alpha\beta} \cong 0 
&
D_{3,\gamma}E_{1,\beta\gamma} &  
-\sum\limits_{\alpha}V_{2,\alpha\gamma} \cong 0  	
\\
D_{2,\beta}E_{2,\alpha\gamma} & 0 
&
D_{3,\gamma}E_{2,\alpha\gamma} & 
-\sum\limits_{\beta}V_{1,\beta\gamma} \cong 0 	
\\	
D_{2,\beta}E_{3,\alpha\beta} & 
-\sum\limits_{\gamma}V_{1,\beta\gamma} \cong 0 
&
D_{3,\gamma}E_{3,\alpha\beta} & 0 
\\\hline 
R_{i}E_{i,\rho\sigma} & 2V_{i,\rho\sigma} \cong 0 
&
R_{i}R_{j} & 0
\\
R_{i}D_{j,\rho} & 
-\sum\limits_{\sigma}V_{i,\rho\sigma} \cong 0 
~~(i \neq j)
& &
\\\hline 
}
\]
\renewcommand{\arraystretch}{1} 
\caption{The flux quantization 
$\int_C \cF / (2 \gp)$ on curves $C$ using triangulation ``$E_1$'' at all \ztwo orbifold fixed points.}
\label{table:Z2_Quant_E1}
\end{table}

\subsection{Consistent line bundles and spectra within specific triangulations}

As explained in subsection \ref{sc:Int_Num}, the dependence of the triple intersection numbers on the specific triangulation (cf. table \ref{table:Z2_Intersection_Numbers}) represents a serious obstacle to a fully generic bundle construction, since both the flux quantization conditions \eqref{eq:quant_raw} and the Bianchi identities \eqref{eq:BIs} are strongly triangulation dependent. Due to the level of this complication, we only consider resolutions of $T^6/\ztwo$ where the same triangulation is taken at each \ztwo singularity. Moreover, we only consider the triangulations ``$E_1$'' and ``$S$'' here, because the triangulations ``$E_1$'', ``$E_2$'' and ``$E_3$'' are the same up to permutations of the various labels.

\subsubsection[Triangulation ``$E_1$'']{Triangulation ``$\boldsymbol{E_1}$''}
\label{sc:line_bundle_E1} 

\paragraph{Conditions on the line bundles.} The 360 flux quantization conditions have been summarized in table \ref{table:Z2_Quant_E1}. The other $3\cdot 64 + 3$ conditions in \eqref{eq:quant_raw} are trivially fulfilled, as indicated by $0$,  because the curves $E_{2,\alpha\gamma} E_{3,\alpha\beta}$, $D_{2,\beta}E_{2,\alpha\gamma}$ and $D_{3,\gamma}E_{3,\alpha\beta}$ do not exist in triangulation ``$E_1$'' and the three curves $R_i R_j (i\neq j)$  never intersect any exceptional divisor.  The quantization conditions on the curves $R_i E_{i,\gr\gs}$ tell us that the bundle vectors are half lattice vectors. The other conditions then become various sum rules for the bundle vectors.

Carrying out the integration over the three $R_{i}$ and the $48$ $E_{r}$ the Bianchi identities \eqref{eq:BIs} read:
\begin{subequations}
\label{eq:ex_Z2_BIeqns}
\begin{eqnarray}
\sum\limits_{\beta,\gamma=1}^{4} V_{1,\beta\gamma}^{2} = \sum\limits_{\alpha,\gamma=1}^{4} V_{2,\alpha\gamma}^{2} = \sum\limits_{\alpha,\beta=1}^{4} V_{3,\alpha\beta}^{2}	& =  24 &,	\label{eq:Z2_BIeqnsA}\\[-1.1mm]
\sum\limits_{\alpha=1}^{4} (V_{2,\alpha\gamma}^{2} + V_{3,\alpha\beta}^{2}) 											& =  12 	&,~~~ \beta, \gamma \in \{1,\ldots 4\}\;,		\label{eq:Z2_BIeqnsB}\\[-1.1mm]
2 V_{2,\alpha\gamma}^{2} - V_{2,\alpha\gamma} \cdot \sum\limits_{\beta=1}^{4} V_{1,\beta\gamma}				& =  2	&,~~~ \alpha, \gamma \in \{1,\ldots 4\}\;,	\label{eq:Z2_BIeqnsC}\\[-1.1mm]
2 V_{3,\alpha\beta}^{2} - V_{3,\alpha\beta} \cdot \sum\limits_{\gamma=1}^{4} V_{1,\beta\gamma} 					& =  2	&,~~~ \alpha, \beta \in \{1,\ldots 4\}\;.		\label{eq:Z2_BIeqnsD}
\end{eqnarray}
\end{subequations}
The three equations listed in \eqref{eq:Z2_BIeqnsA} result from integration over the three inherited divisors and correspond to the Bianchi identities on a K3. The $3 \!\cdot\! 16$ equations \eqref{eq:Z2_BIeqnsB} - \eqref{eq:Z2_BIeqnsD} come from integrating over $E_{1,\beta\gamma}$, $E_{2,\alpha\gamma}$, and $E_{3,\alpha\beta}$ respectively. In this case, we end up with $51$ equations for $48$ vectors, each with $16$ unknowns, so altogether there are $768$ unknowns. 

\paragraph{Spectra computation.} Using triangulation ``$E_1$'' at all $\ztwo$ fixed points we find that  the multiplicity operator \eqref{eq:Multiplicity_Operator_Raw} takes the explicit form 
\equa{ 
N_\text{``$E_1$''}  = & 
 \hskip4pt \sum\limits_{\beta,\gamma=1}^{4}H_{1,\beta\gamma}-\frac{1}{3}\Big[\sum\limits_{\alpha,\gamma=1}^{4}(H_{2,\alpha\gamma}-4H_{2,\alpha\gamma}^{3})+\sum\limits_{\alpha,\beta=1}^{4}(H_{3,\alpha\beta}-4H_{3,\alpha\beta}^{3})\Big]
\non \\[1ex] 
 & -  \sum\limits_{\alpha,\beta,\gamma=1}^{4}H_{1,\beta\gamma}(H_{2,\alpha\gamma}^{2}+H_{3,\alpha\beta}^{2})\;.
\label{eq:ex_Z2_Multiplicity_Operator}
}

\subsubsection[Triangulation ``$S$'']{Triangulation ``$\boldsymbol{S}$''}
\label{sc:line_bundle_S}

\paragraph{Conditions on the line bundles.} For the symmetric triangulation, the flux quantization conditions are summarized in table \ref{table:Z2_Quant_S}. In this triangulation, the curves $D_{i,\rho} E_{i,\sigma\tau}$ do not exist. Again we find that all bundle vectors are multiples of half lattice vectors, but the sum conditions are slightly different in this case.

The Bianchi identities \eqref{eq:BIs} for the symmetric triangulation on the divisor basis $(R_i,E_r)$ also become somewhat more complicated
\begin{subequations}
\label{eq:ex_Z2_BIeqnsS}
\begin{eqnarray}
\sum\limits_{\beta,\gamma=1}^{4} V_{1,\beta\gamma}^{2} = \sum\limits_{\alpha,\gamma=1}^{4} V_{2,\alpha\gamma}^{2} = \sum\limits_{\alpha,\beta=1}^{4} V_{3,\alpha\beta}^{2}\!\!\!\!	& =&\!\!\!\!  24\;,	\label{eq:Z2_BIeqnsAS}\\[-1.1mm]
\sum\limits_{\alpha=1}^{4} (-V_{1,\beta\gamma}^{2}\!+\!V_{2,\alpha\gamma}^{2}\!+\!V_{3,\alpha\beta}^{2})\!+\!2 \sum\limits_{\alpha=1}^{4} (~\,\,\,V_{1,\beta\gamma}V_{2,\alpha\gamma} \!+\! V_{1,\beta\gamma}V_{3,\alpha\beta} \!-\! V_{2,\alpha\gamma}V_{3,\alpha\beta})\!\!\!\! & =&\!\!\!\! 8 ,~~\beta,\! \gamma \!\in\! \{1,\ldots 4\},~~~~		\label{eq:Z2_BIeqnsBS}\\[-1.1mm]
\sum\limits_{\beta=1}^{4}  (~\,\,\,V_{1,\beta\gamma}^{2}\!-\!V_{2,\alpha\gamma}^{2}\!+\!V_{3,\alpha\beta}^{2})\!+\!2 \sum\limits_{\beta=1}^{4}(~\,\,\,V_{1,\beta\gamma}V_{2,\alpha\gamma} \!-\! V_{1,\beta\gamma}V_{3,\alpha\beta} \!+\! V_{2,\alpha\gamma}V_{3,\alpha\beta})\!\!\!\! & =&\!\!\!\! 8 ,~~\alpha,\! \gamma \!\in\! \{1,\ldots 4\},~~~~		\label{eq:Z2_BIeqnsCS}\\[-1.1mm]
\sum\limits_{\gamma=1}^{4} (~\,\,\,V_{1,\beta\gamma}^{2}\!+\!V_{2,\alpha\gamma}^{2}\!-\!V_{3,\alpha\beta}^{2})\!+\!2 \sum\limits_{\gamma=1}^{4} (-V_{1,\beta\gamma}V_{2,\alpha\gamma} \!+\! V_{1,\beta\gamma}V_{3,\alpha\beta} \!+\! V_{2,\alpha\gamma}V_{3,\alpha\beta})\!\!\!\! & =&\!\!\!\! 8 ,~~\alpha,\! \beta\! \in \!\{1,\ldots 4\}.~~~~		\label{eq:Z2_BIeqnsDS}
\end{eqnarray}
\end{subequations}
As the intersection numbers including the inherited divisors are triangulation independent, the Bianchi identities \eqref{eq:Z2_BIeqnsA} and \eqref{eq:Z2_BIeqnsAS} are identical. The $3\cdot 16$ equations \eqref{eq:Z2_BIeqnsBS}-\eqref{eq:Z2_BIeqnsDS} result from integration over the $3\cdot 16$ exceptional divisors $E_{i,\rho\sigma}$. Due to the symmetric structure, they are all similar. However, owing to the fact that all exceptional divisors intersect, the equations contain more terms and are stronger coupled as those obtained for the asymmetric triangulations.

\paragraph{Spectra computation.} For the multiplicity operator \eqref{eq:Multiplicity_Operator_Raw} in the symmetric triangulation we obtain
\equa{ 
\label{eq:ex_Z2_Multiplicity_OperatorS}
N_\text{``$S$''}  = & 
\frac{1}{3} ~~\!\sum\limits_{i=1}^{3}\sum\limits_{\rho,\sigma=1}^{4} \Big(2H_{i,\rho\sigma}^3 + H_{i,\rho\sigma}\Big)+\sum\limits_{\alpha,\beta,\gamma=1}^{4}\Big(H_{1,\beta\gamma} H_{2,\alpha\gamma}H_{3,\alpha\beta}\Big)
\\[1ex] 
 & - \frac{1}{2} \sum\limits_{\alpha,\beta,\gamma=1}^{4}\Big(H_{1,\beta\gamma}^{2} H_{2,\alpha\gamma}\!+\!H_{1,\beta\gamma} H_{2,\alpha\gamma}^{2}\!+\!H_{1,\beta\gamma}^{2} H_{3,\alpha\beta}\!+\!H_{1,\beta\gamma} H_{3,\alpha\beta}^{2}\!+\!H_{2,\alpha\gamma}^{2} H_{3,\alpha\beta}\!+\!H_{2,\alpha\gamma} H_{3,\alpha\beta}^{2}\Big)\;.
\non 
}

\begin{table}[t]
\[
\renewcommand{\arraystretch}{\stretcharry} 
\arry{|c|c||c|c|}{
\hline 
\multicolumn{4}{|c|}{\text{Flux quantization in triangulation ``$S$''}}
\\ \hline\hline 
%
%
E_{1,\beta\gamma} E_{2,\alpha\gamma} & -V_{1,\beta\gamma}-V_{2,\alpha\gamma}+V_{3,\alpha\beta} \cong 0  
& 
D_{1,\alpha}E_{1,\beta\gamma} 	&  0 
\\
E_{1,\beta\gamma} E_{3,\alpha\beta} & -V_{1,\beta\gamma}+V_{2,\alpha\gamma}-V_{3,\alpha\beta} \cong 0  
&
D_{1,\alpha}E_{2,\alpha\gamma} & \sum\limits_{\beta}V_{3,\alpha\beta} \cong 0 
\\
E_{2,\alpha\gamma} E_{3,\alpha\beta} & ~V_{1,\beta\gamma}-V_{2,\alpha\gamma}-V_{3,\alpha\beta} \cong 0  
&
D_{1,\alpha}E_{3,\alpha\beta} & \sum\limits_{\gamma}V_{2,\alpha\gamma} \cong 0 
\\\hline 
D_{2,\beta}E_{1,\beta\gamma} & \sum\limits_{\alpha}V_{3,\alpha\beta} \cong 0 
&
D_{3,\gamma}E_{1,\beta\gamma} & \sum\limits_{\alpha}V_{2,\alpha\gamma} \cong 0 
\\
D_{2,\beta}E_{2,\alpha\gamma} & 0 
&
D_{3,\gamma}E_{2,\alpha\gamma} & \sum\limits_{\beta}V_{1,\beta\gamma} \cong 0 	
\\	
D_{2,\beta}E_{3,\alpha\beta} & \sum\limits_{\gamma}V_{1,\beta\gamma} \cong 0 
&
D_{3,\gamma}E_{3,\alpha\beta} & 0 
\\\hline 
R_{i}E_{i,\rho\sigma} & 2V_{i,\rho\sigma} \cong 0 
&
R_{i}R_{j} & 
0 
\\
R_{i}D_{j,\rho} & 
-\sum\limits_{\sigma}V_{i,\rho\sigma} \cong 0 
~~(i \neq j)
& &
\\\hline 
}
\]
\renewcommand{\arraystretch}{1} 
\caption{The flux quantization $\int_C \cF / (2 \gp)$ on curves $C$ using the triangulation ``$S$'' at all \ztwo orbifold fixed points. As we will see in section \ref{sc:LineBundleOrbiShift}, the resulting set of conditions turns out to be equivalent to the ones obtained from the other triangulations, e.g. from ``$E_1$'' in table \ref{table:Z2_Quant_E1}.}
\label{table:Z2_Quant_S}
\end{table}

\subsection{Donaldson-Uhlenbeck-Yau equations}
\label{sc:DUY}

In addition to the topological conditions on the gauge flux, we have to impose the so-called Donaldson-Uhlenbeck-Yau (DUY) equations. These equations are obtained by integrating the Hermitian Yang-Mills (HYM) equations over the whole manifold. The one-loop corrected DUY equations \cite{Blumenhagen:2005} read
\begin{equation}
\label{eq:DUY_Equation}
\frac{1}{2} \int\limits_{X} J \wedge J \wedge \frac{\mathcal{F}}{2\pi} 
=
\gx_{1L} 
= \frac{e^{2\phi}}{16 \pi} \int\limits_{X} \frac{1}{(2 \pi)^{3}} \left[ \left( \tr \mathcal{F}^{\prime 2}-\frac{1}{2} \tr \mathcal{R}^{2} \right) \mathcal{F}^{\prime} 
+ (\cF' \ra \cF'') 
\right]\;,
\end{equation}
where $\mathcal{F}^{\prime}$ and $\mathcal{F}^{\prime\prime}$ are the (internal) gauge fluxes in the first and second $\text{E}_{8}$, respectively. These are $16$ equations involving the components of the line bundle vectors $V_{r}^{I}$. Inserting the expression for the Abelian gauge flux \eqref{eq:F_Definition}, we can rewrite the HYM as conditions on  the volumes of the exceptional divisors as
\begin{equation}
\label{eq:DUY_in_Volumes}
\text{vol}(E_{r})\, V_{r} = \gx_{1L}\;, 
\end{equation}
using \eqref{eq:Volume_Definitions} and Poincar\'e duality. The right hand side is generically non-zero, which implies that it is generically impossible that all divisor volumes vanish simultaneously.

These equations can be viewed as dynamical conditions on the K\"ahler moduli. A geometrical description in the supergravity approximation is only valid when the volumes of all divisors are positive and large. Consequently, the DUY equations can be in serious conflict with the validity of supergravity. Indeed, if we consider the case in which entries of the $V$'s all have (semi)definite sign, the corresponding volumes must all be zero in absence of the loop-corrections, i.e.\ when $\gx_{1L} =0$. The loop corrections can move some of the volumes away from zero. But as a perturbative loop effect, these corrections will be relatively small compared to the string scale. This suggest that certain resolution models have a geometry which is only partially in blow-up.

In light of this, one may question whether the computation of spectra (using the methods reviewed in previous subsections) is under control at all if the DUY equations confine the resolved space to a complete or partial blow-down. On the other hand,  one can argue in the following way that the computation makes sense: The DUY equations arise as absolute minima of D-term potentials which are proportional to the respective gauge couplings, i.e.\ the string coupling up to volume factors. In the limit of vanishing string coupling the D-term potential vanishes. Consequently, the DUY equations as well as the constraints on the volumes of the divisors become obsolete. Hence, in this limit we can make the volumes sufficiently large so that the supergravity approximation is valid, and the spectrum computation can be trusted. As the string coupling controls the genus expansion, changing the string coupling does not induce topological changes of the geometry which have to do with the worldsheet theories on the corresponding Riemann surfaces. Consequently, since the spectra we compute are chiral, they are preserved when the string coupling is subsequently switched on.

A possible way to avoid ending up in the (partial) blow-down regime at finite string coupling might also be provided by viewing the DUY equations as a collection of D-flatness conditions: If a certain set of charged matter fields take VEVs, thus allowing for large volumes, the supergravity approximation can be saved. This clearly indicates that the true vacuum is not described by a combination of line bundles as we have assumed, but rather by some more complicated non-Abelian bundle. A study of such configuration will not be pursued in the present work.

\subsection{Spectra computations on the resolution with free involution}
\label{sc:freeWLresolution}

We have described how to compute chiral spectra for SUGRA models built on resolved $T^6/\ztwo$ spaces, by using a representation-dependent chiral index. In this subsection we comment how one can extend this procedure after modding out the freely action $\zf$.

The computation of the spectrum on the $T^6/\ztwo\times \zf$ orbifold can be done straightforwardly using standard CFT techniques for orbifolds \cite{Blaszczyk:2009in}. They, in particular, require adding new twisted sectors associated with $\zf$ and build invariant combinations of the states that already exist in the spectrum before modding out the $\zf$. In practice, for the massless spectrum this boils down to the following: The novel twisted sectors are irrelevant as they only produce massive winding modes. The gauge symmetry gets broken by the Wilson line associated with the $\zf$ and only untwisted states that are inert under this Wilson line remain massless. And finally, the fixed tori, on which the twisted matter lives, are identified in pairs under $\zf$, so that the twisted spectrum simply becomes halved.

However, in this paper we are primarily concerned with model building on resolutions, hence we need to determine the spectrum on a $T^6/\ztwo\times \zf$ resolution. To determine such a spectrum one can proceed in two ways: One can either (i)  take half of the chiral spectrum determined on a $\zf$-compatible resolution of $T^6/\ztwo$, or (ii) directly compute the chiral spectrum using the multiplicity operator defined on the resolution of $T^6_{\rm non}/\ztwo = (T^6/\zf)/\ztwo$.

The first approach is heavily motivated by orbifold knowledge: Given that the untwisted orbifold spectrum is always non-chiral for $\ztwo$ orbifolds, chirality only arises from the twisted sectors. The fixed tori, where the twisted matter states are localized, are identified in pairs under $\zf$. 
This seems to suggest that the chiral spectrum on a resolution of $T^6/(\ztwo \times \zf)$ is simply half of the one obtained on the corresponding resolution of $T^6/\ztwo$, provided the branching of representations due to the gauge symmetry breaking induced by the $\zf$ action is taken into account. This procedure has some potential serious flaws: Firstly, one assumes that one can uniquely associate a resolution model with an orbifold construction. Secondly, one assumes that the orbifold spectrum after branching will always be larger than the one of the resolution model. Finally, one makes assumptions about where the chiral matter states are localized on the resolved orbifold, in particular that the chiral resolution matter originates from the twisted sectors. However, the extended chiral multiplicity operator \eqref{eq:Multiplicity_Operator_Raw} only gives the multiplicity of states but does not specify where the states are originating from. As we will discuss in the next section, the first two issues are more subtle: The matching between orbifold and resolution models is complicated for various technical reasons, and it turns out that even in very simple resolutions the spectrum after branching can be larger than the one of the corresponding orbifold.

Consequently, only the second approach can be trusted. It is direct computation, which does not rely in any way on the orbifold knowledge or any a priori assumptions about the resolution spectrum. Nevertheless one can show that the intuition of the former approach, i.e.\ that the chiral spectrum gets halved, is correct. The details of the computation of the multiplicity operator $N_{\rm non}$ are given in appendix \ref{sc:MultiplNonFact}. In particular one can show that $N_{\rm non}=N/2$, where $N$ is the multiplicity operator on the corresponding $\zf$-compatible resolution of $T^6/\ztwo$. For this reason, we simply divide the resolution spectrum on $\text{Res}(T^6/\ztwo)$ by two in the main text of the paper, and refer for the detailed discussion to appendix \ref{sc:MultiplNonFact}.

%
%
%
%
%
%
%
%
%
%
%
%
%
%
%
%
%
%
%
%
%
%
%
%
%
%
%
%
%
%
%
\section{Matching of supergravity and orbifold constructions}
\label{sec:Relation_Orbifold_CY}

This section is devoted to the question to which extent one can make a direct and complete identification between heterotic orbifold and resolution models based on $\ztwo$ orbifolds.

\subsection[\eeight embeddings of line bundles and orbifold actions]{$\boldsymbol{\eeight}$ embeddings of line bundles and orbifold actions}
\label{sc:LineBundleOrbiShift}

Given any element $g$ of the space group of an orbifold, its action is embedded in the gauge degrees of freedom via a local shift vector $V_{g}$. This induces a gauge symmetry breaking localized in the $g$-fixed points, whose details are encoded in $V_{g}$.
The counterpart of this gauge symmetry breaking in the supergravity model built on the resolved orbifold is the very presence of a gauge background: in the blow-down limit the gauge flux $\cF$ gets indeed squeezed inside the orbifold singularities and its effects are just monodromies around them. 
The identification of these monodromies in the CFT and supergravity constructions provides the key to the matching: We can identify the Abelian bundle vectors $V_{r}$ with the local shifts $V_g$  (see e.g.\ \cite{SGN:2008,SGN:2009}) up to $\text{E}_8\times \text{E}_8$ lattice elements:
\begin{equation}
\label{eq:V_indentification}
V_{g}\cong \int_C \left. \frac{\mathcal{F}}{2 \pi}\right|_{r}=V_{r}\;,
\end{equation}
locally at each singularity. Here $C$ denotes a curve of an ordinary divisor that intersects with the exceptional divisor inside the singularity, and the vertical bar means restriction to the local fixed point (labeled by $r$) that is being investigated, i.e. all other exceptional divisors in $\mathcal{F}$ are not considered. 
In this paper we have described a different way of obtaining these identifications: Recall that in subsection \ref{sc:LineBundleRes} we showed that the orbifold boundary conditions \eqref{OrbiBound} are lifted to $\Cplx^*$-scaling actions \eqref{ResBound} in blow-up.

In the $T^6/\ztwo$ case the resulting relations between the line bundle vectors and the local orbifold shifts and Wilson lines are given by 
\begin{equation}
V_{1,\beta\gamma} \cong  V_1+\sum_{i\neq 1,2} n_{1,\gb\gg}^i W_i\;,
\quad 
V_{2,\ga\gamma} \cong  V_2+\sum_{i\neq3,4} n_{2,\ga\gg}^i W_i\;,
\quad 
V_{3,\ga\beta} \cong  V_3+\sum_{i\neq5,6} n_{3,\ga\gb}^i W_i\;,
\label{eq:ex_Z2_Identification}
\end{equation}
where the dependence of the $n$'s on the indexes $\alpha$, $\beta$ and $\gamma$ is defined in equation \eqref{Def_n} and table \ref{tb:Z2_sectors}. This leads to the seemingly asymmetric situation: On one hand we have 48 bundle vectors, while on the other hand the identification with the orbifold boundary conditions allows for only eight independent input vectors up to lattice vectors (i.e. two shifts and six Wilson lines). To understand that there is no paradox, we have to consider carefully the consequences of the flux quantization conditions.

The expressions for the flux quantization conditions are strongly triangulation dependent, as can be seen e.g.\ from tables \ref{table:Z2_Quant_E1} and \ref{table:Z2_Quant_S}. However, for any choice of triangulations of the 64 local $\ztwo$ resolutions these quantization conditions are always equivalent to the following set of conditions: 
\begin{subequations}
\equ{
 2 V_{k,\rho\sigma} \cong 0 \;, \qquad k=1,2,3\;, \quad \rho,\sigma = 1,\ldots,4 \;,
\label{FQordertwo}
\\[2ex] 
\sum_\rho V_{k,\rho\sigma} \cong 0\;,
\qquad 
 \sum_\rho V_{k,\sigma\rho} \cong 0 \;, \qquad k=1,2,3\;, \sigma=1,\ldots,4 \;.
\label{FQsum}
\\[0ex] 
V_{1,\beta\gamma} + V_{2,\alpha\gamma} + V_{3,\alpha\beta} \cong 0 \;, \qquad \alpha,\beta,\gamma=1,\ldots,4 \;, 
\label{FQfixedpoints}
}
\label{FQ}
\end{subequations}

The first two conditions are obtained from the flux quantization on the curves $R_k E_{k,\rho\sigma}$ and $R_k D_{j,\sigma}$, 
respectively, and are therefore triangulation independent. Depending on the triangulation of the resolution of the fixed point $(\ga,\gb,\gg)$ either the curve $D_{1,\alpha}E_{1,\beta\gamma}$ or the curve $E_{2,\ga\gg}E_{3,\ga\gb}$ exists (and similarly for cyclic permutation of the labels 1, 2, 3). However, in all the cases the resulting set of flux quantization conditions is equivalent to \eqref{FQfixedpoints}, upon using \eqref{FQordertwo}. Hence, the set of quantization conditions \eqref{FQ} is in fact universal, i.e.\ valid for any triangulation. Moreover, the universal flux quantization conditions \eqref{FQ} are automatically fulfilled by bundle vectors fulfilling the matching identities of eq.~\eqref{eq:ex_Z2_Identification}.

In addition, using the flux quantization conditions \eqref{FQ} the 48 line bundle vectors can be expressed in terms of eight fundamental ones. These can be chosen to be $V_{1,1i}, V_{2,i1}$ and $V_{3,1i}$ with $i=1,2,3$, with a single constraint 
\equ{
V_{1,11} + V_{2,11} + V_{3,11} \cong 0\;, 
\label{FQconstraint} 
}
which follows immediately from \eqref{FQfixedpoints} for $(\ga,\gb,\gg) =(1,1,1)$. In appendix \ref{sc:FluxQuant} we show this constructively. Out of this set of eight fundamental line bundles we can reconstruct the orbifold shifts and Wilson lines as 
\begin{subequations}
\begin{align}
V_1 := V_{1,11}\;, \qquad V_2 := V_{2,11} \,, \qquad V_3 := V_{3,11} \;,
\\[1ex]  
W_1 := V_{2,31} - V_{2,11} \;, \qquad W_2 := V_{2,21} - V_{2,11} \;, 
\\[1ex] 
W_3 := V_{3,13} - V_{3,11} \;, \qquad W_4 := V_{3,12} - V_{3,11} \;, 
\\[1ex] 
W_5 := V_{1,13} - V_{1,11} \;, \qquad W_6 := V_{1,12} - V_{1,11} \;.
\end{align}
\end{subequations}
with $V_3 \cong V_1 + V_2$ because of the constraint \eqref{FQconstraint}. Hence, the characterization of the gauge flux in terms of the 48 line bundles contains the same amount of information up to lattice vectors as the two shifts and six orbifold Wilson lines that a heterotic $T^6/\ztwo$ model can maximally be equipped with.

\subsection{Open points in the matching: the problem of brothers}
\label{sc:OpenIssues}

From the orbifold perspective the blow-up is generated by giving VEVs to twisted states. A complete blow-up of an orbifold therefore requires that in all twisted sectors at least one twisted state, corresponding to the blow-up mode\footnote{Generically not all twisted states correspond to blow-up modes: some of them may be linked to deformations of the resolved space, i.e. changes in the complex structure moduli, or to bundle deformations etc.} takes a non-vanishing VEV. When a complex twisted scalar $\gf_r$ takes a VEV then the field redefinition
\begin{equation}
\label{eq:Superfield}
\gf_{r}=M_{s} e^{2\pi U_{r}}=M_{s} e^{2\pi(b_r+i\beta_{r})}\;, 
\end{equation}
can be used to transmute it into an axion $\gb_r$ and a K\"ahler modulus $b_r$. The shifted (gauge) momentum $P_{sh}$ of a blow-up mode $\gf_{r}$, being localized at a singularity corresponding to $r$ and developing a VEV, can be identified with a line bundle vector $V_{r}$ in the corresponding patch $r$ of the resolution
\begin{equation}
V_{r} = P_{sh}\;.
\end{equation}
Therefore, specifying all twisted states that take VEVs defines in principle the resolution model precisely.

\begin{table}[t]
\begin{center}
\begin{tabular}{|ll|l|l|l|}
\hline
&&&&\\[-1.5ex]
\multicolumn{2}{|c|}{U sector} &  $\gth_1$-sector &  $\gth_2$-sector &  $\gth_3$-sector \\[2pt]
\hline
\hline
&&&&\\[-2.2ex]
$(\boldsymbol{27},\boldsymbol{1})_{(-2,-2)}$ & $(\overline{\boldsymbol{27}},\boldsymbol{1})_{(2,2)}$  & 
$16 (\overline{\boldsymbol{27}},\boldsymbol{1})_{(2,0)}$  &
$16 (\overline{\boldsymbol{27}},\boldsymbol{1})_{(-1,1)}$ &
$16 (\overline{\boldsymbol{27}},\boldsymbol{1})_{(-1,-1)}$ \\[2pt]
$(\boldsymbol{27},\boldsymbol{1})_{(-2,2)}$ & $(\overline{\boldsymbol{27}},\boldsymbol{1})_{(2,-2)}$ & 
$16 (\boldsymbol{1},\boldsymbol{1})_{(-6,0)}$ &
$16 (\boldsymbol{1},\boldsymbol{1})_{(3,-3)}$ &
$16 (\boldsymbol{1},\boldsymbol{1})_{(3,3)}$ \\[2pt]
$(\boldsymbol{27},\boldsymbol{1})_{(4,0)}$ & $(\overline{\boldsymbol{27}},\boldsymbol{1})_{(-4,0)}$ &
$32 (\boldsymbol{1},\boldsymbol{1})_{{0,2}}$ &
$32 (\boldsymbol{1},\boldsymbol{1})_{(3,1)}$ &
$32 (\boldsymbol{1},\boldsymbol{1})_{(3,-1)}$ \\[2pt]
$(\boldsymbol{1},\boldsymbol{1})_{(6,-2)}$ & $(\boldsymbol{1},\boldsymbol{1})_{(-6,2)}$ &
$32 (\boldsymbol{1},\boldsymbol{1})_{(0,-2)}$ &
$32 (\boldsymbol{1},\boldsymbol{1})_{(-3,-1)}$ & 
$32 (\boldsymbol{1},\boldsymbol{1})_{(-3,1)}$ \\[2pt]
$(\boldsymbol{1},\boldsymbol{1})_{(6,2)}$ & $(\boldsymbol{1},\boldsymbol{1})_{(-6,-2)}$ & & & \\[2pt]
$(\boldsymbol{1},\boldsymbol{1})_{(0,4)}$ & $(\boldsymbol{1},\boldsymbol{1})_{(0,-4)}$ & & & \\[2pt]
\hline
\end{tabular}
\end{center}
\caption{The massless spectrum of the \ztwo standard embedding orbifold model with gauge group $\text{E}_6\times\text{U}(1)^2\times\text{E}_8$.}
\label{table:Z2xZ2_StandardEmbeddingSpectrum}
\end{table}

Furthermore, the description of the resolution model crucially depends on the triangulations chosen at the singularities. From the orbifold point of view, the triangulation can be determined by the relative sizes of the VEVs of the three twisted states at a given \ztwo singularity. For example, if 
\equ{
\left| 
\frac {\gf_{1,\gb\gg} M_s}{\gf_{2,\ga\gg} \gf_{3,\ga\gb}} 
\right| = e^{2\pi (b_{1,\gb\gg}-b_{2,\ga\gg}-b_{3,\ga\gb})}\; >\; 1\;,
}
we need to use triangulation ``$E_1$'' for the resolution of the $\ztwo$ fixed point $(\ga,\gb,\gg)$. To see this notice that in triangulation ``$E_1$'' the curve $D_{1,\ga} E_{1,\gb\gg}$ has positive volume only when $b_{1,\gb\gg}-b_{2,\ga\gg}-b_{3,\ga\gb} > 0$, cf. table \ref{table:Z2_Volume_E1}.

A different problem arises when starting form a \ztwo orbifold resolution model. Given a resolution model defined by a set of line bundle vectors,  it is clear, in principle, what happens in the blow-down limit. The problem in this case is that the \ztwo orbifold can be equipped with additional modular invariant phases under which the various partition functions of different twisted sectors can be combined. One way of generating these phases is by adding $\text{E}_8\times \text{E}_8$ lattice elements to the defining shifts and Wilson lines of the orbifold. Naively one expects that this does not make any difference, but in fact, it might lead to different projections. This means that one orbifold model is in fact part of a whole collection of ``brother'' models. Finding the correct blow-down model associated with a resolution model means finding the right brother model, and this cannot be done by just relying on the matching procedure sketched in \ref{sc:LineBundleOrbiShift}, precisely since, as noted there, that procedure is blind to $\text{E}_8\times \text{E}_8$ lattice elements. We want to stress that this kind of issue is specific to orbifold models whose geometry admits the presence of ``brothers,'' so in particular it affects the matching treated in \cite{SGN:2009} in the $\Intr_{6-II}$ case, without any consequence in the cases studied in 
\cite{Honecker:2006qz,SGN:2007,Nibbelink:2008tv,SGN:2008}.

\subsection{Novel states on the resolution: the standard embedding example}
\label{sc:NovelStates}

Since going from the orbifold point to a resolution model is done by giving a non-trivial VEV to the blow-up modes (i.e.\  twisted orbifold fields), this procedure can generate mass terms for some fields just in the way the Higgs VEV does in the Standard Model. Thus naively one would expect to see at most as many chiral states on the resolution as on the orbifold. However, we observe that there are resolution states with no counterpart on the orbifold and that the number of these states jumps between the different triangulations. These states turn out to be invisible from the orbifold perspective which may indicate that they are non-perturbative in nature. Indeed, it is possible to explain the disappearance of these states by blow-up moduli dependent mass terms.

A concrete and simple setup to study this phenomenon is the standard embedding. The orbifold model is specified by the shifts
\begin{equation}
V_1 = \Big(0,\frac{1}{2},-\frac{1}{2},0^5\Big)\Big(0^8\Big) \quad\text{and}\quad V_2 = \Big(-\frac{1}{2},0,\frac{1}{2},0^5\Big)\Big(0^8\Big)
\end{equation}
and vanishing Wilson lines $W_i = 0$, $i=1,\dots,6$. The resulting 4d model has $\text{E}_6\times\text{U}(1)^2\times\text{E}_8$ gauge group and the charged matter spectrum consists of $3 (\boldsymbol{27},\boldsymbol{1}) + 51 (\overline{\boldsymbol{27}},\boldsymbol{1})$ and 246 singlets (charged under $\text{U}(1)^2$). It is listed in detail in table \ref{table:Z2xZ2_StandardEmbeddingSpectrum}.

The blow-up model is obtained with the bundle vectors
\begin{eqnarray}
\label{eq:StdEmb_BundleVectors}
V_{1,\beta\gamma} = \Big( 0, -\frac{1}{2}, -\frac{1}{2}, 1, 0, 0, 0, 0\Big) \Big( 0,0,0,0,0,0,0,0 \Big) \;,\\ 
V_{2,\alpha\gamma} = \Big(  -\frac{1}{2},0  -\frac{1}{2}, 0, 1, 0, 0, 0\Big) \Big( 0,0,0,0,0,0,0,0 \Big) \;, \\ 
V_{3,\alpha\beta} = \Big(  -\frac{1}{2}, -\frac{1}{2}, 0 , 0, 0, 1, 0, 0\Big) \Big( 0,0,0,0,0,0,0,0 \Big) \;,
\end{eqnarray}
which fulfill the Bianchi identities for any triangulation. This corresponds to choosing the blow-up modes in three different directions inside $\boldsymbol{\overline{27}}$ of $\text{E}_6$ which induces a gauge symmetry breaking $\text{E}_6 \times \text{U}(1)^2 \rightarrow \text{SU}(3) \times \text{SU}(2) \times \text{U}(1)^5$. (In detail: The blow-up modes associated to $V_{1,\gb\gg}$ lead to a breaking to $\text{SO}(10)\times \text{U}(1)$ and all $\rep{\overline{27}}$ are branched to $\rep{\overline{16}}+\rep{10}+\rep{1}$. The blow-up modes associated to $V_{2,\ga\gg}$ acquire a VEV in the $\crep{16}$. This induces further breaking $\text{SO}(10) \ra \text{SU}(5) \times \text{U}(1)$, and the $\rep{\overline{16}}$ branches to $\crep{10} + \rep{5} + \rep{1}$. And finally the blow-up modes associated to $V_{3,\ga\gb}$ branches the $\crep{10}$ of $\text{SU}(5)$.) Since we chose the same blow-up mode in each fixed plane of a given twisted sector, the discussion is the same for all $64$ local resolutions. Thus, here we may drop the fixed point labels $\alpha$, $\beta$ and $\gamma$. The $\text{U}(1)$'s are chosen such that the charge vectors of the blow-up modes \eqref{eq:Superfield} are $Q(\phi_1) = (10,0,0,0,0)$, $Q(\phi_2) = (0,10,0,0,0)$ and $Q(\phi_3) = (0,0,10,0,0)$. In this way the axions corresponding to the blow-up modes have no effect in anomaly cancellation for the last two $\text{U}(1)$ factors. In fact, these $\text{U}(1)$'s have no anomaly at all, as can be checked by directly inspecting the anomaly polynomial. This matches with the fact that the two $\text{U}(1)$'s present in the orbifold model are non-anomalous, but notice that the non-anomalous $\text{U}(1)$'s in the orbifold and those in the resolution {\it cannot} be identified with each other directly, since the blow-up modes are generically charged under the orbifold $\text{U}(1)$'s.

In table~\ref{tab:Doublet_Matching} we list the twisted spectrum of the standard embedding model, after branching it in representations of $\text{SU}(3)\times \text{SU}(2)$, and we match it with the spectra obtained in the resolutions ``$E_i$'' and ``$S$''. The untwisted spectrum has no chiral counterpart in the resolution, so we do not consider it. We provide the $\text{U}(1)$ charges in the basis discussed above, that differs from the basis used in table~\ref{table:Z2xZ2_StandardEmbeddingSpectrum} for the reasons explained above. In what follows we also face the details of the matching in the specific case of states in $({\bf 1},{\bf 2})$ representations, but the mechanisms at work for the other representation are essentially the same.

 \subsubsection[Flopping states: the $(1,2)$ case]{Flopping states: the $(\boldsymbol{1},\boldsymbol{2})$ case}

To face the phenomenon of extra states appearing on the resolutions, we discuss here as an example the matching and appearance of novel states for the $\text{SU}(2)$ doublets $d_i$ listed in  table~\ref{tab:Doublet_Matching}. We focus in particular on the states named $d_1$ and $d_2$. The multiplicities of the orbifold states and of the states appearing in each of the four triangulations of the 64 $\ztwo$ fixed points are shown in table \ref{tab:Doublet_Matching}. As the charges of the orbifold states and the states on the resolution do not completely match we need to perform field redefinitions of the orbifold states using the blow-up modes of the respective sectors given by (cf. \eqref{eq:Superfield}): 
\equ{ 
d_i = e^{2\pi U_r} d_i^{\rm orbi} 
~\mbox{for}~ 
i=1\ldots6
~\text{and}~  
d_7 = e^{-2\pi U_r} d_7^{\rm orbi}\;. 
}

Table \ref{tab:Doublet_Matching} indicates that orbifold and resolution multiplicities of the states $d_1$ and $d_2$ are identical, except for resolution ``$E_2$'', where the multiplicity is $-48$ rather then $16$. The negative multiplicity means that on that resolution one does not see the $d_1$ and $d_2$ states, but rather their charge conjugates which we call $\underline{d}_1$ and $\underline{d}_2$. In order to explain this we first consider the (lowest order) superpotential terms for $d_1$ and $d_2$ that can be written from the orbifold perspective, namely $W=d_1^{orbi}d_2^{orbi}\phi_2$. This term indicates that all states get a mass term in blow-up. From the blow-up perspective the corresponding superpotential can be obtained after field redefinition, and reads $ W = d_1 d_2 e^{-2\pi \left( U_1 + U_3 - U_2 \right)}$. We observe that in all triangulations but ``$E_2$'' the conditions on the blow-up moduli are such that we can interpret this superpotential term as instantonic mass terms. Thus, $d_1$ and $d_2$ have the same multiplicity both in the orbifold point and in resolution, since they are massless modes in a perturbative expansion of the theory, receiving instantonic mass corrections.
When we pass to triangulation ``$E_2$'' from any other triangulation, the twisted moduli fulfill the condition $b_2>b_1+b_3$ and the $d_1 d_2$ mass term cannot be thought of as an instantonic correction to a well defined perturbative theory any more. 

In other words, the supergravity construction fails as soon as  $b_2$ is not smaller than $b_1+b_3$, and we loose control on the ``perturbative'' computation of the spectrum: the non-perturbative corrections take over, and a new ``perturbative'' computation comes at hand, i.e. the supergravity construction made in resolution ``$E_2$'', where the states $d_1$ and $d_2$ indeed disappear from the spectrum. 
This argument holds in the very same way for the pairs $d_3, d_4$ and $d_5, d_6$, disappearing in resolution ``$E_1$'' and ``$E_3$'', respectively. For the $d_7$ states the orbifold mass term is such that in no triangulation it can be seen as an instantonic correction to a perturbatively massless set of states: in supergravity, independently of the resolution type, these states have a large $O(M_s)$ mass and are removed from the massless spectrum.

We explained the fate of the orbifold states when passing in blow-up and when passing from one triangulation to the other. This is not all, since we have underlined states to explain as well. Their fate is somewhat dual to that of non-underlined states.
Let us consider the $\underline{d}_1$ and $\underline{d}_2$ states in triangulation ``$E_2$'': it is reasonable to assume that their superpotential is
\begin{equation}
 W =
\underline{d}_1 \underline{d}_2 e^{2\pi \left( U_1 + U_3 - U_2 \right)}
\end{equation}
and these states are present as massless states with instantonic mass terms only if the moduli are chosen such that we are in triangulation ``$E_2$'', in all the other cases the instantonic correction grows, a perturbative perspective is non-tenable, and the underlined states drop from the massless spectrum.
 
\begin{landscape}
\begin{table}
\begin{center}{\small
\begin{tabular}{cc}
\begin{tabular}{|c|ccc|cccc|c|}
\hline
&&&&&&&&\\[-1.1ex]
{\bf State} & \multicolumn{3}{|c}{\bf Orb. Mult.}  &  \multicolumn{4}{|c|}{\bf Resolution Mult.}     & {\bf $\text{U}(1)$ charges }\\[1ex]
      & $\gth_1$ & $\gth_2$ & $\gth_3$   & ``$E_1$'' & ``$E_2$'' & ``$E_3$'' & ``$S$'' & \\[1ex]
\hline
\hline
&&&&&&&&\\[-1.5ex]
$d_1$ &  16 &    &    &  16 & -48 &  16 &  16 & ( 7,-5, 3, 1, 2) \\[.5ex]\hline &&&&&&&&\\[-1.5ex]
$d_2$ &     &    & 16 &  16 & -48 &  16 &  16 & ( 3,-5, 7,-1,-2) \\[.5ex]\hline &&&&&&&&\\[-1.5ex]
$d_3$ &     & 16 &    &  16 &  16 & -48 &  16 & ( 3, 7,-5,-2,-1) \\[.5ex]\hline &&&&&&&&\\[-1.5ex]
$d_4$ &  16 &    &    &  16 &  16 & -48 &  16 & ( 7, 3,-5, 2, 1) \\[.5ex]\hline &&&&&&&&\\[-1.5ex]
$d_5$ &     &    & 16 & -48 &  16 &  16 &  16 & (-5, 3, 7, 1,-1) \\[.5ex]\hline &&&&&&&&\\[-1.5ex]
$d_6$ &     & 16 &    & -48 &  16 &  16 &  16 & (-5, 7, 3,-1, 1) \\[.5ex]\hline &&&&&&&&\\[-1.5ex]
$d_7$ &  16 & 16 & 16 & -80 & -80 & -80 & -80 & (-5,-5,-5, 0, 0) \\[.5ex]\hline\hline &&&&&&&&\\[-1.5ex]
$\overline{t}_1$ & 16 &    &    &  16 & -48 & -48 & -48 & ( 8,-4,-4, 1, 1) \\[.5ex]\hline &&&&&&&&\\[-1.5ex]
           $t_2$ &    & 16 & 16 &  32 & -32 & -32 & -32 & ( 2,-6,-6,-1,-1) \\[.5ex]\hline &&&&&&&&\\[-1.5ex]
$\overline{t}_3$ &    &    & 16 & -48 & -48 &  16 & -48 & (-4,-4, 8, 0,-1) \\[.5ex]\hline &&&&&&&&\\[-1.5ex]
           $t_4$ & 16 & 16 &    & -32 & -32 &  32 & -32 & (-6,-6, 2, 0, 1) \\[.5ex]\hline &&&&&&&&\\[-1.5ex]
$\overline{t}_5$ &    & 16 &    & -48 &  16 & -48 & -48 & (-4, 8,-4,-1, 0) \\[.5ex]\hline &&&&&&&&\\[-1.5ex]
           $t_6$ & 16 &    & 16 & -32 &  32 & -32 & -32 & (-6, 2,-6, 1, 0) \\[.5ex]\hline &&&&&&&&\\[-1.5ex]
           $t_7$ & 16 &    &    &  16 &  16 &  16 &  16 & ( 6, 2, 2, 2, 2) \\[.5ex]\hline &&&&&&&&\\[-1.5ex]
           $t_8$ &    & 16 &    &  16 &  16 &  16 &  16 & ( 2, 6, 2,-2, 0) \\[.5ex]\hline &&&&&&&&\\[-1.5ex]
           $t_9$ &    &    & 16 &  16 &  16 &  16 &  16 & ( 2, 2, 6, 0,-2) \\[.5ex]\hline
\end{tabular}
&
\begin{tabular}{|c|ccc|cccc|c|}\hline &&&&&&&&\\[-1.5ex]
$\overline{q}_1$ &      16 &         &        & 16 & 16 & 16 & 16 & (-7, 1, 1, 1, 1) \\[.5ex]\hline &&&&&&&&\\[-1.5ex]
$\overline{q}_2$ &         & 16      &        & 16 & 16 & 16 & 16 & ( 1,-7, 1,-1, 0) \\[.5ex]\hline &&&&&&&&\\[-1.5ex]
$\overline{q}_3$ &         &         & 16     & 16 & 16 & 16 & 16 & ( 1, 1,-7, 0,-1) \\[.5ex]\hline\hline &&&&&&&&\\[-1.5ex]
$\phi_1$         &      16 &         &        & \multicolumn{4}{|c|}{1$^{st}$ Blow-up mode} & (10, 0, 0, 0, 0) \\[.5ex]\hline &&&&&&&&\\[-1.5ex]
$\phi_2$         &         & 16      &        & \multicolumn{4}{|c|}{2$^{nd}$ Blow-up mode} & ( 0,10, 0, 0, 0) \\[.5ex]\hline &&&&&&&&\\[-1.5ex]
$\phi_3$         &         &         & 16     & \multicolumn{4}{|c|}{3$^{rd}$ Blow-up mode} & ( 0, 0,10, 0, 0) \\[.5ex]\hline\hline &&&&&&&&\\[-1.5ex]
$s_1$            &  16     &         &        &  16 & 16 &  16 & 16 & ( 6, 2, 2,-3,-3) \\[.5ex]\hline &&&&&&&&\\[-1.5ex]
$s_2$            &         & 16      &        &  16 & 16 &  16 & 16 & ( 2, 6, 2, 3, 0) \\[.5ex]\hline &&&&&&&&\\[-1.5ex]
$s_3$            &         &         & 16     &  16 & 16 &  16 & 16 & ( 2, 2, 6, 0, 3) \\[.5ex]\hline &&&&&&&&\\[-1.5ex]
$s_4$            &  16     &         &-16     & -64 &  0 &  64 &  0 & (-8, 0, 8, 1, 2) \\[.5ex]\hline &&&&&&&&\\[-1.5ex]
$s_5$            &         &  32$^*$ &        & -64 &  0 &   0 &  0 & (-2,10, 2,-1,-2) \\[.5ex]\hline &&&&&&&&\\[-1.5ex]
$s_6$            &         &  32$^*$ &        &   0 &  0 & -64 &  0 & ( 2,10,-2, 1, 2) \\[.5ex]\hline &&&&&&&&\\[-1.5ex]
$s_7$            &  16     &-16      &        & -64 &  64 &  0 &  0 & (-8, 8, 0, 2, 1) \\[.5ex]\hline &&&&&&&&\\[-1.5ex]
$s_8$            &         &         & 32$^*$ & -64 &   0 &  0 &  0 & (-2, 2,10,-2,-1) \\[.5ex]\hline &&&&&&&&\\[-1.5ex]
$s_9$            &         &         & 32$^*$ &   0 &  64 &  0 &  0 & ( 2,-2,10, 2, 1) \\[.5ex]\hline &&&&&&&&\\[-1.5ex]
$s_{10}$         &         & 16      &-16     &   0 & -64 & 64 &  0 & ( 0,-8, 8,-1, 1) \\[.5ex]\hline &&&&&&&&\\[-1.5ex]
$s_{11}$         &  32$^*$ &         &        &   0 & -64 &  0 &  0 & (10,-2, 2, 1,-1) \\[.5ex]\hline &&&&&&&&\\[-1.5ex]
$s_{12}$         &  32$^*$ &         &        &   0 &   0 & 64 &  0 & (10, 2,-2,-1, 1) \\[.5ex]\hline
\end{tabular}
\end{tabular}
}\end{center}
\caption{\label{tab:Doublet_Matching}
Orbifold and resolution multiplicities of the states in the standard embedding (but in non-standard blow-up).
The first column lists the labels of $\text{SU}(3)\times \text{SU}(2)$ representations: ``$d_x$'' for $(\boldsymbol{1},\boldsymbol{2})$, ``$t_x$'' for $(\boldsymbol{3},\boldsymbol{1})$, ``$q_x$'' for $(\boldsymbol{3},\boldsymbol{2})$, $\phi_x$ for the blow-up modes, and ``$s_x$'' denote the singlets. An overlined label indicates the corresponding conjugate representation. In giving the multiplicities in the orbifold we indicate the twisted sector to which they belong ($\gth_1$, $\gth_2$, and $\gth_3$). When the resolution the multiplicities are calculated for all four triangulations (using the same triangulation at each singularity). Some singlets are marked with a $^*$ sign to indicate that their matching is ambiguous since they can be redefined in two possible ways, both having a viable physical interpretation.}
\end{table}
\end{landscape}

\subsubsection{A ``mother'' theory including all states}
The description given above explains how states ``disappear'', but does not account for the presence of states from the very beginning.
This we can try to face by considering a unified description where both underlined and non-underlined states live {\it at the same time} (it is of course arguable whether such a description has the status of a physical theory, or whether it is just a convenient bookkeeping construction.)

Restricting to the case of the doublets $d_1$, $d_2$, $\underline{d}_1$, and $\underline{d}_2$, we can consider a ``mother'' theory containing 16 copies of the non-underlined states, and 48 copies of the underlined ones.
In such a theory the dynamics would be described, at the lowest level, by a superpotential of the form
\begin{equation}
W_M=d_1 d_2 e^{-2\pi \left( U_1 + U_3 - U_2 \right)}+\underline{d}_1 \underline{d}_2 e^{2\pi \left( U_1 + U_3 - U_2 \right)}+d_1 \underline{d}_1+d_2\underline{d}_2
\end{equation}
inducing a $64\times 64$ mass matrix. Such mass matrix would have $O(1)$ blocks linking $d_i$'s with $\underline{d}_i$'s states, and 
$O( e^{\pm 2\pi \left( U_1 + U_3 - U_2 \right)})$ ``diagonal'' blocks. This mass matrix indicates how the underlined states decouple in all triangulation but ``$E_2$'', while the non-underlined states would receive small mass terms, in a sort of see-saw mechanism. In triangulation``$E_2$'' the reverse see-saw would instead be at work, decoupling the non-underlined states and re-coupling the underlined ones, with small mass term.

A similar mechanism, with minor modifications, would be at hand for the other doublet states, as well as for all states in the model.

%
%
%
%
%
%
%
%
%
%
%
%
%
%
%
%
%
%
%
%
%
%
%
%
%
%
%
%
%
%
%
\section{MSSM on the orbifold and on the resolution}
\label{ch:MSSM_in_Blowup}

The main objective of this paper is to construct resolution models of heterotic orbifolds. In the previous sections we have discussed the geometrical resolutions and how to investigate supergravity and super Yang-Mills on them. In this section we first recall some basic facts of how to describe heterotic string theory on orbifolds, then we consider an orbifold model with the MSSM spectrum, and, building on this model, we finally consider a model built on the resolved orbifold having the MSSM spectrum, and comment on matching the orbifold and the supergravity model.

\subsection[Heterotic \ztwo orbifolds with a free involution]{Heterotic $\boldsymbol{\ztwo}$ orbifolds with a free involution}
\label{sc:OrbiCFT}

In CFT constructions of  heterotic orbifolds the space group acts on the gauge degrees of freedom as well. The orbifold rotations $\theta_i$ are embedded in the gauge degrees of freedom via the three gauge shifts $V_i$. The translations over the six torus basis vectors $e_i$ are embedded via  six Wilson lines $W_i$, and finally the \zf\ involution is characterized by an additional Wilson line $W$.

This data defines the heterotic orbifoldand has to satisfy certain consistency conditions: In order to define a proper representation of the space group, one has to impose 
\equ{
2V_i\cong 0\;, 
\quad 
V_3 \cong -V_1-V_2\;,
\quad 
2 W_j \cong 0\;,
\quad 
2W \cong  W_{2}+W_{4}+W_{6}\;,
\quad 
W_{2}\cong W_{4} \cong W_{6}\;.
\label{eq:Const_V_W}
}
Modular invariance of the corresponding one loop string partition function and the existence of well-defined projections imposes further requirements on the shift vectors and on the Wilson lines (see e.g.\ in \cite{Ploger:2007} and appendix \ref{ch:Derivation_MI_Free_WL}):
\begin{equation}
\label{eq:Orbifold_Consistency_Requirements}
V_i\cdot V_j-  \varphi_i\cdot \varphi_j
  \equiv 
W_{p} \cdot W_{q} 
\equiv 
 V_i \cdot W_{p=2i,2i-1}
\equiv 0\;,
\end{equation}
where ``$\equiv$" means equal up to integers. The same requirements  also constrains the embedding of the freely acting Wilson line 
\begin{equation}
\label{eq:10D_Modulari_Invariance}
W\cdot W_p \equiv 2\, W^2 \equiv 0\;. 
\end{equation}
Details of the derivation can be found in appendix \ref{ch:Derivation_MI_Free_WL}. As a side remark we note that these conditions do not seem to have an analog in supergravity.  As the action is free, the Bianchi identities, which are the central consistency requirement of the CY manifold, are not modified. So it would be very interesting to investigate how the orbifold consistency requirement derived above can be understood as string consistency condition on a CY.

\begin{table}[t]
\centering
\begin{tabular}{cc}
\begin{tabular}{|c|c|c|c||c|l|c|}
\hline
\multicolumn{5}{|c|}{}                        &        &       \\[-1.5ex]
\multicolumn{5}{|c|}{Mult.}                   & Irrep. & Label \\
U & $\gth_1$ & $\gth_2$ & $\gth_3$ & $\Sigma$ &        &       \\[2pt]
\hline
\hline &&&&&&\\[-2.2ex]
3 & 4 &   & 8 & 15 & $\left(\overline{\mathbf{5}};\mathbf{1},\mathbf{1}\right)$		& $\overline{D}$\\[2pt]
3 & 2 &   & 4 &  9 & $\left(\mathbf{5};\mathbf{1},\mathbf{1}\right)$			& $D$		\\[2pt]
\hline
\hline &&&&&&\\[-2.2ex]
  &  4 & 8 &    & 12 & $\left(\mathbf{1};\mathbf{4},\mathbf{1}\right)$			& $T^{(1)}$	\\[2pt]
  &  4 & 8 &    & 12 & $\left(\mathbf{1};\overline{\mathbf{4}},\mathbf{1}\right)$	& $\overline{T}^{(1)}$	\\[2pt]
  &  2 &   &    &  2 & $\left(\mathbf{1};\mathbf{6},\mathbf{1}\right)$			& $V^{(1)}$	\\[2pt]
6 & 22 &   & 28 & 56 & $\left(\mathbf{1};\mathbf{1},\mathbf{1}\right)$			& $S$		\\[2pt]
\hline
\end{tabular} &
\begin{tabular}{|c|c|c|c||c|l|c|}
\hline
\multicolumn{5}{|c|}{}                        &        &       \\[-1.5ex]
\multicolumn{5}{|c|}{Mult.}                   & Irrep. & Label \\
U & $\gth_1$ & $\gth_2$ & $\gth_3$ & $\Sigma$ &        &       \\[2pt]
\hline
\hline &&&&&&\\[-2.2ex]
  & 2 &   & 4 &  6 & $\left(\mathbf{10};\mathbf{1},\mathbf{1}\right)$			& $Q$		\\[2pt]
  &   &   &   &    & & \\[2pt]
\hline
\hline &&&&&&\\[-2.2ex]
   & 4 & 4 & 4 & 12 & $\left(\mathbf{1};\mathbf{1},\mathbf{4}\right)$			& $T^{(2)}$	\\[2pt]
   & 4 & 4 & 4 & 12 & $\left(\mathbf{1};\mathbf{1},\overline{\mathbf{4}}\right)$	& $\overline{T}^{(2)}$	\\[2pt]
   & 2 &   &   &  2 & $\left(\mathbf{1};\mathbf{1},\mathbf{6}\right)$			& $V^{(2)}$		\\[2pt]
   &   &   &   &    & & \\[2pt]
\hline
\end{tabular}
\end{tabular}
\caption{The massless $\text{SU}(5)$ GUT spectrum of the \ztwo can accommodate a net number of six generations of quarks and leptons. The representations with respect to $[\text{SU}(5)] \times [\text{SU}(4) \times \text{SU}(4)]$ together with their multiplicities (separated by: untwisted sector U, twisted sectors $\gth_1$, $\gth_2$, $\gth_3$ and the total multiplicity $\Sigma$) and the labels are listed.}
\label{table:Z2_Orbifold_Spectrum}
\end{table}
\subsection{An MSSM orbifold}
\label{sc:MSSMorbifold}

To obtain a concrete realization of an orbifold model with an MSSM spectrum after modding out the freely acting $\zf$, we make the following choices. First of all, we take the Wilson lines 
\equ{
W_{2}=W_{4}=W_{6} = 2 W
\label{EqualWL}
}
equal on the orbifold to allow for a straightforward \zf\ identification \eqref{eq:Const_V_W}, and set one Wilson line to zero, i.e.\ $W_1 =0$. For the shifts and the other Wilson lines we take 
\begin{subequations}
\label{eq:Z2_Model_Input}
\begin{eqnarray}
V_{1} & = & \Big(\frac{1}{4},-\frac{1}{4},-\frac{1}{4},\frac{1}{4},-\frac{3}{4},-\frac{3}{4},\frac{1}{4},\frac{1}{4}\Big)\hskip0.248cm\Big(1,0,0,0,0,0,0,0\Big)\;,\\
V_{2} & = & \Big(\frac{3}{4},\frac{1}{4},-\frac{1}{4},\frac{1}{4},\frac{1}{4},-\frac{3}{4},\frac{1}{4},\frac{1}{4}\Big)\hskip0.848cm\Big(1,0,0,0,0,0,0,0\Big)\;,\\
W_{2}\!=\!W_{4}\!=\!W_{6}  & = & \Big(\!\!-\!\frac{5}{4},\frac{3}{4},-\frac{3}{4},\frac{9}{4},-\frac{7}{4},-\frac{3}{4},\frac{5}{4},-\frac{3}{4}\Big) \Big(\!\!-\!\frac{1}{4},\frac{11}{4},\frac{3}{4},-\frac{3}{4},-\frac{7}{4},-\frac{3}{4},\frac{5}{4},\frac{3}{4}\Big)\;,\\
W_{3} & = & \Big(\!\!-\!1,-1,0,-2,0,-2,2,-3\Big)\hskip0.69cm\Big(\!\!-\!\frac{7}{4},-\frac{1}{4},\frac{3}{4},-\frac{1}{4},-\frac{5}{4},\frac{1}{4},\frac{1}{4},\frac{5}{4}\Big)\;,\\
W_{5} & = & \Big(\frac{1}{4},\frac{9}{4},-\frac{13}{4},\frac{11}{4},\frac{3}{4},\frac{11}{4},-\frac{1}{4},-\frac{1}{4}\Big) \Big(\frac{3}{4},\frac{1}{4},-\frac{11}{4},-\frac{3}{4},-\frac{3}{4},-\frac{1}{4},-\frac{5}{4},\frac{3}{4}\Big)\;.
\end{eqnarray}
\end{subequations}
This set satisfies all orbifold consistency conditions described in subsection \ref{sc:OrbiCFT}.

Before modding out the freely acting \zf, the gauge group of the model is 
\begin{equation}
\label{eqn:GUTgroup}
G_\text{GUT}=[\text{SU}(5) \times \text{U}(1)^{4}]\times [\text{SU}(4) \times \text{SU}(4) \times \text{U}(1)^{2}]\;,
\end{equation}
and the full massless spectrum is given in table \ref{table:Z2_Orbifold_Spectrum}. The GUT part of the spectrum contains six $\mathbf{10}$-plets. For the $\overline{\mathbf{5}}$-plets, we find a net number of $15-9=6$ . Hence it can accommodate six SM generations and a certain number of Higgses and vector-like exotics. The remaining states are charged under the hidden gauge group or are singlets. 

In a second step, we divide out the freely acting \zf\ together with its associated freely acting Wilson line $W$, resulting in a gauge symmetry breaking from the GUT group of equation (\ref{eqn:GUTgroup}) to 
\begin{equation}
G_\text{SM}=[\text{SU}(3) \times \text{SU}(2) \times \text{U}(1)^{5}]\times [\text{SU}(4) \times \text{SU}(2)^2 \times \text{U}(1)^{3}]\;,
\end{equation}
and the charged massless spectrum is given in table \ref{table:Z2_MSSMOrbifold_Spectrum}.

In principle this model allows for a full blow-up of all 24 fix tori without breaking the SM gauge group, since there is at least one twisted SM singlet per fixed torus that can be chosen as a blow-up mode along an F-flat direction (of the cubic superpotential). This is in contrast to the models discussed in \cite{SGN:2009,Blaszczyk:2009in}. However, compared to the model in \cite{Blaszczyk:2009in}, the present model has a big phenomenological drawback as there exists one $\text{U}(1)$ that cannot be broken along D-flat directions of SM singlets. This $\text{U}(1)$ turns out to be $B-L$. Consequently, this model suffers under a $\text{U}(1)_{B-L}$ symmetry remaining unbroken down to the SUSY-breaking scale -- an undesired fact, e.g. for light neutrinos via the see-saw mechanism.

\begin{table}[t]\small
\centering
\begin{tabular}{cc}
\begin{tabular}{|c|c|c|c||c|l|c|}
\hline
\multicolumn{5}{|c|}{}                        &        &       \\[-1.5ex]
\multicolumn{5}{|c|}{Mult.}                   & Irrep. & Label \\
U & $\gth_1$ & $\gth_2$ & $\gth_3$ & $\Sigma$ &        &       \\[2pt]
\hline
\hline &&&&&&\\[-2.2ex]
 3 & 2 &   & 4 & 9 & $(\overline{\mathbf{3}},\mathbf{1};\mathbf{1},\mathbf{1},\mathbf{1})_{ 1/3}$ & $\overline{d}$ \\
   & 2 &   & 4 & 6 & $(\mathbf{1},\mathbf{2};\mathbf{1},\mathbf{1},\mathbf{1})_{-1/2}$            & $\ell$ \\
 3 & 1 &   & 2 & 6 & $(\mathbf{3},\mathbf{1};\mathbf{1},\mathbf{1},\mathbf{1})_{-1/3}$            & $d$ \\
   & 1 &   & 2 & 3 & $(\mathbf{1},\mathbf{2};\mathbf{1},\mathbf{1},\mathbf{1})_{ 1/2}$            & $\overline{\ell}$ \\
\hline
\hline &&&&&&\\[-2.2ex]
   &  2 & 4 &    &  6 & $(\mathbf{1},\mathbf{1};\mathbf{4},\mathbf{1},\mathbf{1})_{0}$            & $t$ \\
   &  2 & 4 &    &  6 & $(\mathbf{1},\mathbf{1};\overline{\mathbf{4}},\mathbf{1},\mathbf{1})_{0}$ & $\overline{t}$ \\
   &  1 &   &    &  1 & $(\mathbf{1},\mathbf{1};\mathbf{6},\mathbf{1},\mathbf{1})_{0}$            & $v$ \\
 6 & 13 &   & 14 & 33 & $(\mathbf{1},\mathbf{1};\mathbf{1},\mathbf{1},\mathbf{1})_{0}$            & $s$ \\
\hline
\end{tabular}
 &
\begin{tabular}{|c|c|c|c||c|l|c|}
\hline
\multicolumn{5}{|c|}{}                        &        &       \\[-1.5ex]
\multicolumn{5}{|c|}{Mult.}                   & Irrep. & Label \\
U & $\gth_1$ & $\gth_2$ & $\gth_3$ & $\Sigma$ &        &       \\[2pt]
\hline
\hline &&&&&&\\[-2.2ex]
   & 1 &   & 2 & 3 & $(\mathbf{3},\mathbf{2};\mathbf{1},\mathbf{1},\mathbf{1})_{ 1/6}$ & $q$ \\
   & 1 &   & 2 & 3 & $(\overline{\mathbf{3}},1;1,1,1)_{-2/3}$                          & $\overline{u}$ \\
   & 1 &   & 2 & 3 & $(\mathbf{1},\mathbf{1};\mathbf{1},\mathbf{1},\mathbf{1})_{1}$    & $\overline{e}$ \\
   &   &   &   &   & & \\[1pt]
\hline
\hline &&&&&&\\[-2.2ex]
   & 4 & 4 & 4 & 12 & $(\mathbf{1},\mathbf{1};\mathbf{1},\mathbf{2},\mathbf{1})_{0}$   & $x$ \\
   & 4 & 4 & 4 & 12 & $(\mathbf{1},\mathbf{1};\mathbf{1},\mathbf{1},\mathbf{2})_{0}$   & $y$ \\
   & 1 &   &   &  1 & $(\mathbf{1},\mathbf{1};\mathbf{1},\mathbf{2},\mathbf{2})_{0}$   & $z$ \\
   &   &   &   &    & & \\[0pt]
\hline
\end{tabular}
\end{tabular}
\caption{The massless MSSM spectrum of the \ztwo with freely acting \zf. The representations with respect to $[\text{SU}(3) \times \text{SU}(2) \times \text{U}(1)_Y]\times [\text{SU}(4) \times \text{SU}(2)^2]$ together with their multiplicities (separated by: untwisted sector U, twisted sectors $\gth_1$, $\gth_2$, $\gth_3$ and the total multiplicity $\Sigma$) and the labels are listed.}
\label{table:Z2_MSSMOrbifold_Spectrum}
\end{table}

\subsection{MSSM on a resolved orbifold}

In this section we consider an explicit model obtained by wrapping line bundles built as described in section~\ref{ch:Preliminaries} on the smooth resolution of $T^6/\ztwo$ built in section~\ref{ch:Het_Sugra_on Resolution}. As we have seen in subsection \ref{sc:line_bundle_E1} the choice of using triangulation ``$E_1$" at all \ztwo fixed points is very convenient because it leads to considerable simplifications of the Bianchi identities. Even with this choice of triangulation solving the Bianchi identities is far from trivial. We therefore make some additional simplifying assumptions. These assumptions are inspired by the structure of heterotic orbifold CFTs, which we discussed in subsection~\ref{sc:OrbiCFT}.

\renewcommand{\arraystretch}{1.1}
\begin{table}[t]
\centering
\begin{tabular}{|c||r@{}r@{\hskip8pt}r@{\hskip8pt}r@{\hskip8pt}r@{\hskip8pt}r@{\hskip8pt}r@{\hskip8pt}r@{\hskip4pt}r@{}c@{}r@{\hskip8pt}r@{\hskip8pt}r@{\hskip8pt}r@{\hskip8pt}r@{\hskip8pt}r@{\hskip8pt}r@{\hskip8pt}r@{}|}
\hline
$V_{r}$ & \multicolumn{18}{c|}{expression for the bundle vector $V_r$}\\
\hline
\hline
$V_{1,11}, V_{1,22}$ & $\!\!($ & $-\frac{1}{4}$ & $\frac{1}{4}$ & $\frac{1}{4}$ & $-\frac{1}{4}$ & $-\frac{1}{4}$ & $-\frac{1}{4}$ & $-\frac{1}{4}$ & $-\frac{1}{4}$ & $)$ 
$\!\!($ & $0$ & $0$ & $0$ & $0$ & $0$ & $0$ & $-1$ & $0)$ \\
$V_{1,12}, V_{1,21}$ & $\!\!($ & $-\frac{1}{2}$ & $0$ & $\frac{1}{2}$ & $0$ & $0$ & $0$ & $0$ & $0$ & $)$ 
$\!\!($ & $-\frac{3}{4}$ & $\frac{1}{4}$ & $\frac{1}{4}$ & $-\frac{1}{4}$ & $-\frac{1}{4}$ & $-\frac{1}{4}$ & $-\frac{1}{4}$ & $\frac{1}{4})$ \\
$V_{1,13}, V_{1,24}$ & $\!\!($ & $-\frac{1}{2}$ & $0$ & $\frac{1}{2}$ & $0$ & $0$ & $0$ & $0$ & $0$ & $)$ 
$\!\!($ & $\frac{1}{4}$ & $-\frac{1}{4}$ & $-\frac{1}{4}$ & $-\frac{1}{4}$ & $-\frac{1}{4}$ & $\frac{1}{4}$ & $-\frac{3}{4}$ & $\frac{1}{4})$ \\
$V_{1,14}, V_{1,23}$ & $\!\!($ & $-\frac{1}{4}$ & $\frac{1}{4}$ & $\frac{1}{4}$ & $-\frac{1}{4}$ & $-\frac{1}{4}$ & $-\frac{1}{4}$ & $-\frac{1}{4}$ & $-\frac{1}{4}$ & $)$ 
$\!\!($ & $0$ & $\frac{1}{2}$ & $\frac{1}{2}$ & $0$ & $0$ & $-\frac{1}{2}$ & $\frac{1}{2}$ & $0)$ \\
$V_{1,31}, V_{1,42}$ & $\!\!($ & $-\frac{1}{4}$ & $\frac{1}{4}$ & $-\frac{3}{4}$ & $-\frac{1}{4}$ & $-\frac{1}{4}$ & $-\frac{1}{4}$ & $-\frac{1}{4}$ & $-\frac{1}{4}$ & $)$ 
$\!\!($ & $-\frac{1}{4}$ & $\frac{1}{4}$ & $\frac{1}{4}$ & $\frac{1}{4}$ & $\frac{1}{4}$ & $-\frac{1}{4}$ & $-\frac{1}{4}$ & $-\frac{1}{4})$ \\
$V_{1,32}, V_{1,41}$ & $\!\!($ & $-\frac{1}{2}$ & $0$ & $-\frac{1}{2}$ & $0$ & $0$ & $0$ & $0$ & $0$ & $)$ 
$\!\!($ & $0$ & $-\frac{1}{2}$ & $\frac{1}{2}$ & $0$ & $0$ & $-\frac{1}{2}$ & $\frac{1}{2}$ & $0)$ \\
$V_{1,33}, V_{1,44}$ & $\!\!($ & $\frac{1}{2}$ & $-1$ & $-\frac{1}{2}$ & $0$ & $0$ & $0$ & $0$ & $0$ & $)$ 
$\!\!($ & $0$ & $0$ & $0$ & $0$ & $0$ & $0$ & $0$ & $0)$ \\
$V_{1,34}, V_{1,43}$ & $\!\!($ & $-\frac{1}{4}$ & $-\frac{3}{4}$ & $\frac{1}{4}$ & $-\frac{1}{4}$ & $-\frac{1}{4}$ & $-\frac{1}{4}$ & $-\frac{1}{4}$ & $-\frac{1}{4}$ & $)$ 
$\!\!($ & $-\frac{1}{4}$ & $-\frac{1}{4}$ & $-\frac{1}{4}$ & $\frac{1}{4}$ & $\frac{1}{4}$ & $\frac{1}{4}$ & $\frac{1}{4}$ & $-\frac{1}{4})$ \\
\hline
\hline
$V_{2,11}, V_{2,22}$ & \multirow{2}{*}{$\!\!($} & \multirow{2}{*}{$\frac{1}{4}$} & \multirow{2}{*}{$-\frac{1}{4}$} & \multirow{2}{*}{$\frac{1}{4}$} & \multirow{2}{*}{$-\frac{1}{4}$} & \multirow{2}{*}{$-\frac{1}{4}$} & \multirow{2}{*}{$-\frac{1}{4}$} & \multirow{2}{*}{$-\frac{1}{4}$} & \multirow{2}{*}{$-\frac{1}{4}$} & \multirow{2}{*}{$)$ 
$\!\!($} & \multirow{2}{*}{$0$} & \multirow{2}{*}{$0$} & \multirow{2}{*}{$0$} & \multirow{2}{*}{$0$} & \multirow{2}{*}{$0$} & \multirow{2}{*}{$0$} & \multirow{2}{*}{$-1$} & \multirow{2}{*}{$0)$} \\
$V_{2,31}, V_{2,42}$ & & & & & & & & & & & & & & & & & & \\
\hline
$V_{2,12}, V_{2,21}$ & \multirow{2}{*}{$\!\!($} & \multirow{2}{*}{$0$} & \multirow{2}{*}{$\frac{1}{2}$} & \multirow{2}{*}{$-\frac{1}{2}$} & \multirow{2}{*}{$0$} & \multirow{2}{*}{$0$} & \multirow{2}{*}{$0$} & \multirow{2}{*}{$0$} & \multirow{2}{*}{$0$} & \multirow{2}{*}{$)$ 
$\!\!($} & \multirow{2}{*}{$-\frac{1}{4}$} & \multirow{2}{*}{$-\frac{1}{4}$} & \multirow{2}{*}{$-\frac{1}{4}$} & \multirow{2}{*}{$\frac{1}{4}$} & \multirow{2}{*}{$\frac{1}{4}$} & \multirow{2}{*}{$\frac{1}{4}$} & \multirow{2}{*}{$-\frac{3}{4}$} & \multirow{2}{*}{$-\frac{1}{4})$} \\
$V_{2,32}, V_{2,41}$ & & & & & & & & & & & & & & & & & & \\
\hline
$V_{2,13}, V_{2,24}$ & \multirow{2}{*}{$\!\!($} & \multirow{2}{*}{$0$} & \multirow{2}{*}{$-\frac{1}{2}$} & \multirow{2}{*}{$\frac{1}{2}$} & \multirow{2}{*}{$0$} & \multirow{2}{*}{$0$} & \multirow{2}{*}{$0$} & \multirow{2}{*}{$0$} & \multirow{2}{*}{$0$} & \multirow{2}{*}{$)$ 
$\!\!($} & \multirow{2}{*}{$\frac{1}{4}$} & \multirow{2}{*}{$\frac{3}{4}$} & \multirow{2}{*}{$-\frac{1}{4}$} & \multirow{2}{*}{$-\frac{1}{4}$} & \multirow{2}{*}{$-\frac{1}{4}$} & \multirow{2}{*}{$\frac{1}{4}$} & \multirow{2}{*}{$\frac{1}{4}$} & \multirow{2}{*}{$\frac{1}{4})$} \\
$V_{2,33}, V_{2,44}$ & & & & & & & & & & & & & & & & & & \\
\hline
$V_{2,14}, V_{2,23}$ & \multirow{2}{*}{$\!\!($} & \multirow{2}{*}{$\frac{1}{4}$} & \multirow{2}{*}{$-\frac{1}{4}$} & \multirow{2}{*}{$\frac{1}{4}$} & \multirow{2}{*}{$-\frac{1}{4}$} & \multirow{2}{*}{$-\frac{1}{4}$} & \multirow{2}{*}{$-\frac{1}{4}$} & \multirow{2}{*}{$-\frac{1}{4}$} & \multirow{2}{*}{$-\frac{1}{4}$} & \multirow{2}{*}{$)$ 
$\!\!($} & \multirow{2}{*}{$0$} & \multirow{2}{*}{$\frac{1}{2}$} & \multirow{2}{*}{$\frac{1}{2}$} & \multirow{2}{*}{$0$} & \multirow{2}{*}{$0$} & \multirow{2}{*}{$-\frac{1}{2}$} & \multirow{2}{*}{$\frac{1}{2}$} & \multirow{2}{*}{$0)$} \\
$V_{2,34}, V_{2,43}$ & & & & & & & & & & & & & & & & & & \\
\hline
\hline
$V_{3,11}, V_{3,22}$ & \multirow{2}{*}{$\!\!($} & \multirow{2}{*}{$\frac{1}{2}$} & \multirow{2}{*}{$\frac{1}{2}$} & \multirow{2}{*}{$1$} & \multirow{2}{*}{$0$} & \multirow{2}{*}{$0$} & \multirow{2}{*}{$0$} & \multirow{2}{*}{$0$} & \multirow{2}{*}{$0$} & \multirow{2}{*}{$)$ 
$\!\!($} & \multirow{2}{*}{$0$} & \multirow{2}{*}{$0$} & \multirow{2}{*}{$0$} & \multirow{2}{*}{$0$} & \multirow{2}{*}{$0$} & \multirow{2}{*}{$0$} & \multirow{2}{*}{$0$} & \multirow{2}{*}{$0)$} \\
$V_{3,31}, V_{3,42}$ & & & & & & & & & & & & & & & & & & \\
\hline
$V_{3,12}, V_{3,21}$ & \multirow{2}{*}{$\!\!($} & \multirow{2}{*}{$-\frac{1}{4}$} & \multirow{2}{*}{$-\frac{1}{4}$} & \multirow{2}{*}{$\frac{3}{4}$} & \multirow{2}{*}{$-\frac{1}{4}$} & \multirow{2}{*}{$-\frac{1}{4}$} & \multirow{2}{*}{$-\frac{1}{4}$} & \multirow{2}{*}{$-\frac{1}{4}$} & \multirow{2}{*}{$-\frac{1}{4}$} & \multirow{2}{*}{$)$ 
$\!\!($} & \multirow{2}{*}{$-\frac{1}{4}$} & \multirow{2}{*}{$-\frac{1}{4}$} & \multirow{2}{*}{$-\frac{1}{4}$} & \multirow{2}{*}{$\frac{1}{4}$} & \multirow{2}{*}{$\frac{1}{4}$} & \multirow{2}{*}{$\frac{1}{4}$} & \multirow{2}{*}{$\frac{1}{4}$} & \multirow{2}{*}{$-\frac{1}{4})$} \\
$V_{3,32}, V_{3,41}$ & & & & & & & & & & & & & & & & & & \\
\hline
$V_{3,13}, V_{3,24}$ & \multirow{2}{*}{$\!\!($} & \multirow{2}{*}{$-\frac{1}{2}$} & \multirow{2}{*}{$-\frac{1}{2}$} & \multirow{2}{*}{$0$} & \multirow{2}{*}{$0$} & \multirow{2}{*}{$0$} & \multirow{2}{*}{$0$} & \multirow{2}{*}{$0$} & \multirow{2}{*}{$0$} & \multirow{2}{*}{$)$ 
$\!\!($} & \multirow{2}{*}{$\frac{3}{4}$} & \multirow{2}{*}{$\frac{1}{4}$} & \multirow{2}{*}{$\frac{1}{4}$} & \multirow{2}{*}{$\frac{1}{4}$} & \multirow{2}{*}{$\frac{1}{4}$} & \multirow{2}{*}{$-\frac{1}{4}$} & \multirow{2}{*}{$-\frac{1}{4}$} & \multirow{2}{*}{$-\frac{1}{4})$} \\
$V_{3,33}, V_{3,44}$ & & & & & & & & & & & & & & & & & & \\
\hline
$V_{3,14}, V_{3,23}$ & \multirow{2}{*}{$\!\!($} & \multirow{2}{*}{$-\frac{1}{4}$} & \multirow{2}{*}{$-\frac{1}{4}$} & \multirow{2}{*}{$-\frac{1}{4}$} & \multirow{2}{*}{$-\frac{1}{4}$} & \multirow{2}{*}{$-\frac{1}{4}$} & \multirow{2}{*}{$-\frac{1}{4}$} & \multirow{2}{*}{$-\frac{1}{4}$} & \multirow{2}{*}{$-\frac{1}{4}$} & \multirow{2}{*}{$)$ 
$\!\!($} & \multirow{2}{*}{$-\frac{1}{2}$} & \multirow{2}{*}{$0$} & \multirow{2}{*}{$0$} & \multirow{2}{*}{$-\frac{1}{2}$} & \multirow{2}{*}{$-\frac{1}{2}$} & \multirow{2}{*}{$0$} & \multirow{2}{*}{$0$} & \multirow{2}{*}{$\frac{1}{2})$} \\
$V_{3,34}, V_{3,43}$ & & & & & & & & & & & & & & & & & & \\
\hline
\end{tabular}
\caption{Set of 48 line bundle vectors such that they solve the Bianchi identities \eqref{eq:ex_Z2_BIeqns} obtained by using resolution ``$E_1$" for all 64 $\mathbbm{C}^{3}/\mathbbm{Z}_{2}\times\mathbbm{Z}_{2}$ fixed points.}
\label{table:Z2_BI_Solution_Vectors}
\end{table}
\renewcommand{\arraystretch}{1.0}

\subsubsection{Line bundle vector ansatz}
\label{sc:LineBundleAnsatz}

By putting the orbifold consistency conditions, discussed in subsection \ref{sc:OrbiCFT}, and connections between orbifold line bundle vectors and the local orbifold twists together as explained in subsection \ref{sc:LineBundleOrbiShift}, and making some simplifying further assumption,  we can obtain a simplifying ansatz for the line bundle vectors. As the objective here is not to obtain a full classification, but rather to find at least some solutions, we do not mind that this ansatz will be far from generic.

In subsection \ref{sc:MSSMorbifold} we construct an MSSM orbifold where on the torus several Wilson lines are equal and $W_1 = 0$, see \eqref{EqualWL}. By the identification \eqref{eq:ex_Z2_Identification} this means that many line bundle vectors are equal up to \eeight lattice vectors. We assume that all line bundle vectors that are equal up to lattice elements, are strictly equal, i.e.\ 
\begin{subequations}
\equ{
\label{eq:ex_Z2_Identification_Simplification}
\arry{c}{ 
V_{i,22} = V_{i,11}\;, 
\quad 
V_{i,21} = V_{i,12}\;,
\quad 
V_{i,24}  = V_{i,13}\;,
\quad 
V_{i,23} = V_{i,14}\;, 
\\[1ex] 
V_{i,42} = V_{i,31}\;,
\quad 
V_{i,41} = V_{i,32}\;, 
\quad 
V_{i,44} = V_{i,33}\;, 
\quad 
V_{i,43} = V_{i,34}\;,
}
\qquad i = 1,2,3, 
\\[2ex]  
V_{j,3\gs} = V_{j,1\gs}\;, 
\qquad 
V_{j,4\gs} = V_{j,2\gs}\;, 
\qquad j = 2,3; \quad \gs =1,\ldots,4\;, 
}
\end{subequations} 
consequently the number of independent bundle vectors reduces from $48$ to $16$.

A final simplification comes from the following observation: The relation between a line bundle vectors $V_r$ and a local orbifold twist $V_g$ is  same as the definition of the shifted momentum $P_{sh} = V_g + P$, with $P \in \gL$, of a twisted state that string theory predicts at the orbifold fixed tori. Massless twisted states satisfy $P_{sh}^2 = 3/2$ on the \ztwo orbifold when no oscillator excitations are switched on. Inspired by this, we assume that vectors from the same sector all have length 3/2. This assumption automatically ensures that the Bianchi identities \eqref{eq:Z2_BIeqnsA}, obtained by integrating over the inherited divisors, are satisfied.

By applying all these assumptions simultaneously, the complete system of Bianchi identities \eqref{eq:ex_Z2_BIeqns} simplifies tremendously:  
\begin{subequations}
\label{eq:ex_Z2_BIeqns_Simplified}
\begin{gather}
V_{i,\rho\sigma}^{2} = \frac{3}{2}\;,~~ i \in \left\{1,2,3\right\} , ~ \rho,\sigma \in \left\{1,2,3,4\right\} \;,\label{eq:Z2_BIeqnsSimplifiedA}\\[1pt]
(V_{1,11}+V_{1,12}+V_{1,31}+V_{1,32}) \cdot V_{2,11}=1\;,\qquad
(V_{1,11}+V_{1,12}+V_{1,31}+V_{1,32}) \cdot V_{2,12}=1\;,\label{eq:Z2_BIeqnsSimplifiedC}\\[6pt]
(V_{1,13}+V_{1,14}+V_{1,33}+V_{1,34}) \cdot V_{2,13}=1\;,\qquad
(V_{1,13}+V_{1,14}+V_{1,33}+V_{1,34}) \cdot V_{2,14}=1\;,\label{eq:Z2_BIeqnsSimplifiedE}\\[6pt]
(V_{1,11}+V_{1,12}+V_{1,13}+V_{1,14}) \cdot V_{3,11}=1\;,\qquad
(V_{1,11}+V_{1,12}+V_{1,13}+V_{1,14}) \cdot V_{3,12}=1\;,\label{eq:Z2_BIeqnsSimplifiedG}\\[6pt]
(V_{1,31}+V_{1,32}+V_{1,33}+V_{1,34}) \cdot V_{3,13}=1\;,\qquad
(V_{1,31}+V_{1,32}+V_{1,33}+V_{1,34}) \cdot V_{3,14}=1\;.\label{eq:Z2_BIeqnsSimplifiedI}
\end{gather}
\end{subequations}
In particular, instead of 48 Diophantine equations involving scalar products of different bundle vectors, (cf. equations \eqref{eq:Z2_BIeqnsB}-\eqref{eq:Z2_BIeqnsD}), only eight of them remain.

\subsubsection[A six generation GUT on triangulation ``$E_1$"]{A six generation GUT on triangulation ``$\boldsymbol{E_1}$"}
\label{sec:Explicit_SU5_Blowup_GUT}

To solve the combined system of flux quantization conditions and Bianchi identities we use the strategies put forward in the previous subsection. From the orbifold data \eqref{eq:Z2_Model_Input} one possible solution of the reduced set of Bianchi equations \eqref{eq:ex_Z2_BIeqns_Simplified} in triangulation ``$E_1$'' is given in table \ref{table:Z2_BI_Solution_Vectors}.

To ensure that the resolution model can ultimately be interpreted as an MSSM  candidate, the bundle vectors were chosen such that the unbroken gauge group contains an $\text{SU}(5)$ factor. Indeed, the commutant of the line bundle  background reads
\equ{
[\text{SU}(5)] \times [\text{SU}(3) \times \text{SU}(2)]\;; 
}
the $\text{SU}(3)$ and $\text{SU}(2)$ are part of the second (hidden) $\text{E}_8$ gauge group. By applying the multiplicity operator \eqref{eq:ex_Z2_Multiplicity_Operator} to all \eeight states we can determine the complete chiral spectrum of the resolution model. The multiplicities are integral for the \eeight roots and there is no non-Abelian anomaly. The resulting particle spectrum is given in table \ref{table:Z2_Complete_Spectrum}. The spectrum contains six $\mathbf{10}$'s, twelve $\overline{\mathbf{5}}$'s and five $\rep{5}$'s, hence the model can describe a net number of six GUT generations. In addition some of the vector-like $\crep{5} + \rep{5}$ can be interpreted as Higgs pairs. The distinction between the 70 scalar in the first $\text{E}_8$ and 80 in the second $\text{E}_8$ might seem somewhat arbitrary, but is fixed by the fact they are charged under different $\text{U}(1)$ factors coming either from the first or the second $\text{E}_8$.

\renewcommand{\arraystretch}{1.2}
\begin{table}
\centering
\subfloat[Massless spectrum of the first $\text{E}_{8}$\label{table:Z2_E81SpectrumSmall}]{
\begin{tabular}[t]{|c|l@{\hskip.7cm}||c|l@{\hskip.7cm}|}
\hline
$\#$	& irrep 									& $\#$	& irrep 										\\
\hline
\hline
6	& $(\mathbf{10}; \rep{1},\rep{1})$							&  
70	& $(\mathbf{1}; \rep{1},\rep{1})$ 
\\ 
12	& $(\mathbf{\overline{5}}; \rep{1},\rep{1})$  & 
6	& $(\mathbf{5}; \rep{1},\rep{1})$	
\\ 
\hline
\end{tabular}
}
\qquad
\subfloat[Massless spectrum of the second $\text{E}_{8}$\label{table:Z2_E82SpectrumSmall}]{
\begin{tabular}[t]{|c|l@{\hskip.7cm}||c|l@{\hskip.7cm}|}
\hline
$\#$	& irrep 									& $\#$	& irrep										\\
\hline
\hline
16	& $(\rep{1}; \mathbf{3},\mathbf{1})$					& 16		& $(\rep{1};\overline{\mathbf{3}},\mathbf{1})$			\\
32	& $(\rep{1};\mathbf{1},\mathbf{2})$					& 80		& $(\rep{1};\mathbf{1},\mathbf{1})$						\\
\hline
\end{tabular}
}
\caption{Chiral massless spectrum of the \ztwo model. The multiplicities are calculated using \eqref{eq:ex_Z2_Multiplicity_Operator}. The representations under $\text{SU}(5)$ of the first $\text{E}_{8}$ and $\text{SU}(3) \times \text{SU}(2)$ of the second $\text{E}_{8}$ are given in boldface.
}
\label{table:Z2_Complete_Spectrum}
\end{table}
\renewcommand{\arraystretch}{1.0}

\subsubsection[The MSSM after modding out the $\zf$ involution]{The MSSM after modding out the $\boldsymbol{\zf}$ involution}

After having obtained an $\text{SU}(5)$ GUT model with six generations we define a \zf~Wilson line in the $\gt$ direction, as in the orbifold limit, that simultaneously  breaks the $\text{SU}(5)$ GUT group down to the standard model gauge group, $\text{SU}(3) \times \text{SU}(2)\times \text{U}(1)_Y$, and reduces the number of states by a factor of two. This naive approach to the computation of the spectrum provides the correct result, as discussed in section~\ref{sc:freeWLresolution} and in appendix~\ref{sc:NonFact}.

Hence the model describes three generations of quarks and lepton of the Standard Model. As the hypercharge is embedded in a canonical way in the $\text{SU}(5)$ and all $48$ line bundle vectors are $\text{SU}(5)$ singlets, the hypercharge is orthogonal to all blow-up modes, and hence not broken.

\subsubsection[DUY and moduli of the \ztwo orbifold resolution]{DUY and moduli of the $\boldsymbol{\ztwo}$ orbifold resolution}

To see whether the solution is consistent with the DUY equations, we insert the gauge flux into equation \eqref{eq:DUY_Equation}. Before giving these equations explicitly, we first make some simplifying observations: First of all in order to ensure that the freely acting involution \zf\ can be modded out, the volumes need to be identified pairwise. Since the Wilson line $W_1$ is not switched on, groups of four volumes associated with the exceptional divisors $E_{2,\gs\gr}$ and $E_{3,\gs\gr}$ satisfy the same equations because they get contracted with identical line bundle vectors. As we merely want to understand to what extend a full blow-up is possible, we simply take all volumes equal that are related to each other in these ways. This means that we have 16 independent volumes, $\text{vol}_1,\ldots, \text{vol}_{16}$, one for each of the different bundle vector given as the rows in table \ref{table:Z2_BI_Solution_Vectors}.

If we ignore the loop corrections to the DUY equations, we find that following relations 
\begin{subequations}
\equ{
\text{vol}_{2} = 0\;,  \quad
\text{vol}_{15} = \text{vol}_{16}\;, \quad 
\text{vol}_{4} + \text{vol}_{5} + \text{vol}_{12} = 0\;, \quad 
\text{vol}_{1} + \text{vol}_{3} + \text{vol}_{9} + \text{vol}_{16} = 0\;, 
\label{eq:DUY_exA}\\[1ex] 
\text{vol}_{10} = \text{vol}_{4}+ \text{vol}_{11} + \text{vol}_{12} + \text{vol}_{16}\;, 
\quad 
\text{vol}_{7} = \frac {\text{vol}_{13} + \text{vol}_{14} + \text{vol}_{4} - \text{vol}_{9} - \text{vol}_{16}}2\;, 
\label{eq:DUY_exB}\\[1ex] 
\text{vol}_{6} = \text{vol}_{11}\;, \quad 
\text{vol}_{8} = \text{vol}_{14}\;,
\label{eq:DUY_exC}\quad 
\text{vol}_{11} = \frac {3 \text{vol}_{13} + \text{vol}_{14} - \text{vol}_{4} - 2 \text{vol}_{5} + \text{vol}_{9} - 3 \text{vol}_{16}}2\;.
}
\end{subequations} 
As the volumes need to be positive, the equations \eqref{eq:DUY_exA} imply that all the nine volumes that occur in these equations have to vanish simultaneously. The remaining equations in \eqref{eq:DUY_exB} and \eqref{eq:DUY_exC} express that seven volumes can be made large simultaneously, and these volumes are functions of $\text{vol}_{13}$ and $\text{vol}_{14}$. When loop corrections to the DUY equations are taken into account all volumes can be made positive.

\subsection{Matching with MSSM orbifold}

Table \ref{table:Z2_vectorIdentifications} summarizes the identification between twisted orbifold states and line bundle vectors. As it turns out, the states which are projected out in four dimensions have shifted momenta that are the negatives of those that survive the projections. Hence this table shows that the present orbifold model cannot be the blow-down model of the resolution model described in the previous section. 

\begin{table}
\centering
$
\begin{array}{|ll||ll||ll|}
\hline
\multicolumn{2}{|c||}{\theta_{1} \text{-sector}} & \multicolumn{2}{|c||}{\theta_{2} \text{-sector}} & \multicolumn{2}{|c|}{\theta_{3} \text{-sector}}						\\
\hline
\hline
V_{1,11} \leftrightarrow T^{(1)}_{1}			& V_{1,31} \leftrightarrow S_{8}			&
V_{2,11} \leftrightarrow ~\,~T^{(1)}_{5}		& V_{2,31} \leftrightarrow ~\,~T^{(1)}_{9}		&
V_{3,11} \leftrightarrow S_{31}				& V_{3,31} \leftrightarrow S_{45}		
\\
V_{1,12} \leftrightarrow \overline{T}^{(2)}_{1}		& V_{1,32} \leftrightarrow V^{(1)}_{1}			&
V_{2,12} \leftrightarrow -\overline{T}^{(1)}_{5}	& V_{2,32} \leftrightarrow -\overline{T}^{(1)}_{9}	&
V_{3,12} \leftrightarrow S_{35}				& V_{3,32} \leftrightarrow S_{49}			
\\
V_{1,13} \leftrightarrow T^{(1)}_{2} 			& V_{1,33} \leftrightarrow S_{13}			&
V_{2,13} \leftrightarrow ~\,~T^{(1)}_{6}		& V_{2,33} \leftrightarrow ~\,~T^{(1)}_{10}		& 
V_{3,13} \leftrightarrow T^{(2)}_{9}			& V_{3,33} \leftrightarrow T^{(2)}_{11}	
\\
V_{1,14} \leftrightarrow \overline{T}^{(1)}_{2}		& V_{1,34} \leftrightarrow S_{16}			&
V_{2,14} \leftrightarrow ~\,~\overline{T}^{(1)}_{6}	& V_{2,34} \leftrightarrow ~\,~\overline{T}^{(1)}_{10}	&
V_{3,14} \leftrightarrow \overline{T}^{(2)}_{9}		& V_{3,34} \leftrightarrow \overline{T}^{(2)}_{11}
\\[1ex] 
V_{1,21} \leftrightarrow \overline{T}^{(2)}_{3}		& V_{1,41} \leftrightarrow V^{(1)}_{2}			&
V_{2,21} \leftrightarrow -\overline{T}^{(1)}_{7}	& V_{2,41} \leftrightarrow -\overline{T}^{(1)}_{11}	& 
V_{3,21} \leftrightarrow S_{37}				& V_{3,41} \leftrightarrow S_{51}		
\\
V_{1,22} \leftrightarrow T^{(1)}_{3}			& V_{1,42} \leftrightarrow S_{19}			&
V_{2,22} \leftrightarrow ~\,~T^{(1)}_{7}		& V_{2,42} \leftrightarrow ~\,~T^{(1)}_{11}		&
V_{3,22} \leftrightarrow S_{40}				& V_{3,42} \leftrightarrow S_{54}	
\\
V_{1,23} \leftrightarrow \overline{T}^{(1)}_{4}		& V_{1,43} \leftrightarrow S_{22}			&
V_{2,23} \leftrightarrow ~\,~\overline{T}^{(1)}_{8}	& V_{2,43} \leftrightarrow ~\,~\overline{T}^{(1)}_{12}	&
V_{3,23} \leftrightarrow \overline{T}^{(2)}_{10}	& V_{3,43} \leftrightarrow \overline{T}^{(2)}_{12}
\\
V_{1,24} \leftrightarrow T^{(1)}_{4}			& V_{1,44} \leftrightarrow S_{26}			&
V_{2,24} \leftrightarrow ~\,~T^{(1)}_{8}		& V_{2,44} \leftrightarrow ~\,~T^{(1)}_{12}		& 
V_{3,24} \leftrightarrow T^{(2)}_{10}			& V_{3,44} \leftrightarrow T^{(2)}_{12}		
\\
\hline
\end{array}
$
\caption{Identification between the orbifold states and the line bundle vectors. The nomenclature of the twisted states is summarized in table \ref{table:Z2_Orbifold_Spectrum}. The $4$ vectors in the $\theta_{2} \text{-sector}$ with a minus sign are present in the six-dimensional theory but are projected out in four dimensions. The negative of these vectors are present in four dimensions.}
\label{table:Z2_vectorIdentifications}
\end{table} 
%
%
%
%
%
%
%
%
%
%
%
%
%
%
%
%
%
%
%
%
%
%
%
%
%
%
%
%
%
%
%
%
%
%
\section{Conclusions}
\label{sec:Conclusions}

%
%
The main result of our paper is the construction of a smooth compactification of \eeight heterotic supergravity producing a four dimensional model that contains the MSSM spectrum. The smooth Calabi-Yau manifold is obtained by resolving the orbifold $T^6/\ztwo$ and we embed line bundles into the \eeight gauge group in order to fulfill the Bianchi identity. Furthermore, this embedding is chosen such that the first $\text{E}_8$ factor is broken to an $\text{SU}(5)$ GUT gauge group and we obtain six chiral matter generations of $\rep{10} + \crep{5}$. Both the $T^6/\ztwo$ orbifold and a large class of its resolutions admit a freely acting $\zf$ involution. The breaking of $\text{SU}(5)$ to the standard model gauge group is achieved by associating an appropriately chosen Wilson line to this involution. This allows for the GUT breaking to happen at a scale different from the string scale, while the hypercharge remains non-anomalous and therefore unbroken in the resolution model. Moreover, the involution halves the chiral GUT spectrum, resulting in three generations of quarks and leptons.

%
%
Maybe the biggest surprise of this work is that there can be more massless chiral states on a resolution than the orbifold spectrum can account for (after branching and Higgsing, induced by the blow-up modes). This is remarkable because the orbifold is a point of enhanced symmetry in the moduli space where one expects the massless spectrum to be the largest. The existence of the extra resolution states seems to be intimately connected with the fact that the smooth Calabi-Yau resolution is not unique. The different resolutions, corresponding to varying the triangulations of the 64 $\Cplx^3/\ztwo$ singularities, are related to each other via a web of flop transitions. During these transitions, parts of the spectra (which are chiral only w.r.t.\ some anomalous $\text{U}(1)$ factors) can jump. The flops occur when certain combinations of the twisted K\"ahler moduli, corresponding to volumes of imaginable curves $C$, change sign. In a simple toy model we can explain the disappearance of some states after a flop by observing that possible instantonic mass terms of the schematic form $\exp(-\text{Vol}(C))$ explode as soon as one passes to a flopped geometry.  As the novel states are discovered using a representation dependent index theorem on the resolution, a truly string theoretical interpretation of these states is lacking as far as we know. Fortunately, the chiral MSSM spectrum is unaffected by flop transitions, since it is impossible to write down instantonic mass terms for any of the chiral MSSM states. However, the Higgs sector might be sensitive to this.

The main findings discussed above are obtained from a number of interesting technical results:
\begin{itemize}
\item 
%
%
We give a complete characterization of the resolved $T^6/\ztwo$ space embedded in a large toric variety. This has been achieved by viewing two-tori as elliptic curves, so that this orbifold can be described as an algebraic hypersurface in a certain toric variety \cite{Vafa:1994}. Combining this with the knowledge about the non-compact toric resolutions of the local $\Cplx^3/\ztwo$ singularities and the topological gluing conditions (see e.g. \cite{Denef:2005mm,Lust:2006} and references therein) we obtain a complete geometrical description. This construction in particular explains why the local compactifications of resolved ${\Cplx^3/\ztwo}$ singularities using compact toric varieties based on certain ``auxiliary polyhedra'' give the correct intersection numbers. 
\item 
%
%
A novel characterization of line bundles on the resolved orbifold is obtained in terms of the scalings that define the resolved singularities as toric varieties. This allows us in particular to re-obtain the matching between gauge fluxes on the resolved singularities and the embedding of the orbifold action into the gauge degrees of freedom (as shifts and Wilson lines) in the singular case.
\item 
%
%
We show that the embedding of the 48 line bundles into the \eeight gauge group possesses the same amount of information as the two shifts and six orbifold Wilson lines (up to lattice vectors) that a heterotic $T^6/\ztwo$ model can maximally be equipped with. The reason is that the line bundle flux quantization conditions in any triangulation are equivalent to a universal set of requirements. This set of conditions implies that the 48 line bundle vectors can be expressed in terms of eight fundamental ones (up to lattice vectors), just like local orbifold gauge shift vectors are determined by two shifts and six Wilson lines. 
\item 
%
%
The orbifold $T^6/\ztwo\times \zf$ can be constructed in two equivalent ways: Either one starts from the orbifold $T^6/\ztwo$ based on a factorizable torus $T^6$ and then mods out the $\zf$ action, or one first defines the non-factorizable torus $T_{\rm non} = T^6/\zf$ which is subsequently $\ztwo$-orbifolded. Accordingly, there are also two equivalent ways of defining the resolved geometry. To determine the spectrum on the resolution of $\ztwo\times\zf$  in the former approach one needs to use the orbifold information telling us that the twisted chiral states are identified pairwise. However, given that we find additional states on the resolutions it is not clear whether we can make use of a similar identification for them as well. Therefore, chiral spectra can only be reliably computed on the resolution when the resolved space is regarded as $\text{Res}(T^6_{\rm non}/\ztwo)$. 
\item
%
%
The exact CFT description of the heterotic $T^6/\ztwo \times \zf$ orbifold implies that modular invariance consistency requirements have to be imposed on the freely acting Wilson line as well. Therefore, it is important to note that the full orbifold partition function includes new twisted winding sectors involving the translation $\gt$, i.e.~with boundary conditions of the form $\left(\theta_i, \tau \right)$. These six dimensional states are localized in between the $\ztwo$ fixed points in the directions in which the $\theta_i$ rotation acts.
\end{itemize}

%
%
To conclude, let us mention one important issue which needs to be resolved in further studies. The supergravity construction of the MSSM model on the resolution is strongly inspired by a specific MSSM orbifold model. Nevertheless, we cannot state that the supergravity model is a ``blow-up'' of the orbifold model. The matching is hampered by the fact that the gauge fluxes are identified with the local orbifold action up to \eeight lattice elements. The latter are not irrelevant in orbifold model building, since they map a model into a so-called ``brother'' model having the same gauge group but possibly a different spectrum.\footnote{In this work we ignored the alternative description in which orbifold ``brother'' models are characterized by different discrete torsion phases.} This is reflected in the fact that the identification of line bundle vectors with twisted blow-up modes seems to indicate that some of the blow-up modes are not part of the four dimensional massless spectrum.

\subsection*{Acknowledgments}

We would like to thank Wilfried Buchm\"uller, Andres Collinucci, Arthur Hebecker, Johannes Held, Adrian Mertens, Hans-Peter Nilles, Michael Ratz, Sa\'ul Ramos-S\'anchez, Emanuel Scheidegger and Angel Uranga for valuable discussions. 
The work of M.B., F.R., and M.T. was partially supported by the SFB-Tansregio TR33 ``The Dark Universe'' (Deutsche Froschungsgemeinschaft) and 
the European Union 7th network program ``Unification in the LHC era'' (PITN-GA-2009-237920). The work of P.V. is supported by LMU Excellent.

%
%
%
%
%
%
%
%
%
%
%
%
%
%
%
%
%
%
%
%
%
%
%
%
%
%
%
%
%
%
%
%
%
%
\appendix 
\def\theequation{\thesection.\arabic{equation}} 
\setcounter{equation}{0}
%
\setcounter{equation}{0}
\section{Two-torus as an elliptic curve}
\label{sc:Weierstrass}

In this appendix we exploit some basic properties of Weierstrass functions that can be found in textbooks like \cite{Ahlfors,Koblitz} which can be used to represent a two-torus as an elliptic curve.

The Weierstrass function $\cP(z)$ is an even double periodic function 
\equ{
\cP(z+1) = \cP(z+\gr) = \cP(z), 
}
i.e.\ a function on the two-torus $T^2 = \Cplx/\gL$ with lattice $\gL$ generated by $1$ and the complex structure $\gr\in\Cplx$. (In the main text we mostly take $\gr = i$ for simplicity.) The Weierstrass function $\cP(z)$ and its derivative $\cP'(z)$ have pole expansions given by 
\equ{
\cP(z) = \frac 1{z^2} + \sum_{\go\in\gL_{\neq 0}} 
\Big\{ 
\frac 1{(z-\go)^2} - \frac1{\go^2}
\Big\}, 
\qquad
\cP'(z) = -2 \sum_{\go \in \gL}\frac 1{(z-\go)^3}.  
\label{WeierstrassPOLES} 
}
The Weierstrass function satisfies the differential equation
\equ{
\cP'(z)^2 = 4 
\big( \cP(z) - \gve_1 \big)\big( \cP(z) - \gve_2 \big)\big( \cP(z) - \gve_3 \big).
\label{WeierstrassDIF}
}
The three zeros $\gve_1 = \cP(\frac 12)$,  $\gve_2 = \cP(\frac \gr2)$ and $\gve_3 = \cP(\frac {1+\gr}2)$ of $\cP'(z)$ satisfy 
\equ{
\gve_1 + \gve_2 + \gve_3 = 0. 
\label{WeierstrassROOTS}
}
Because of the differential equation \eqref{WeierstrassDIF} the Weierstrass function defines a mapping between a two-torus and an elliptic curve in $(u,y) \in \Cplx^2$
\equ{
y^2 = 4 (u-\gve_1)(u-\gve_2) (u-\gve_3)
\label{elliptic} 
}
via $(u, y) = (\cP(z), \cP'(z))$.

The addition of two points $z_1$ and $z_2$ in $T^2$ leads to the sum formula  
\equ{
\cP(z_1+z_2) = \frac 14 \Big( \frac{\cP'(z_1)-\cP'(z_2)}{\cP(z_1)-\cP(z_2)} \Big)^2 
- \cP(z_1) - \cP(z_2). 
\label{WeierstrassADD}
}
The fact that the sum of the three roots $\gve_1$, $\gve_2$ and $\gve_3$ vanishes in \eqref{WeierstrassROOTS} is a particular consequence of this addition formula. The addition law for the derivative of the Weierstrass function can be cast in the form 
\equ{
\cP'(z_1+z_2) =  -\frac 14 \Big( \frac{\cP'(z_1)-\cP'(z_2)}{\cP(z_1)-\cP(z_2)} \Big)^3
+ 3\, \frac{\cP'(z_1)-\cP'(z_2)}{\cP(z_1)-\cP(z_2)} \, \cP(z_2) +  \cP'(z_1) - 2\, \cP(z_2). 
}
The image $(u',y') = \gt(u,y)$ of a point $(u,y)$ on the elliptic curve under translation $\gt$ over $\frac \gr2$ on $T^2$ can be computed using these addition formulae. Because $\gve_2=\cP(\frac \gr2)$ is one on the roots of $\cP'(z)$ they simplify considerably when taking $z_2 = \frac \gr2$. For the factors in the elliptic equation \eqref{elliptic} we find 
\equ{
\arry{ll}{ \dsp 
u' - \gve_1 = - \frac {\gve_1-\gve_2}{u-\gve_2} \, (u-\gve_3), 
\quad & \quad  \dsp 
u' - \gve_2 = \frac {(\gve_1-\gve_2)(\gve_3-\gve_2)}{u-\gve_2},
\\[2ex] \dsp 
u' - \gve_3 = - \frac {\gve_3-\gve_2}{u-\gve_2} \, (u-\gve_1),
\quad & \quad  \dsp 
y' = -  \frac {(\gve_1-\gve_2)(\gve_3-\gve_2)}{(u-\gve_2)^2}\, y. 
}
\label{ROOTmapping}
}
This shows in particular that $(u',y')$ lies on the elliptic curve whenever $(u,y)$ does.

It is possible to construct a linear realization of the $\gt$-translation by extending the Weierstrass function to a mapping $T^2 \ra \cO(2;\CP^1)$. The homogeneous coordinates $(u,v;y)$ of $\cO(2; \CP^1)$ are identified up to $\gm \in \Cplx^*$ scalings given by 
\equ{
(u, v; y) \sim (\gm\, u, \gm\, v; \gm^2\, y ). 
\label{WeierstrassSCALINGS}
}
This ensures that the mapping   
\equ{
\left\{ 
\arry{lcll}{
z & \mapsto & \big( \cP(z), 1; \cP'(z) \big), 
& \quad \text{away from lattice points,}
\\[1ex] 
0 & \mapsto & \big( 1, 0; 0\big), 
& \quad \text{at lattice points,}
\\[1ex] 
z & \mapsto & \big( 1, 1/\cP(z); \cP'(z)/\cP^2(z) \big),
& \quad \text{near lattice points,}
}
\right. 
}
is analytic on the whole torus including the lattice points. The translation $\gt$ can be represented on the homogeneous coordinates as 
\equ{
(u,v;y) \ra(u',v';y') = \gt(u,v;y) = 
\big( 
\gve_2\, u + (\gve_1\gve_3 + \gve_2^2)v, u-\gve_2v; -(\gve_1-\gve_2)(\gve_3-\gve_2) y
\big)
}
by making a $\Cplx^*$-scaling with $\gm = u-\gve_2$ after the $\gt$-translation \eqref{ROOTmapping} in local coordinates. Hence on the homogeneous coordinates of $\CP^1$ the translation $\gt$ acts as matrix multiplication 
\equ{
U = \pmtrx{u \\ v} \ra \gt (U) = M_\gt U, 
\qquad 
M_\gt = \pmtrx{\gve_2 & \gve_1\gve_3 + \gve_2^2 \\ 1 & -\gve_2}. 
\label{CPtau}
}
In this representation it is straightforward to check (using \eqref{WeierstrassROOTS}) that $\gt^2$ acts as  a factor $-\det M_\gt = (\gve_1-\gve_2)(\gve_3-\gve_2)$ times the identity on $U$ and as $(\det M_\gt)^2$ on $y$. Hence, by a $\Cplx^*$-scaling with $\gm = - \det M_\gt$ we bring the homogeneous coordinates $(u,v; y)$ back to themselves, i.e.\ $\gt^2$ acts as the identity.

The elliptic equation can be written in terms of the homogeneous coordinates of $\cO(2;\CP^1)$ as 
\equ{
y^2 = 4 \prod_{\ga=1}^4 N_\ga\cdot U
}
where $N_\ga$ are lying vectors given by 
$N_1 = \pmtrx{0 & 1}$, $N_{2}= \pmtrx{1 & - \gve_2}$, 
 $N_{3}= \pmtrx{1 & - \gve_1}$
 and  $N_{4}= \pmtrx{1 & - \gve_3}$. It is not difficult to check that the $\gt$ action on them $\gt(N_\ga) = N_\ga M_\gt$ leads to 
\equ{
\arry{ll}{
\gt(N_1) = N_2, 
\quad & \quad 
\gt(N_3) = (\gve_2-\gve_1) \, N_4, 
\\[1ex] 
\gt(N_2) = (\gve_1-\gve_2)(\gve_3-\gve_2)\, N_1, 
\quad & \quad 
\gt(N_4) = (\gve_2-\gve_3) \, N_3. 
}
\label{FixedPointMapping}
}
This shows explicitly that the $\Intr_2$ fixed points are mapped onto each other pairwise.

%
%
%
%
%
%
%
%
%
%
%
%
%
%
%
%
%
%
%
%
%
%
%
%
%
%
%
%
%
%
%
%
%
%
\setcounter{equation}{0}
\section{Resolution of a non-factorizable torus orbifold}
\label{sc:NonFact}

In the main part of the paper we have divided out the freely acting \zf\ involution more or less by hand by saying  that its net effect is halving the number of exceptional divisors. This procedure was motivated by noting that on the orbifold the involution \zf\ merely leads to an identification of twisted sectors. In this appendix we  give more detailed arguments why this procedure is allowed.

Our arguments follow essentially two independent lines of reasoning which are summarized in figure~\ref{fg:NonFact}. The two approaches arise as we can either consider the factorizable torus $T_{\rm fac}^6$  on which we subsequently perform the $\ztwo$ orbifolding, then the resolution procedure and finally apply the $\zf$ involution. This is the route followed in the main text since there we have always assumed that the torus is factorizable. Alternatively, we note that applying the freely acting involution $\zf$ to the factorizable torus $T^6_{\rm fac}$ leads to a non-factorizable one, called $T^6_{\rm non}$. This non-factorizable torus can be orbifolded under $\ztwo$ and resolved subsequently. The fact that these two procedures result in the same space is mathematically phrased by saying that the diagram in figure \ref{fg:NonFact} commutes.

\begin{figure}[t]  
\[
\begin{CD}
T^{6}_{\rm fac}
~@>~~\ztwo~~>>~
T^6_{\rm fac}/\ztwo
~@>~~\text{resolution}~~>>~ 
\text{Res}\left(T^6_{\rm fac}/\ztwo\right)
\\[1ex] 
@V\zf VV @V\zf VV @V\zf VV
\\[1ex] 
T^6_{\rm fac}/\zf
~@>~~\ztwo~~>> ~
(T^6_{\rm fac}/\ztwo)/\zf
~@>~~\text{resolution}~~>> ~ 
\text{Res}\left( (T^6_{\rm fac}/\ztwo)\right)/\zf
\\[1ex]
@| @| @| 
\\[1ex]
T^{6}_{\rm non}
~@>~~\ztwo~~>> ~
T^6_{\rm non}/\ztwo
~@>~~\text{resolution}~~>> ~ 
\text{Res}\left(T^6_{\rm non}/\ztwo\right) 
\end{CD}
\non
\]
\caption{This commutative diagram indicates different routes that lead to the resolution of the $T^6/\ztwo\times\zf$ orbifold. The first column indicates that orbifolding a factorizable torus $T^6_{\rm fac}$ by the freely acting involution $\zf$ can be thought of as a non-factorizable torus.  The lower line indicates that our procedure can be understood as the resolution of a $\ztwo$ orbifold on a non-factorizable lattice.
\label{fg:NonFact}}
\end{figure}

\subsection{The freely acting involution on the factorizable orbifold and its resolution}
\label{sec:T6Z2Z2Z2free}

In the first approach one starts with the results for the geometry of the factorizable orbifold $T^6_{\rm fac}/\ztwo$ and its resolution $\text{Res}(T^6_{\rm fac}/\ztwo)$ discussed at length in section \ref{ch:Het_Sugra_on Resolution}. For concreteness we fix the complex structure of the three two-tori that make up $T^6_{\rm fac}$ to be $i$. This means that on the complex coordinates $z_1,z_2,z_3$ of the three two-tori the $\zf$ acts as 
\equ{
\tau: ~(z_1,z_2,z_3) ~\longrightarrow~(z_1+i/2, z_2+i/2, z_3+i/2)\;.
\label{Z2free_mapping} 
}
This involution is modded out by identifying the images of fixed tori and fixed points, divisors and intersections under the action of this involution. This procedure is therefore very similar to the construction of standard non-prime orbifold resolutions where invariant divisor combinations are formed, as discussed in \cite{Lust:2006}. All objects that are defined after applying the freely acting $\zf$ involution are denoted by tildes on them, i.e.\ $\tR_i$, etc. (Note that this convention is in a certain sense opposite to the one of ref.~\cite{Lust:2006}.)

We begin with the inherited divisors. We represent the inherited divisors of $\text{Res}(T^6_{\rm fac}/\ztwo)/\zf$ as 
\equ{
 \label{eq:Nonfac_R} 
 \tR_i := \{z_i = c_i \} \cup  \{z_i = -c_i \} \cup  \{z_i = c_i + i/2 \} \cup  \{z_i = -c + i/2 \}
} 
on the covering space $T^6_{\rm fac}$, i.e.\ we consider all the $\ztwo$ and $\zf$ images of the defining equation $z_i = c_i$. This means that if we think of the divisors $\tR_i$ in the covering space $T^6_{\rm fac}/\ztwo$ we have 
\begin{align}
 \tR_i = R_i(c_i) + R_i(c_i+i/2) \sim 2 \; R_i\;.  
 \label{eqn:match_R}
\end{align}
The first equality indicates that on this cover $\tR_i$ corresponds to two disjoint branches. In the second step we used that the divisors $R_i(c_i) := \{ z_i = c_i\}$ are so-called sliding divisors, so that we can continuously change the value of $c_i+i/2$ to become equal $c_i$.

The triple intersection number of the inherited divisors are determined as follows. Since on the covering space $T^6_{\rm fac}$ each $\tR_i$ has four parallel branches, it follows that the divisors $\tR_1$, $\tR_2$ and $\tR_3$ intersect in $4^3 = 64$ points. To find the intersection number on $\text{Res}(T^6_{\rm fac}/\ztwo)/\zf$ we mod out the finite groups \ztwo and $\zf$ one after the other (or, according to figure \ref{fg:NonFact}, equivalently simultaneously) which gives 
\(
\tR_1 \tR_2 \tR_3 = {64}/{2^3} = 8\;. 
\)
Notice that this intersection number is {\em not} the same as the intersection number of the inherited divisors $R_1, R_2$ and $R_3$ on the resolution of the factorizable orbifold $T^6_{\rm fac}/\ztwo$, see table \ref{table:Z2_Intersection_Numbers}. Nevertheless this result is compatible with the procedure outlined in \cite{Lust:2006} resulting in their equation (4.9) provided one interprets the $n_i$ as the number of different images of $c_i$ under the discrete group $G$.

This result for $\tR_1 \tR_2 \tR_3$ is also obtained by taking $\text{Res}(T^6_{\rm fac}/\ztwo)$ as the covering space of $\text{Res}(T^6_{\rm fac}/\ztwo)/\zf$. Indeed, given that one knows the intersection ring on the covering space, one expresses all divisors in terms of those on this cover, then computes the intersection numbers by using those on the cover, and finally divide the result by 2 (the order of $\zf$). Following this strategy one obtains 
\equ{
\tR_1 \tR_2 \tR_3 = \frac{2^3 \cdot 2}2 = 8\;, 
\label{Int_tR}
}
using that each $\tR_i \sim 2 \; R_i$ and $R_1R_2R_3 = 2$ on the cover. This method has an advantage over taking $T^6_{\rm fac}$ as the covering space as it can be applied to the ordinary and exceptional divisors in resolution as well.

On $\text{Res}(T^6_{\rm fac}/\ztwo)/\zf$ we define the ordinary divisors
\begin{align}
 \label{eq:Nonfac_D}
 \tD_{i,1} := \{ z_i = 0 \} \cup \{ z_i = i/2 \} \;, 
 \qquad  
 \tD_{i,3} := \{ z_i = 1/2 \} \cup \{ z_i = 1/2 + i/2 \} \;, 
\end{align}
so that in terms of the ordinary divisors on the cover $T^6_{\rm fac}/\ztwo$ (or its resolution) we have 
\begin{align}
 \tD_{i,a} = D_{i,a} +  D_{i,a+1} \;, 
 \qquad a =1,3\;. 
 \label{eqn:match_D}
\end{align}
Following the strategy outlined above, the intersection numbers of the $\tR$'s with the $\tD$'s are computed using those of the covering space $\text{Res}(T^6_{\rm fac}/\ztwo)$ resulting in 
\equ{
 \tD_{i, a} \tR_j \tR_k = 4\;, 
\qquad 
\tD_{i,a} \tD_{j,b} \tR_k = 0\;, 
\label{Int_tDtR} 
}
with $i\neq j\neq k$.

To understand the exceptional divisors on $\text{Res}(T^6_{\rm fac}/\ztwo)/\zf$ we first consider the $\gt$ action \eqref{Z2free_mapping} for the $\ztwo$ fixed point labels $\ga, \gb$ on the factorizable torus $T^6_{\rm fac}$. Given the labeling scheme defined in table \ref{tb:Z2_sectors} we see that this mapping takes $1 \leftrightarrow 2$ and $3 \leftrightarrow 4$. This means that the number of fixed points and consequently exceptional divisors is halved. On the space 
$\text{Res}(T^6_{\rm fac}/\ztwo)/\zf$ only $\tau$-invariant combinations survive. Therefore we define
\begin{align}
\tE_{k,ab}^+ := E_{k,ab} + E_{k,(a+1)(b+1)} \;, 
\qquad 
\tE_{k,ab}^- := E_{k,(a+1)b} + E_{k,a(b+1)}\;.  
\label{eqn:match_E}
\end{align}
Since these objects are both \zf-symmetric, they also exist on the resolution. Collectively we write $\tE_{k,ab}^{\go}$ where $\go = \pm$ and $a,b=1,3$. As the exceptional divisors are associated with fixed tori on the resolution, this also fixes the labels of the fixed tori of the orbifold $(T^6_{\rm fac}/\ztwo)/\zf$.

Before we get to the intersection numbers involving these exceptional divisors, we first note that the exceptional divisors $\tE$ can only intersect in the 32 $\ztwo$ fixed point resolutions. We label these fixed points by $(a,b,c;\gO)$ where $\gO = (\go_1,\go_2,\go_3) = \{ (+++), (+--), (-+-), (--+)\}$, i.e.\ there are no $\ztwo$ fixed points with an odd number of minuses in $\gO$: Since intersections on the cover $\text{Res}(T^6_{\rm fac}/\ztwo)$ are local, it follows that all intersections with three $\tE^-$ or with one $\tE^-$ and two $\tE^+$ are always all necessarily zero. (This can be verified easily by writing out these divisors in terms of  the exceptional divisors $E$ on the cover $\text{Res}(T^6_{\rm fac}/\ztwo)$, since then independently of the value of the intersection numbers of the $E$, one can never find compatible fixed point labels on this cover.) From this one can determine the correspondence between the $\ztwo$ fixed points of $\text{Res}(T^6_{\rm fac}/\ztwo)/\zf$ and its cover $\text{Res}(T^6_{\rm fac}/\ztwo)$: 
\equ{ 
\arry{c}{
(a,b,c; +++) \sim \{ (a+1,b+1,c+1)\;;~(a,b,c) \}\;, 
\\[1ex] 
(a,b,c; +--) \sim  \{ (a+1,b,c)\;;~ (a,b+1,c+1) \}\;, 
\\[1ex] 
(a,b,c; -+-) \sim  \{ (a,b+1,c)\;;~ (a+1,b,c+1) \}\;, 
\\[1ex]  
(a,b,c; --+) \sim  \{ (a,b,c+1)\;;~ (a+1,b+1,c) \}\;. 
}
\label{eq:LocalRes_match} 
}

The intersection numbers involving these exceptional divisors on $\text{Res}(T^6_{\rm fac}/\ztwo)/\zf$ have the following properties: To begin with, in the light of the above statement at the resolution of the $\ztwo$ fixed point $(a,b,c;\gO)$ only the divisors 
\equ{
\tR_i\;, \quad 
\tD_{1,a}\;, \quad \tD_{2,b}\;, \quad \tD_{3,c}\;, \quad 
\tE_{1,bc}^{\go_1}\;, \quad \tE_{2,ac}^{\go_2}\;, \quad \tE_{3,ab}^{\go_3}
\non 
}
are relevant. For them we have the following non-vanishing triangulation-independent intersections numbers:
\equ{
\tR_3 \tE_{3,ab}^\go \tD_{1,a} = \tR_3 \tE_{3,ab}^\go \tD_{2,b} = 2\;, 
\label{Int_tRtEtD} 
}
with $\go = \pm$. As for the intersections $\tR_1\tR_2\tR_3$ and $\tD_{i,a} \tR_j \tR_k$, these intersection numbers are not equal to those on $\text{Res}(T^6_{\rm fac}/\ztwo)$.

The intersections between $\tD_{i,a}$ and $\tE_{k,ab}^\go$ depend on the triangulation chosen for this fixed point. As locally the singularities of the factorizable and non-factorizable orbifolds are the same ($\mathbb{C}^3/ \ztwo$), their resolutions correspond to one of the four triangulations of the non-compact toric diagrams depicted in figure \ref{fig:Z2_Noncompact_Toric_Diagrams}. Hence from these diagrams one can immediately read off the corresponding intersections.  Consequently, contrary to the other non-vanishing intersection numbers discussed above, these intersections are equal to those of $\text{Res}(T^6_{\rm fac}/\ztwo)$. One can also compute these intersections from the covering space $\text{Res}(T^6_{\rm fac}/\ztwo)$ using the expressions \eqref{eqn:match_D} and \eqref{eqn:match_E}. The fixed point $(a,b,c;\gO)$ corresponds to two $\ztwo$ fixed points on the cover, hence one should of course take identical triangulations at both of them when preforming the computation in this way.

As usual from these fundamental intersection numbers we can compute all other (self-)intersection numbers by exploiting the linear equivalence relations. In analogy to the factorizable blow-up \eqref{eq:Z2_Compact_Equivalences} we find
\begin{align}
 2 \tD_{1,a} ~\sim~ \tR_i - \sum_{b=1,3 \atop \go=\pm} \tE^{\go}_{3,ab} - \sum_{c=1,3 \atop \go=\pm} \tE^{\go}_{2,ac} \;,
\end{align}
and analogous equations for $\tD_{2,\beta}$ and $\tD_{3,\gamma}$. Using the matching equations \eqref{eqn:match_R}, \eqref{eqn:match_D} and \eqref{eqn:match_E} we see that the linear equivalence relations can also be deduced from the factorizable ones. Alternatively, one can compute all other (self-)intersection by using these matching relations to compute them directly from the intersections on the cover.

Next, let us discuss the Chern classes of 
$\text{Res}(T^6_{\rm fac}/\ztwo)/\zf$. There are two ways to obtain them: First, as we did in the factorizable case, we make the ansatz
\begin{align}
 c^{\rm non} = \prod_D (1+D)^{n_D} \cdot \prod_E (1+E)^{n_E} \cdot \prod_R (1+R)^{n_R} \;.
\end{align}
From the toric geometry construction of the local resolutions we know that $n_D=n_E=1$. Furthermore, $n_R$ was determined by requiring the cancellation of all $R$-divisors in $c_1$ which is achieved here by setting $n_R=-1$. The other possibility is to take $c^{\rm fac}$ as given in \eqref{eq:Z2_Total_Chern_Class} and perform the identifications \eqref{eqn:match_R}, \eqref{eqn:match_D}, and \eqref{eqn:match_E}. In either case we get the same result:  $c^{\rm non}_1 = 0$. The Euler number equals 
\equ{ 
\chi\left(\text{Res}(T^6_{\rm fac}/\ztwo)/\zf\right) = c_3^{\rm non} =48\;, 
}
as expected in agreement with \cite{Donagi:2004}.

Using these Chern classes we can also determine the topology of the inherited divisor $\tR_i$. Its first Chern class vanishes because $c_1(\tR_i) =  - \tR_i$, which gives zero when restricted to $\tR_i$, as $\tR_i^2 S= 0$ for all divisors $S$. Its Euler number, 
\equ{ 
\chi(\tR_i)= c^{\rm non}_2 \tR_i = 24\;, 
}
then tells us that we can interpret $\tR_i$ as a K3 surface. In the blow-down limit $\tR_i$ tends to the orbifold $T^4/\Intr_2$ which has 16 fixed points. From the perspective of the non-factorizable orbifold resolution these fixed points arise from the 8 (and not 16) exceptional divisors $\tE_{i,ab}^\go$, each of which intersects $\tR_i$ two times, see \eqref{Int_tRtEtD}, resulting in the 16 $\Intr_2$ fixed points on $\tR_i$.

\subsection{The non-factorizable torus approach}

An alternative description of the $T^6/\ztwo\times\zf$ orbifold is based on the non-factorizable lattice $\Gamma_{\rm non}$ obtained from $\Gamma_{\rm fac}$ and $\tau$ with an ordinary $\ztwo$ action.

The factorizable lattice $\Gamma_{\rm fac}$ is generated by the lattice vectors $e_1,\ldots, e_6$. We can choose the non-factorizable lattice $\Gamma_{\rm non}$ to be spanned by the basis vectors $e_1, e_3, e_5$ and two of three basis vectors $e_2, e_4, e_6$ augmented by $\tau$, for example 
\begin{equation}
\tilde{e}_1 = e_1\;, \quad 
\tilde{e}_2 = e_2\;,\quad 
\tilde{e}_3 = e_3\;,\quad  
\tilde{e}_4 = e_4\;,\quad 
\tilde{e}_5 = e_5 \quad\text{and}\quad 
\tilde{e}_6 = \tau\;.
\label{NonFacBasis}
\end{equation}
By making some appropriate permutations, sign flips and renormalizations of these basis vectors, we can infer that the lattice $\Gamma_{\rm non}$ is isomorphic to the root lattice of $\text{SU}(4)\times \text{SU}(2)^3$, and therefore the new torus lattice is in particular non-factorizable. Using this lattice we can define a non-factorizable torus $T^6_{\rm non} = \Real^6/\Gamma_{\rm non}$, hence by construction $T^6_{\rm non} = T^6_{\rm fac}/\zf$.

\begin{figure}
\centering 
\subfloat[$\tilde R_i$ on the non-factorizable lattice]{\label{fg:tR_NonFac}\includegraphics[width=0.32\textwidth]{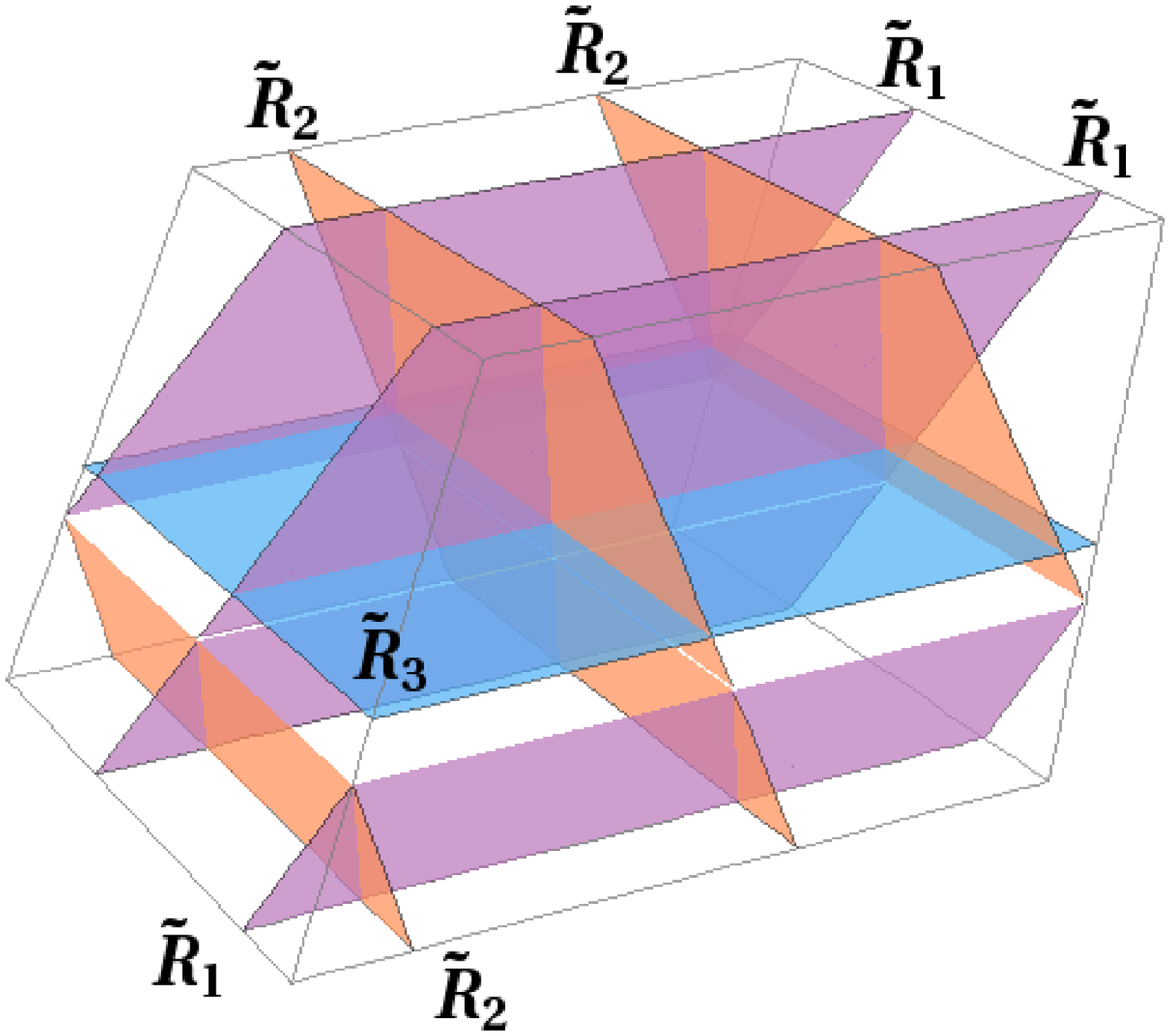}}
\subfloat[$\tilde D_i$ on the non-factorizable lattice]{\label{fg:Fixed_NonFac_D}\includegraphics[width=0.32\textwidth]{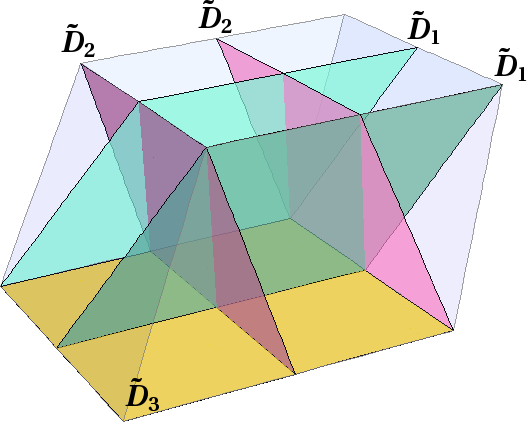}}
\subfloat[$\tilde E_i^\omega$ on the non-factorizable lattice]{\label{fg:Fixed_NonFac}\includegraphics[width=0.32\textwidth]{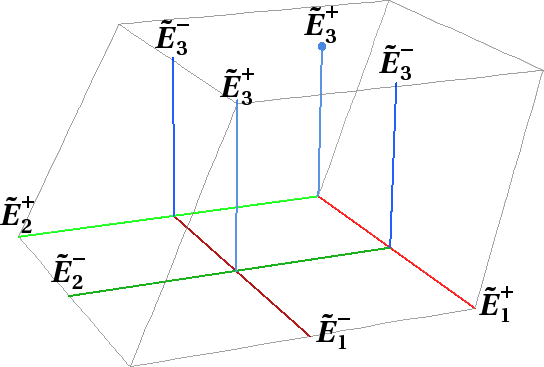}}
\caption{This figure depicts the positions of the inherited, ordinary, and exceptional divisors in the orbifold limit in the fundamental domain \eqref{NonFacFund} of the non-factorizable lattice. By comparing \eqref{eq:Nonfac_R} and \eqref{eq:Nonfac_D}, as well as by comparing (a) and (b), it is obvious that for fixed $c_i=0$ the ``sliding" divisors $\tilde R_i$ go over to the fixed $\tilde D_i$. For the exceptional divisors, it should be noted that both $\tilde E_3^+$ and $\tilde E_3^-$ consist of two branches in the fundamental domain.}
\end{figure}

Even though the lattice has changed when working on the non-factorizable torus, the definitions of the inherited divisors $\tR_i$ have not. The reason for this is that they are defined w.r.t.\ to the complex structure that survives the $\ztwo$ orbifolding. To see the consequence of this, we can define a fundamental domain of the non-factorizable torus as 
\equ{ 
T^6_{\rm non} = 
\Big\{ 
\sum_{i=1}^3 x_{2i-1}\; e_{2i-1} 
+ (x_2+y/2) \; e_2   + (x_4+y/2) \; e_4  + (y/2) \; e_6 \;;~~  
0 \leq x_a, y < 1
\Big\}, 
\label{NonFacFund} 
}
given the choice of basis vectors \eqref{NonFacBasis}. The complex coordinates used in the definition of the inherited divisors $\tR_i$ read 
\equ{
z_1 = x_1 + i \; (x_2+y/2) \;,
\qquad 
z_1 = x_3 + i \; (x_4+y/2) \;,
\qquad 
z_3 = x_5 + i \; (y/2) 
\label{Cmplx_NonFac} 
}
in this parameterization. This has the effect that with the choice of the fundamental domain defined in \eqref{NonFacFund}, the divisors $\tR_1$ and $\tR_2$ consist of two full branches related to each other by a translation of the lattice vector $\te_6=\tau$ of the non-factorizable lattice. (Generically, one of the branches is subdivided into two subbranches inside the fundamental domain). Contrary, the divisor $\tR_3$ has a single branch. This implies that on the non-factorizable torus the intersection $\tR_1\tR_2\tR_3 = 4$ rather than $1$ as it would be on the factorizable torus. This situation is depicted in figure \ref{fg:tR_NonFac}.

From the non-factorizable torus $T^6_{\rm non}$ we can obtain the orbifold $T^6_{\rm non}/\ztwo$ by the standard orbifold procedure. The $\ztwo$ fixed points are then labeled by $(a,b,c; \gO)$. Here $a,b,c = 1,3$ indicate the fixed point positions in the $e_1, e_3$, and $e_5$ directions. Furthermore, $\gO=(+++), (+--), (-+-)$ and $(--+)$ refer to the fixed points at the origin, at $e_2/2$, at $e_4/2$ and $(e_2+e_4)/2$, respectively. In figure \ref{fg:Fixed_NonFac} these four possibilities of $\gO$ are the intersection points of the lines. The lines themselves depict the positions of the exceptional divisors $\tE_{i,ab}^\go$ in blow-down.

The intersection numbers of the divisors $\tR_i$, $\tD_{i,a}$ and $E^\go_{i,bc}$ can also be computed directly on the resolution of $T^6_{\rm non}/\ztwo$. Given the fundamental domain defined in \eqref{NonFacFund}, we see that $\tR_1$ and $\tR_2$ both have four branches inside this fundamental domain, while $\tR_3$ has only two. Similarly, $\tD_1$ and $\tD_2$ have two branches and $\tD_3$ just a single one. From this we infer that 
\equ{
\tR_1 \tR_2 \tR_3 = \frac{4^2 \cdot 2}{4} = 8\;, 
\quad 
\tR_1 \tR_3 \tD_{3,c} = \frac{4^2 \cdot 1}{4} = 4\;, 
\quad 
\tR_1 \tD_{2,b} \tR_3 = \tD_{1,a} \tR_2 \tR_3 = 
\frac{4\cdot 2^2}{4} = 4\;, 
}
because the order of the orbifold group is $2^2=4$. For the intersections between $\tR$, $\tD$ and $\tE$ we note that since in blow-down $\tE$ coincides with $\Intr_2$ fixed lines, the effective orbifold order is two on these intersections. Consequently we find that 
\equ{ 
\tR_3 \tE^\go_{3,ab} \tD_{1,a} = \tR_3 \tE^\go_{3,ab} \tD_{2,b} = 
\frac{2\cdot 2}{2} = 2\;, 
\quad 
\tR_1 \tE^\go_{1,bc} \tD_{2,b} = \tR_1 \tE^\go_{1,bc} \tD_{3,c} = 
\frac{4\cdot 1}{2} = 2\;, 
}
etc. Here we have used in particular that both $\tE^\go_{1,bc} \tD_{2,b}$ and $\tE^\go_{1,bc} \tD_{3,c}$ have just a single branch in the fundamental domain \eqref{NonFacFund}. Clearly, the intersections obtained here are the same as those obtained in the previous subsection of this appendix. 

\subsection{Multiplicities on non-factorizable resolution}
\label{sc:MultiplNonFact}

In order to build a supergravity model on the non-factorizable solutions we put Abelian fluxes on the exceptional divisors, corresponding to the local shifts on the orbifold. The expansion of the internal field strength reads
\begin{align}
 \label{eq:NonFact_Flux} 
{\mathcal{F}} = \sum_r \tE_r H_r\;, 
\qquad H_r = \tV_r^I H^I \;,
\end{align}
where the sum runs over all exceptional divisors. (So that concretely we have  $\tH_{k,ab}^\omega = \tV_{k,ab}^{\omega I}H^I$.) Inserting this and the Chern class into the multiplicity operator \eqref{eq:Multiplicity_Operator_Raw}, we can integrate it over the manifold using the intersection numbers. Concretely, using the ``$E_1$'' triangulation at all 32 $\ztwo$ local resolutions, the multiplicity operator reads
\equa{ 
\tN_\text{``$E_1$''}  & = 
 \hskip4pt \sum\limits_{b,c=1,3} \sum\limits_{\omega=\pm}H_{1,bc}^{\omega}
 -\frac{1}{3}\; 
 \sum\limits_{a,c=1,3}
 \sum\limits_{\omega=\pm}
 \Big\{H_{2,ac}^{\omega}-4(H_{2,ac}^{{\omega}})^3\Big\}
- \frac 13\; 
\sum\limits_{a,b=1,3}\sum\limits_{\omega=\pm}\Big\{
 H_{3,ab}^{\omega}-4(H_{3,ab}^{{\omega}})^3
 \Big\}
\non \\[1ex] 
 & -  \sum\limits_{a,b,c=1,3} \sum\limits_\Omega H_{1,bc}^{\omega_1}
 \Big\{
 (H_{2,ac}^{{\omega_2}})^2 + (H_{3,ab}^{{\omega_3}})^2\
 \Big\}\;.
\label{eq:ex_Z2_Multiplicity_Operator_NonFact}
}
It is this operator we have used for the computation of the MSSM model the non-factorizable resolution in subsection \ref{ch:MSSM_in_Blowup}. 
As before, the expression for the multiplicity operator is very sensitive to the triangulation chosen. For example, in the symmetric triangulation we find 
\equa{ 
\label{eq:ex_Z2_Multiplicity_OperatorS_NonFact}
\tN_\text{``$S$''}  & = 
~~ \frac{1}{3} ~~\!\sum\limits_{i=1}^{3}\sum\limits_{a,b=1,3}\sum\limits_{\omega=\pm} 
\Big\{
2\tH_{i,ab}^{\omega3} + \tH_{i,ab}^\omega 
\Big\}
+\sum\limits_{a,b,c=1,3}\sum\limits_{\Omega}
\tH_{1,bc}^{\omega_1} \tH_{2,ac}^{\omega_2} \tH_{3,ab}^{\omega_3}
\\[1ex] 
 & - \frac{1}{2} \sum\limits_{a,b,c=1,3}\sum\limits_{\Omega}
 \Big\{
[\tH_{2,ac}^{\omega_2} \!+\!\tH_{3,ab}^{\omega_3}] 
(\tH_{1,bc}^{\omega_1})^2 
+ 
[\tH_{1,bc}^{\omega_1} \!+\! \tH_{3,ab}^{\omega_3}]
(\tH_{2,ac}^{\omega_2})^2
+
[\tH_{1,bc}^{\omega_1} \!+\!\tH_{2,ac}^{\omega_2}]
 (\tH_{3,ab}^{\omega_3})^2
 \Big\}\;.
\non 
}

Next we show that the values of the multiplicity operators $N_{\rm non}$ on the non-factorizable resolution and $N_{\rm fac}$ on the corresponding $\zf$-compatible factorizable resolutions are related as 
\equ{
N_{\rm fac} = 2\; N_{\rm non}. 
}
Consider the spectrum of a line bundle model with bundle vectors $V_{k,ab}^\go$, for a given triangulation of the non-factorizable resolution. Then the $\zf$-compatible triangulation resolution has the same local triangulation at the the fixed points which are identified under the $\zf$ action. Moreover, the flux identifications read:
\begin{align}
V_{k,ab} = V_{k,(a+1)(b+1)} = \tV_{k,ab}^+\;, 
\qquad 
V_{k,(a+1)b} = V_{k,a(b+1)} = \tV_{k,ab}^- \;.  
\label{eqn:match_V}
\end{align}
in terms of the fluxes on the non-factorizable resolution. Inserting this in the multiplicity operator $N_{\rm fac}$ on the corresponding $\zf$-compatible resolutions immediately gives the result. 

%
%
%
%
%
%
%
%
%
%
%
%
%
%
%
%
%
%
%
%
%
%
%
%
%
%
%
%
%
%
%
%
%
%
\setcounter{equation}{0}
\section{Linear dependence of the bundle vectors}
\label{sc:FluxQuant}

In this appendix we show that of the 48 vectors that characterize the line bundle embedding in the $\eeight$ gauge group only eight are independent up to lattice vectors due to the flux quantization conditions \eqref{FQ}. More concretely, we show that any bundle vector $V_{k,\gr\gs}$ can be expressed in terms a linear combination of the following nine vectors $V_{1,1i}, V_{2,i1}$ and $V_{3,1i}$. We call this fundamental set of bundle vectors. As three vectors, $V_{1,11},V_{2,11}$ and $V_{3,11}$, are related via \eqref{FQconstraint}, the fundamental set indeed only represents eight independent vectors.

First of all, \eqref{FQsum} with $\gs =4$ leads two expressions for $V_{k,44}$ given by
\equ{
V_{k,44} \cong - \sum_i V_{k,4i}\;
\quad\text{and}\quad
V_{k,44} \cong - \sum_i V_{k,i4}\;,
}
where the sum is over $i=1,2,3$ and remember that $\cong$ means that the vectors are equal up to $\eeight$ root lattice vectors. We can use \eqref{FQsum} again to express $V_{k,4i}$ and $V_{k,i4}$ in terms of $V_{k,ij}$ as
\equ{
V_{k,4i} \cong - \sum_i V_{k,ji}\;,
\quad\text{and}\quad
V_{k,i4} \cong - \sum_i V_{k,ij}\;. 
}
Summing either of these expressions over $i$ shows that the two expressions for $V_{k,44}$ above are in fact equal, and therefore do not lead to an additional consistency requirement. Note that this exhausts all possibilities to use the relations \eqref{FQsum}, because employing \eqref{FQsum} again will reintroduce the index 4. Next, we use \eqref{FQfixedpoints} to rewrite vectors $V_{k,ij}$ in terms of vectors that have at least one index equal to 1: 
\equ{
V_{1,ij} \cong - V_{2,1j} - V_{3,1i}\;,
\qquad 
V_{2,ij} \cong - V_{1,1j} - V_{3,i1}\;,
\qquad 
V_{3,ij} \cong - V_{1,j1} - V_{2,i1}\;. 
}
The vectors $V_{3,1i}, V_{1,1j}$ and $V_{2,i1}$ are already in the set of fundamental vectors defined above. For the other three we can use \eqref{FQfixedpoints} again to obtain 
\equ{
V_{2,1i} \cong - V_{1,1i} - V_{3,11}\;,
\qquad 
V_{3,i1} \cong - V_{1,11} - V_{2,i1}\;,
\qquad 
V_{1,i1} \cong - V_{2,11} - V_{3,1i}\;. 
}

By combining the various expressions obtained here one infers that all line bundle vectors can be related to the fundamental set. Moreover, one can show that equations \eqref{FQsum} and \eqref{FQfixedpoints} do not give further constraints on the fundamental set, because applying any of these equations to a vector of the fundamental set takes us outside this set. Hence we have shown that the flux quantization conditions imply that out of the 48 line bundle vectors only eight are independent (up to lattice vectors). 
%
%
%
%
%
%
%
%
%
%
%
%
%
%
%
%
%
%
%
%
%
%
%
%
%
%
%
%
%
%
%
%
%
%
\setcounter{equation}{0}
\section{Derivation of modular invariance for freely acting Wilson lines}
\label{ch:Derivation_MI_Free_WL}

In this appendix we investigate the necessary requirements to be imposed in order to be allowed to mod out the freely acting \zf~element in the CFT construction of orbifolds. Similar to the case of the non-free shifts and Wilson lines, we obtain these requirements by considering the string partition functions. Remember that $\gth_{i}$ leaves the $i^{th}$ torus invariant, and that the torus lattice is taken to be spanned by $1$ and $i$. The boundary conditions for the bosonic and fermionic left- and right-movers are
\begin{equation}
\begin{array}{l@{}l@{~}l@{~}}
X^{i}&(z + 1) 								& = e^{-2\pi i (p_{1} \varphi_{1}+p_{2} \varphi_{2})} X^{i}(z) - 2\pi [4 (m_{i}+i n_{i})+2(a_{i} + i b_{i})+i c]\;,\\
X^{i}&(z - \tau)							& = e^{-2\pi i (\overline{p}_{1} \varphi_{1}+\overline{p}_{2} \varphi_{2})} X^{i}(z) - 2\pi [4 (\overline{m}_{i}+i \overline{n}_{i})+2(\overline{a}_{i} + i \overline{b}_{i})+i \overline{c}]\;,\\
\psi^{i}&(z + 1)							& = e^{2 \pi i (p_{1} \varphi_{1}+p_{2} \varphi_{2} + \frac{s}{2})}\psi^{i}(z)\;,\\
\psi^{i}&(z  - \tau)							& = e^{2 \pi i (\overline{p}_{1} \varphi_{1}+\overline{p}_{2} \varphi_{2} + \frac{\overline{s}}{2})}\psi^{i}(z)\;,\\
\lambda_{a}^{I}&(z + 1)					& = e^{2 \pi i (\frac{s_{a}}{2} + p_{1} V_{1}+p_{2} V_{2} + a_{i} W_{2 i -1} + b_{i} W_{2 i} + c W)} \lambda_{a}^{I}(z)\;,\\
\lambda_{a}^{I}&(z  - \tau)					& = e^{2 \pi i (\frac{\overline{s}_{a}}{2} + \overline{p}_{1} V_{1}+\overline{p}_{2} V_{2} + \overline{a}_{i} W_{2 i -1} + \overline{b}_{i} W_{2 i} + \overline{c} W)} \lambda_{a}^{I}(z)\;,\\
\end{array}
\end{equation}
where $p_{1},\overline{p}_{1},p_{2},\overline{p}_{2},a_{i},\overline{a}_{i},b_{i},\overline{b}_{i},c,\overline{c} \in \{0,1\}$, $m_{i},\overline{m}_{i},n_{i},\overline{n}_{i} \in \mathbbm{Z}$, and $i \in \{1,2,3\}$, where the twist and shift vectors $\varphi_i$ and $V_i$, and the Wilson lines $W_i$ and $W$ are defined as in subsection \ref{sc:OrbiCFT}, as well as the lattice translations. 
It is convenient to summarize the boundary conditions into the so-called local orbifold shifts
\equ{ 
\varphi_g = p_1 \varphi_1 + p_2 \varphi_2,\quad V_g= p_1 v_1 + p_2 V_2 + a_i W_{2i-1} + b_i W_{2i} +c W\;.
\label{LocalShifts}
}

The factors of the partition function due to the worldsheet fermions is subdivided into the partition function of the left-movers $Z_{\lambda}$ and right-movers~$Z_{\psi}$: 
\begin{subequations}
\label{eq:Fermionic_Partition_Functions}
\begin{eqnarray}
Z_{\lambda^{(a)}}\db{V_{g}^{(a)}}{V_{\overline{g}}^{(a)}} & \defi & \frac{1}{2}\sum\limits_{s_{a},\overline{s_{a}}} e^{-2 \pi i \frac{\overline{s_{a}}}{2} e_{8} \cdot V_{g}^{(a)}}~ \theta\db{\frac{1-s_{a}}{2}e_{8}-V_{g}^{(a)}}{\frac{1-\overline{s_{a}}}{2}e_{8}-V_{\overline{g}}^{(a)}} \cdot \eta^{-8}\;,\\
Z_{\psi}\db{\varphi_{g}}{\varphi_{\overline{g}}} & \defi & \frac{1}{2}\sum\limits_{s}e^{-\pi i s \overline{s}} \theta\db{\frac{1-s}{2}e_{4}-\varphi_{g}}{\frac{1-\overline{s}}{2}e_{4}-\varphi_{\overline{g}}} \cdot \eta^{-4}\;,
\end{eqnarray}
\end{subequations}
where $s,\bar{s}$ label the spin structure  and the index $a$ labels the first and the second $\text{E}_{8}$. Here we have defined $e_{k} \equiv (1,1,\ldots,1)$ to be a vector having $k$ times the entry $1$, and introduced theta functions:
\begin{equation}
\theta\db{\alpha}{\overline{\alpha}}(z|\tau) \defi 
\sum \limits_{n \in \Intr^N} e^{2 \pi i \tau \frac{1}{2}(n-\alpha)^{2}} e^{2 \pi i (z-\overline{\alpha})\cdot(n-\alpha)}\;, 
\end{equation}
with vector valued characteristics $\alpha,\overline{\alpha} \in \mathbb{R}^{N}$, and the Dedekind $\eta$-function
\begin{equation}
\eta (\tau) \defi e^{2 \pi i \tau \frac{1}{24}}\prod\limits_{n \leq 1} (1-e^{2 \pi i \tau n)}\;. 
\end{equation}
Here $\gt$ refers to the modular parameter of the worldsheet torus. 

The partition function for the bosons is more involved, as we need to distinguish whether the $\theta_{i}$'s act trivially or not, and we can have different structures depending on the fact that a sector lives in ten, six or four dimensions. Defining 
\begin{equation}
\begin{array}{l@{}l@{~}l@{~}}
Z_{0} & \defi & \frac{1}{\sqrt{\tau_{2}}} \frac{1}{|\eta(\tau)|^{2}}\;,
\\[2ex] 
Z_{X}\db{\alpha}{\overline{\alpha}} & \defi & Z_{0} \sum\limits_{m,\overline{m}} e^{-\frac{2 \pi}{2 \alpha^{\prime} \tau_{2}} |(4 \overline{m} + \overline{\alpha})-(4 m + \alpha) (\tau_{1}+i\tau_{2})|^{2}}
\end{array}
\end{equation}
we can write
\begin{subequations}
\label{eq:Bosonic_Partition_Functions}
\begin{eqnarray}
Z_{X}^{10D} &=& Z_{0}^{2} \prod\limits_{i=1}^{3}Z_{X}\db{2 a_{i}}{2 \overline{a}_{i}} Z_{X}\db{2 b_{i} + c}{2 \overline{b}_{i} + \overline{c}}\;,\label{eq:Bosonic_Partition_Function1}\\
Z_{X}^{6D}  &=& Z_{0}^{2} Z_{X}\db{2 a_{i}}{2 \overline{a}_{i}} Z_{X}\db{2 b_{i}+c}{2 \overline{b}_{i}+\overline{c}} \left( \left|\eta(\tau) \theta \db{\frac{1-p_{i}}{2}}{\frac{1-\overline{p}}{2}}^{-1}\right|^{2}\right)^{2}\;,\label{eq:Bosonic_Partition_Function2}\\
Z_{X}^{4D} &=& Z_{0}^{2} \prod\limits_{i=1}^{3} \left|\eta(\tau)\theta \db{\frac{1-p_{i}}{2}}{\frac{1-\overline{p}_{i}}{2}}^{-1} \right|^{2}, \text{ with } p_{3} \defi (p_{1} + p_{2}) \mmod 2\;.\label{eq:Bosonic_Partition_Function3}
\end{eqnarray}
\end{subequations}
The complete partition function is then given by
\begin{equation}
\label{eq:Z2_Complete_Partition_Function}
Z = \sum \limits_{[g,\bg]=0} \, 
e^{2 \pi i \frac{1}{2} \big\{ 
 \varphi_{g} \cdot (\varphi_{\overline{g}}-e_{4})
- V_{g} \cdot (V_{\overline{g}}-e_{16})
\big\}}\, 
Z_{\psi}^* Z_{X} Z_{\lambda^{(1)}} Z_{\lambda^{(2)}}\;,
\end{equation}
where the sum runs over the $10$D, the $6$D, and the $4$D sector of the theory.

\begin{figure}
\centering
\includegraphics[width=.7\textwidth]{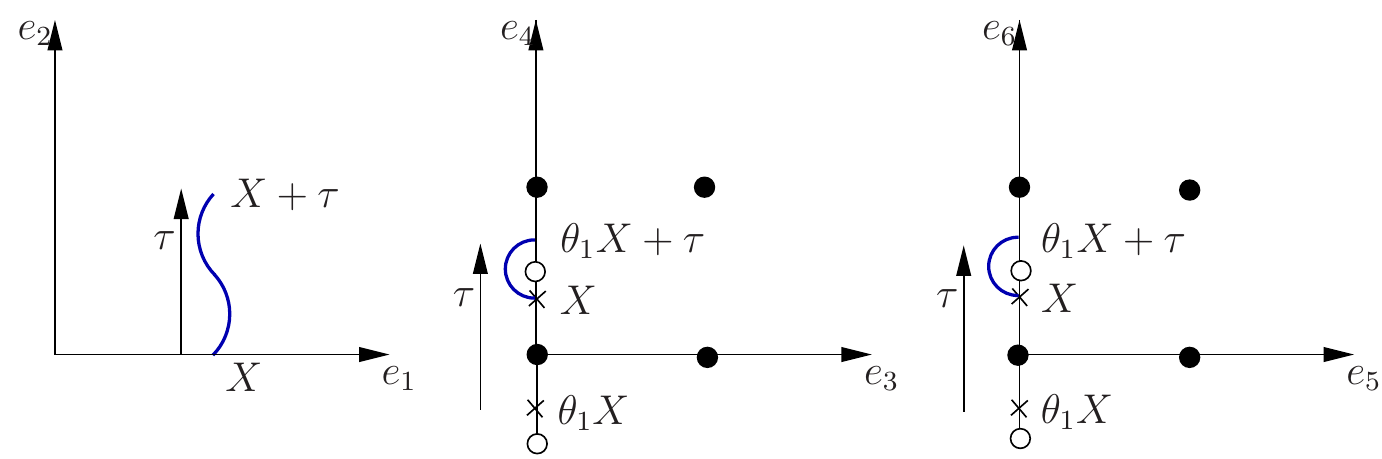}
\caption
{\label{Fig:Extra6d} In this figure we show a string in the $\tau\,\theta_1$ sector, i.e.\ a string closed up to the combined action of $\tau$ and the orbifold rotation $\theta_1$.}
\end{figure}

We remark that besides the standard untwisted and $\theta_i$-twisted sectors, there are also $\tau$- and mixed $\tau\,\theta_i$-twisted sectors. Given that $\tau$ and all mixed $\tau\,\theta_i$ operators are freely acting, these sectors are characterized by strings that are closed after a translation in the internal space. Thus, they have non-zero winding number and their minimal mass is larger than the string scale, given that all the $T^6$ radii are chosen to be larger than the string length. For these reasons these sectors have no role in the phenomenology of the obtained models.
The new $\tau$-twisted sector lives in ten dimendions, while the $\tau\,\theta_i$-twisted states are localized in two of the three tori, given that the action of the mixed operators $\tau\,\theta_i$ is free only in the $i$-th torus, and has fixed points in the other tori. In figure~\ref{Fig:Extra6d} we show the localization of the $\tau\,\theta_1$ sector, by explicitely building a string that is closed upon the  $\tau\,\theta_1$ action.

Let us now investigate modular invariance. Using standard properties of theta functions with vector valued characteristics, we find
\begin{equation}
\label{eq:Theta_Eta_Modular_Invariance}
\begin{array}{l@{}l@{~}l@{~}}
\theta \db{\alpha}{\overline{\alpha}}(\tau+2)~\eta^{-d}(\tau+2)
&=& e^{-2 \pi i (\alpha^{2}+\frac{d}{12})}\theta\db{\alpha}{\overline{\alpha}+2\alpha}(\tau) ~\eta^{-d}(\tau)\;.
\end{array}
\end{equation}
This can only be modular invariant if $\alpha$ contains at most order 2 elements. As \zf~is an order $4$ element, the partition function returns to itself up to phases only for $\tau \rightarrow \tau + 4$ if $c \neq 0$.
Applying \eqref{eq:Theta_Eta_Modular_Invariance} to the fermionic partition functions \eqref{eq:Fermionic_Partition_Functions}, we get 
\begin{subequations}
\label{eq:Fermionic_Modular_Invariances}
\begin{eqnarray}
Z_{\lambda_{a}}(\tau+2) & = & e^{-2 \pi i(V_{g}^{2}+ \frac{2}{3})} Z_{\lambda_{a}}(\tau)\;, \label{eq:Fermionic_Modular_Invariance1}
\\[2ex] 
Z_{\psi}(\tau+2) & = & e^{-2 \pi i(\varphi_g^{2}+ \frac{1}{3})} Z_{\psi}(\tau)\;. \label{eq:Fermionic_Modular_Invariance2} 
\end{eqnarray}
\end{subequations}
Equation \eqref{eq:Fermionic_Modular_Invariance1} follows when we assume that 
\begin{equation}
e_{8}\cdot V_{1} \equiv e_{8} \cdot V_{2} \equiv e_{8} \cdot W_{i} \equiv 2 e_{8} \cdot W \equiv (1-s_{a})/2 e_{8}^{2}\equiv 0\;,
\end{equation}
and \eqref{eq:Fermionic_Modular_Invariance2} when 
\begin{equation}
e_{4}\cdot \varphi_{1} \equiv e_{4} \cdot \varphi_{2} \equiv (1-s)/2 e_{4}^{2}\equiv 0\;,
\end{equation}
where $a\equiv b$ if $a=b\,{\rm mod}\, 1$.  These conditions are automatically fulfilled in any well defined $T^6/\ztwo$ orbifold, since they incorporate the requirement that the orbifold action has order 2 also on the 10D spinors of the original torus (equation~\ref{eq:Fermionic_Modular_Invariance2}), and on the spinorial representations of each $\text{E}_8$   (equation~\ref{eq:Fermionic_Modular_Invariance1}).

Now we can give the constraints on the complete partition function $Z$ to be modular invariant. 
In case $c=0$ we obtain a modular invariance condition for $\tau \rightarrow \tau + 2$, from substituting \eqref{eq:Fermionic_Modular_Invariances} into \eqref{eq:Z2_Complete_Partition_Function}, and using the notation of local shift \eqref{LocalShifts} we may write 
\begin{equation}
\label{eq:4D_6D_Modulari_Invariance_App}
V_g^2 \equiv \varphi_g^{2}\;, 
\end{equation}
when $c=0$. 
However when $c \neq 0$ we have an order 4 element and therefore we  find for $\tau \rightarrow \tau + 4$ the novel condition for $c\neq 0$: 
\begin{equation}
2V_g^2 \equiv 2\varphi_g^{2}\;. 
\end{equation}
As soon as equation~\ref{eq:4D_6D_Modulari_Invariance_App} is satisfied, and the identifications \eqref{eq:Const_V_W}  are imposed, the novel condition can be reduces to $2W^2\equiv 0$.

However, modular invariance is not the only quadratic constraint on shift vectors: In \cite{Ploger:2007} it was argued that  one has to impose the condition
\begin{equation}
\frac{\bN}{2} (V{}_g \cdot V{}_{\bar g}- \varphi_g\cdot  \varphi_{\bar g})\equiv 0\;, 
\end{equation}
where $\bN$ is the order of action of the projecting space group element ${\bar g}$. This condition can be understood from the pre-factor phase 
\(
\exp \big\{ 2 \pi i \frac{1}{2} 
(\varphi_{g} \cdot \varphi_{\overline{g}}
- V_{g} \cdot V_{\overline{g}})
\big\}
\)
in front of the complete partition function \eqref{eq:Z2_Complete_Partition_Function}: In order for the summing over the various $\overline{p}_{1},\overline{p}_{2},\overline{a}_{i},\overline{b}_{i},\overline{c} \in \{0,1\}$, $\overline{m}_{i},\overline{n}_{i} \in \mathbbm{Z}$, and $i \in \{1,2,3\}$, lead to the well-defined projections of appropriate order, the above condition needs to be fulfilled. As the notation in \eqref{eq:Z2_Complete_Partition_Function} indicates, this condition only needs to be enforced when the space group elements $g$ and $\bg$ commute. The resulting conditions on the shifts $V_i$ and Wilson lines $W_p$ and $W$ that need to be fulfilled when combining the various consistency requirements are summarized in subsection \ref{sc:OrbiCFT}. 

\newpage

%
%
%
%
%
%
%
%
%
%
%
%
%
%
%
%
%
%
%
%
%
%
%
%
%
%
%
%
%
%
%
%
%
%
\bibliographystyle{paper}
{\small
\bibliography{bibliography}
}
\end{document}

%% file: texdefs.tex
\renewcommand{\d}{\mathrm{d}}

\newcommand{\TR}{\textsc{tr}}

\DeclareMathSymbol{\mg}{\mathrel}{symbols}{"1D}

\newcommand{\ga}{\alpha}
\newcommand{\gb}{\beta}
\renewcommand{\gg}{\gamma}
\newcommand{\gd}{\delta}

\newcommand{\gve}{\varepsilon}
\newcommand{\gf}{\phi}

\newcommand{\gx}{\xi}
\newcommand{\gm}{\mu}

\newcommand{\gk}{\kappa}
\newcommand{\gl}{\lambda}
\newcommand{\gr}{\rho}
\newcommand{\gth}{\theta}

\newcommand{\gs}{\sigma}
\newcommand{\gt}{\tau}
\newcommand{\go}{\omega}

\newcommand{\gp}{\pi}

\newcommand{\gG}{\Gamma}

\newcommand{\gL}{\Lambda}

\newcommand{\gO}{\Omega}

\newcommand{\cA}{{\cal A}}
\newcommand{\cB}{{\cal B}}

\newcommand{\cF}{{\cal F}}

\newcommand{\cO}{{\cal O}}
\newcommand{\cP}{{\cal P}}

\newcommand{\cR}{{\cal R}}

\newcommand{\te}{{\tilde e}}

\newcommand{\tD}{{\tilde D}}
\newcommand{\tE}{{\tilde E}}

\newcommand{\tH}{{\tilde H}}

\newcommand{\tN}{{\tilde N}}

\newcommand{\tR}{{\tilde R}}

\newcommand{\tV}{{\tilde V}}

\newcommand{\tr}{\text{tr}}

\newcommand{\Id}{\text{\small 1}\hspace{-3.5pt}\text{1}}

\newcommand{\ra}{\rightarrow}

\newcommand{\der}{\partial}

\newcommand{\inv}{^{-1}}
\newcommand{\dsp}{\displaystyle}

\newcommand{\beq}{\begin{equation}}
\newcommand{\eeq}{\end{equation}}
\newcommand{\barr}{\begin{array}}
\newcommand{\earr}{\end{array}}
\newcommand{\equ}[1]{\begin{gather} #1 \end{gather}}
\newcommand{\equa}[1]{\begin{align} #1 \end{align}}

\newcommand{\tabu}[2]{\begin{tabular}{#1} #2 \end{tabular}}
\newcommand{\arry}[2]{\begin{array}{#1} #2 \end{array}}

\newcommand{\pmtrx}[1]{\begin{pmatrix} #1 \end{pmatrix}}

\newcommand{\non}{\nonumber}
\newcounter{oldcounter}

\newcommand{\bder}{\bar\partial}
\newcommand{\bg}{{\bar g}}

\newcommand{\bx}{{\bar x}}

\newcommand{\bz}{{\bar z}}

\newcommand{\bN}{{\bar N}}

\newcommand{\bgG}{{\bar\Gamma}}

\newcommand{\bcA}{{\bar{\cal A}}}

\newcommand{\tgvf}{{\tilde \varphi}}

\newcommand{\Intr}{\mathbbm{Z}}
\newcommand{\Cplx}{\mathbbm{C}}
\newcommand{\Real}{\mathbbm{R}}

\newcommand{\CP}{\mathbbm{CP}}

\newcommand{\ba}[2]{\[\begin{array}{#2}\label{#1}}
\newcommand{\ea}{\end{array}\]}
\newcommand{\be}{\begin{equation}}
\newcommand{\ee}{\end{equation}}
\newcommand{\bea}{\begin{eqnarray}}
\newcommand{\eea}{\end{eqnarray}}

\newcommand{\rep}[1]{\mathbf{#1}}
\newcommand{\crep}[1]{\overline{\rep{#1}}}

\newcommand{\ztwo}[0]{\ensuremath{\mathbbm{Z}_{2} \times \mathbbm{Z}_{2}} }

\newcommand{\eeight}[0]{\ensuremath{\text{E}_{8} \!\times\! \text{E}_{8}} }

\newcommand{\db}[2]{\ensuremath{\!\!\left[\begin{array}{c}#1\\#2\end{array}\right]}}
\newcommand{\defi}[0]{:=}
\newcommand{\mmod}[0]{\text{ mod }}
\newcommand{\zf}[0]{\ensuremath{\mathbbm{Z}_{2,\text{free}}}}

%% file: paper.bbl
\providecommand{\href}[2]{#2}\begingroup\raggedright\begin{thebibliography}{10}

\bibitem{Candelas:1985}
P.~Candelas, G.~T. Horowitz, A.~Strominger, and E.~Witten ``{Vacuum
  Configurations for Superstrings}'' {\em Nucl. Phys.} {\bf B258} (1985)
46--74.

\bibitem{Dixon:1985}
L.~J. Dixon, J.~A. Harvey, C.~Vafa, and E.~Witten ``{Strings on Orbifolds}''
  {\em Nucl. Phys.} {\bf B261} (1985)
678--686.

\bibitem{Dixon:1986}
L.~J. Dixon, J.~A. Harvey, C.~Vafa, and E.~Witten ``{Strings on Orbifolds. 2}''
  {\em Nucl. Phys.} {\bf B274} (1986)
285--314.

\bibitem{Donaldson:1985}
S.~Donalson ``Anti-self-dual {Y}ang--{M}ills connections over complex algebraic
  surfaces and stable vector bundles'' {\em Proc. Londan Math. Soc.} {\bf 50}
  (1985) 1--26.

\bibitem{Uhlenbeck:1986}
K.~Uhlenbeck and S.~Yau ``{On the existence of Hermitian-Yang-Mills connections
  in stable vector bundles}'' {\em Commun. Pure Appl. Math.} {\bf 39} (1986)
  257--293.

\bibitem{Andreas:1999ty}
B.~Andreas, G.~Curio, and A.~Klemm ``{Towards the standard model spectrum from
  elliptic Calabi- Yau}'' {\em Int. J. Mod. Phys.} {\bf A19} (2004) 1987
\href{http://www.arXiv.org/abs/hep-th/9903052}{[{\tt hep-th/9903052}]}.

\bibitem{Donagi:2000zs}
R.~Donagi, B.~A. Ovrut, T.~Pantev, and D.~Waldram ``{Standard-model bundles}''
  {\em Adv. Theor. Math. Phys.} {\bf 5} (2002) 563--615
\href{http://www.arXiv.org/abs/math/0008010}{[{\tt math/0008010}]}.

\bibitem{Candelas:1985en}
P.~Candelas, G.~T. Horowitz, A.~Strominger, and E.~Witten ``{Vacuum
  Configurations for Superstrings}'' {\em Nucl. Phys.} {\bf B258} (1985)
46--74.

\bibitem{Witten:1985xc}
E.~Witten ``{Symmetry Breaking Patterns in Superstring Models}'' {\em Nucl.
  Phys.} {\bf B258} (1985)
75.

\bibitem{Donagi:2000fw}
R.~Donagi, B.~A. Ovrut, T.~Pantev, and D.~Waldram ``{Spectral involutions on
  rational elliptic surfaces}'' {\em Adv. Theor. Math. Phys.} {\bf 5} (2002)
  499--561
\href{http://www.arXiv.org/abs/math/0008011}{[{\tt math/0008011}]}.

\bibitem{Donagi:2000zf}
R.~Donagi, B.~A. Ovrut, T.~Pantev, and D.~Waldram ``{Standard-model bundles on
  non-simply connected Calabi-Yau threefolds}'' {\em JHEP} {\bf 08} (2001) 053
\href{http://www.arXiv.org/abs/hep-th/0008008}{[{\tt hep-th/0008008}]}.

\bibitem{Braun:2005}
V.~Braun, Y.-H. He, B.~A. Ovrut, and T.~Pantev ``{The exact MSSM spectrum from
  string theory}'' {\em JHEP} {\bf 05} (2006) 043
\href{http://www.arXiv.org/abs/hep-th/0512177}{[{\tt hep-th/0512177}]}.

\bibitem{Bouchard:2005}
V.~Bouchard and R.~Donagi ``{An SU(5) heterotic standard model}'' {\em Phys.
  Lett.} {\bf B633} (2006) 783--791
\href{http://www.arXiv.org/abs/hep-th/0512149}{[{\tt hep-th/0512149}]}.

\bibitem{Strominger:1986uh}
A.~Strominger ``{Superstrings with Torsion}'' {\em Nucl. Phys.} {\bf B274}
  (1986)
253.

\bibitem{Dijkstra:2004cc}
T.~P.~T. Dijkstra, L.~R. Huiszoon, and A.~N. Schellekens ``{Supersymmetric
  Standard Model Spectra from RCFT orientifolds}'' {\em Nucl. Phys.} {\bf B710}
  (2005) 3--57
\href{http://www.arXiv.org/abs/hep-th/0411129}{[{\tt hep-th/0411129}]}.

\bibitem{Dijkstra:2004ym}
T.~P.~T. Dijkstra, L.~R. Huiszoon, and A.~N. Schellekens ``{Chiral
  supersymmetric standard model spectra from orientifolds of Gepner models}''
  {\em Phys. Lett.} {\bf B609} (2005) 408--417
\href{http://www.arXiv.org/abs/hep-th/0403196}{[{\tt hep-th/0403196}]}.

\bibitem{GatoRivera:2009yt}
B.~Gato-Rivera and A.~N. Schellekens ``{Heterotic Weight Lifting}'' {\em Nucl.
  Phys.} {\bf B828} (2010) 375--389
\href{http://www.arXiv.org/abs/0910.1526}{[{\tt 0910.1526}]}.

\bibitem{GatoRivera:2010gv}
B.~Gato-Rivera and A.~N. Schellekens ``{Asymmetric Gepner Models (Revisited)}''
\href{http://www.arXiv.org/abs/1003.6075}{[{\tt 1003.6075}]}.

\bibitem{Faraggi:1989ka}
A.~E. Faraggi, D.~V. Nanopoulos, and K.-j. Yuan ``{A Standard Like Model in the
  4D Free Fermionic String Formulation}'' {\em Nucl. Phys.} {\bf B335} (1990)
347.

\bibitem{Kiritsis:2008ry}
E.~Kiritsis, B.~Schellekens, and M.~Tsulaia ``{Discriminating MSSM families in
  (free-field) Gepner Orientifolds}'' {\em JHEP} {\bf 10} (2008) 106
\href{http://www.arXiv.org/abs/0809.0083}{[{\tt 0809.0083}]}.

\bibitem{Ibanez:1986}
L.~E. Ibanez, H.~P. Nilles, and F.~Quevedo ``{Orbifolds and Wilson Lines}''
  {\em Phys. Lett.} {\bf B187} (1987)
25--32.

\bibitem{Ibanez:1987sn}
L.~E. Ibanez, J.~E. Kim, H.~P. Nilles, and F.~Quevedo ``{Orbifold
  Compactifications with Three Families of SU(3) x SU(2) x U(1)$^{\text{n}}$}''
  {\em Phys. Lett.} {\bf B191} (1987)
282--286.

\bibitem{Ibanez:1987pj}
L.~E. Ibanez, J.~Mas, H.-P. Nilles, and F.~Quevedo ``{Heterotic Strings in
  Symmetric and Asymmetric Orbifold Backgrounds}'' {\em Nucl. Phys.} {\bf B301}
  (1988)
157.

\bibitem{Buchmuller:2005}
W.~Buchm\"uller, K.~Hamaguchi, O.~Lebedev, and M.~Ratz ``{Supersymmetric standard
  model from the heterotic string}'' {\em Phys. Rev. Lett.} {\bf 96} (2006)
  121602
\href{http://www.arXiv.org/abs/hep-ph/0511035}{[{\tt hep-ph/0511035}]}.

\bibitem{Buchmuller:2006}
W.~Buchm\"uller, K.~Hamaguchi, O.~Lebedev, and M.~Ratz ``{Supersymmetric standard
  model from the heterotic string. II}'' {\em Nucl. Phys.} {\bf B785} (2007)
  149--209
\href{http://www.arXiv.org/abs/hep-th/0606187}{[{\tt hep-th/0606187}]}.

\bibitem{Lebedev:2006kn}
O.~Lebedev {\em et al.} ``{A mini-landscape of exact MSSM spectra in heterotic
  orbifolds}'' {\em Phys. Lett.} {\bf B645} (2007) 88--94
\href{http://www.arXiv.org/abs/hep-th/0611095}{[{\tt hep-th/0611095}]}.

\bibitem{Lebedev:2008}
O.~Lebedev, H.~P. Nilles, S.~Ramos-S\'anchez, M.~Ratz, and P.~K.~S. Vaudrevange
  ``{Heterotic mini-landscape (II): completing the search for MSSM vacua in a
  Z6 orbifold}'' {\em Phys. Lett.} {\bf B668} (2008) 331--335
\href{http://www.arXiv.org/abs/0807.4384}{[{\tt 0807.4384}]}.

\bibitem{SGN:2009}
S.~Groot~Nibbelink, J.~Held, F.~R\"uhle, M.~Trapletti, and P.~K.~S. Vaudrevange
  ``{Heterotic Z6-II MSSM Orbifolds in Blowup}'' {\em JHEP} {\bf 03} (2009) 005
\href{http://www.arXiv.org/abs/0901.3059}{[{\tt 0901.3059}]}.

\bibitem{Witten:1996mz}
E.~Witten ``{Strong Coupling Expansion Of Calabi-Yau Compactification}'' {\em
  Nucl. Phys.} {\bf B471} (1996) 135--158
\href{http://www.arXiv.org/abs/hep-th/9602070}{[{\tt hep-th/9602070}]}.

\bibitem{Hebecker:2003we}
A.~Hebecker ``{Grand unification in the projective plane}'' {\em JHEP} {\bf 01}
  (2004) 047
\href{http://www.arXiv.org/abs/hep-ph/0309313}{[{\tt hep-ph/0309313}]}.

\bibitem{Hebecker:2004ce}
A.~Hebecker and M.~Trapletti ``{Gauge unification in highly anisotropic string
  compactifications}'' {\em Nucl. Phys.} {\bf B713} (2005) 173--203
\href{http://www.arXiv.org/abs/hep-th/0411131}{[{\tt hep-th/0411131}]}.

\bibitem{Donagi:2008}
R.~Donagi and K.~Wendland ``{On orbifolds and free fermion constructions}''
\href{http://www.arXiv.org/abs/0809.0330}{[{\tt 0809.0330}]}.

\bibitem{Blaszczyk:2009in}
M.~Blaszczyk {\em et al.} ``{A Z2xZ2 standard model}'' {\em Phys. Lett.} {\bf
  B683} (2010) 340--348
\href{http://www.arXiv.org/abs/0911.4905}{[{\tt 0911.4905}]}.

\bibitem{Kappl:2010inprep}
R.~Kappl, B.~Petersen, M.~Ratz, R.~Schieren and P.~Vaudrevange in preparation 2010

\bibitem{Blumenhagen:2005pm}
R.~Blumenhagen, G.~Honecker, and T.~Weigand ``Supersymmetric (non-)abelian
  bundles in the type {I} and {SO(32)} heterotic string'' {\em JHEP} {\bf 08}
  (2005) 009
\href{http://www.arXiv.org/abs/hep-th/0507041}{[{\tt hep-th/0507041}]}.

\bibitem{Blumenhagen:2005}
R.~Blumenhagen, G.~Honecker, and T.~Weigand ``{Loop-corrected compactifications
  of the heterotic string with line bundles}'' {\em JHEP} {\bf 06} (2005) 020
\href{http://www.arXiv.org/abs/hep-th/0504232}{[{\tt hep-th/0504232}]}.

\bibitem{Anderson:2010tc}
L.~B. Anderson, J.~Gray, and B.~Ovrut ``{Yukawa Textures From Heterotic
  Stability Walls}''
\href{http://www.arXiv.org/abs/1001.2317}{[{\tt 1001.2317}]}.

\bibitem{Ploger:2007}
F.~Pl\"oger, S.~Ramos-S\'anchez, M.~Ratz, and P.~K.~S. Vaudrevange ``{Mirage
  Torsion}'' {\em JHEP} {\bf 04} (2007) 063
\href{http://www.arXiv.org/abs/hep-th/0702176}{[{\tt hep-th/0702176}]}.

\bibitem{GrootNibbelink:2007ew}
S.~Groot~Nibbelink, H.~P. Nilles, and M.~Trapletti ``{Multiple anomalous U(1)s
  in heterotic blow-ups}'' {\em Phys. Lett.} {\bf B652} (2007) 124--127
\href{http://www.arXiv.org/abs/hep-th/0703211}{[{\tt hep-th/0703211}]}.

\bibitem{Held:2009}
J.~Held ``{Resolving the Singularities of Compact Heterotic Orbifolds}''
  Master's thesis Ruprecht-Karls Universit{\"a}t Heidelberg Germany 2009.
\newblock available at the Institutsbiliothek Theoretische Physik Heidelberg.

\bibitem{Lust:2006}
D.~L\"ust, S.~Reffert, E.~Scheidegger, and S.~Stieberger ``{Resolved toroidal
  orbifolds and their orientifolds}'' {\em Adv. Theor. Math. Phys.} {\bf 12}
  (2008) 67--183
\href{http://www.arXiv.org/abs/hep-th/0609014}{[{\tt hep-th/0609014}]}.

\bibitem{Reffert:2006}
S.~Reffert ``{Toroidal orbifolds: Resolutions, orientifolds and applications in
  string phenomenology}''
\href{http://www.arXiv.org/abs/hep-th/0609040}{[{\tt hep-th/0609040}]}.

\bibitem{SGN:2007}
S.~Groot~Nibbelink, M.~Trapletti, and M.~Walter ``{Resolutions of $C^n/Z_n$
  Orbifolds, their U(1) Bundles, and Applications to String Model Building}''
  {\em JHEP} {\bf 03} (2007) 035
\href{http://www.arXiv.org/abs/hep-th/0701227}{[{\tt hep-th/0701227}]}.

\bibitem{SGN:2007xq}
S.~Groot~Nibbelink ``{Blowups of Heterotic Orbifolds using Toric Geometry}''
\href{http://www.arXiv.org/abs/0708.1875}{[{\tt 0708.1875}]}.

\bibitem{SGN:2008}
S.~Groot~Nibbelink, T.-W. Ha, and M.~Trapletti ``{Toric Resolutions of
  Heterotic Orbifolds}'' {\em Phys. Rev.} {\bf D77} (2008) 026002
\href{http://www.arXiv.org/abs/0707.1597}{[{\tt 0707.1597}]}.

\bibitem{Vafa:1994}
C.~Vafa and E.~Witten ``{On orbifolds with discrete torsion}'' {\em J. Geom.
  Phys.} {\bf 15} (1995) 189--214
\href{http://www.arXiv.org/abs/hep-th/9409188}{[{\tt hep-th/9409188}]}.

\bibitem{Fulton}
W.~Fulton {\em {Introduction to toric varieties}}.
\newblock Annals of mathematics studies ; 131, The William H. Roever lectures
  in geometry. Princeton Univ. Pr. 1997.

\bibitem{Nakahara}
M.~Nakahara {\em {Geometry, topology and physics}}.
\newblock Taylor \& Francis New York [u.a.] 2003.

\bibitem{Griffiths}
P.~A. Griffiths and J.~Harris {\em {Principles of algebraic geometry}}.
\newblock Wiley New York [u.a.] 1994.

\bibitem{Aspinwall:1993}
P.~S. Aspinwall, B.~R. Greene, and D.~R. Morrison ``{Measuring small distances
  in N=2 sigma models}'' {\em Nucl. Phys.} {\bf B420} (1994) 184--242
\href{http://www.arXiv.org/abs/hep-th/9311042}{[{\tt hep-th/9311042}]}.

\bibitem{Witten:1993}
E.~Witten ``{Phases of N = 2 theories in two dimensions}'' {\em Nucl. Phys.}
  {\bf B403} (1993) 159--222
\href{http://www.arXiv.org/abs/hep-th/9301042}{[{\tt hep-th/9301042}]}.

\bibitem{last}
M. Fischer, M. Ratz and P.K.S. Vaudrevange in preparation

\bibitem{Faraggi:2006}
A.~E. Faraggi, S.~F\"orste, and C.~Timirgaziu ``{Z(2) x Z(2) heterotic orbifold
  models of non factorisable six dimensional toroidal manifolds}'' {\em JHEP}
  {\bf 08} (2006) 057
\href{http://www.arXiv.org/abs/hep-th/0605117}{[{\tt hep-th/0605117}]}.

\bibitem{Hull:1997kk}
C.~M. Hull ``{Actions for (2,1) sigma models and strings}'' {\em Nucl. Phys.}
  {\bf B509} (1998) 252--272
\href{http://www.arXiv.org/abs/hep-th/9702067}{[{\tt hep-th/9702067}]}.

\bibitem{Honecker:2006qz}
G.~Honecker and M.~Trapletti ``{Merging heterotic orbifolds and K3
  compactifications with line bundles}'' {\em JHEP} {\bf 01} (2007) 051
\href{http://www.arXiv.org/abs/hep-th/0612030}{[{\tt hep-th/0612030}]}.

\bibitem{Nibbelink:2008tv}
S.~Groot~Nibbelink, D.~Klevers, F.~Pl\"oger, M.~Trapletti, and P.~K.~S.
  Vaudrevange ``{Compact heterotic orbifolds in blow-up}'' {\em JHEP} {\bf 04}
  (2008) 060
\href{http://www.arXiv.org/abs/0802.2809}{[{\tt 0802.2809}]}.

\bibitem{Denef:2005mm}
F.~Denef, M.~R. Douglas, B.~Florea, A.~Grassi, and S.~Kachru ``{Fixing all
  moduli in a simple F-theory compactification}'' {\em Adv. Theor. Math. Phys.}
  {\bf 9} (2005) 861--929
\href{http://www.arXiv.org/abs/hep-th/0503124}{[{\tt hep-th/0503124}]}.

\bibitem{Ahlfors}
L.~Ahlfors {\em {Complex Analysis}}.
\newblock McGraw-Hill Book Company 1953.

\bibitem{Koblitz}
N.~Koblitz {\em {Introduction to Elliptic Curves and Modular Forms}}.
\newblock {Graduate texts in mathematics; 97}. Cambridge University Press 1993.

\bibitem{Donagi:2004}
R.~Donagi and A.~E. Faraggi ``{On the number of chiral generations in Z(2) x
  Z(2) orbifolds}'' {\em Nucl. Phys.} {\bf B694} (2004) 187--205
\href{http://www.arXiv.org/abs/hep-th/0403272}{[{\tt hep-th/0403272}]}.

\end{thebibliography}\endgroup
